\documentclass[3p]{elsarticle}

\usepackage{lineno,hyperref}
\usepackage{graphicx}
\usepackage{subfig}
\usepackage{array}

\usepackage{amsmath}
\usepackage{algorithmic}
\usepackage{algorithm}
\usepackage{doi}
\usepackage{url}

\usepackage{setspace}
\usepackage{graphicx,lipsum}
\usepackage{caption}
\usepackage{tikz}

\newcolumntype{P}[1]{>{\centering\arraybackslash}p{#1}} 
\newcolumntype{M}[1]{>{\centering\arraybackslash}m{#1}} 
\journal{Journal of Network and Computer Applications}

\begin{document}

\begin{frontmatter}

\title{A Socio-inspired CALM Approach to Channel Assignment Performance Prediction and WMN Capacity Estimation}


\author{Srikant Manas Kala$^\dag$, Vanlin Sathya$^*$, M Pavan Kumar Reddy$^\dag$, Betty Lala$^\P$, and~Bheemarjuna~Reddy~Tamma$^\dag$}
\address{$^\dag$Indian Institute of Technology Hyderabad, India\\$^*$University of Chicago, Illinois, USA\\$^\P$Kyushu University, Fukuoka, Japan.}
\fntext[myfootnote]{Email: cs12m1012@iith.ac.in (Srikant Manas Kala), vanlin@uchicago.edu (Vanlin Sathya), cs12b1025@iith.ac.in (M Pavan Kumar Reddy), 5ES18005K@s.kyushu-u.ac.jp (Betty Lala), tbr@iith.ac.in (Bheemarjuna Reddy Tamma).}
 



  %
  %

\begin{abstract}
A significant amount of research literature is dedicated to interference mitigation in Wireless Mesh Networks (WMN), with a special emphasis on designing channel allocation (CA) schemes which alleviate the impact of interference on WMN performance. But having countless CA schemes at one's disposal makes the task of choosing a suitable CA for a given WMN extremely tedious and time consuming. There is a conspicuous absence of reliable CA performance prediction metrics to assist in the selection of a high performing CA scheme for a WMN. 
Popularly used theoretical \textit{interference estimation metrics} \emph{viz.}, $CDAL_{cost}$, and $CXLS_{wt}$, have certain flaws which we discuss in this work. We also elucidate the shortcomings of \textit{Total Interference Degree} (TID) and propose a hypothesis explaining why it is not a reliable CA performance prediction tool. Besides, these metrics are unable to fulfill our ultimate objective of theoretically predicting the \textit{expected network capacity} of a CA scheme deployed in a WMN, with high confidence. In this work, we propose a new interference estimation and CA performance prediction algorithm called \textit{CALM}, which is inspired by social theory. We borrow the sociological idea of \textquotedblleft a \textit{sui generis} social reality\textquotedblright, and apply it to WMNs with significant success. To achieve this, we devise a novel \textit{Sociological Idea Borrowing Mechanism} that facilitates easy operationalization of sociological concepts in other domains. Further, we formulate 
a Mixed Integer Non-linear Programming (MINLP) optimization model to determine maximal network capacity of a WMN. Since the MINLP model does not run in polynomial time due to non-linear constraints, we design a heuristic Mixed Integer Programming (MIP) model called \textit{NETCAP} which makes use of link quality estimates generated by CALM to offer a reliable framework for network capacity prediction. We demonstrate the efficacy of CALM by evaluating its theoretical estimates against experimental data obtained through exhaustive simulations on ns-3 802.11g environment, for a comprehensive CA test-set of forty CA schemes consisting of topology preserving, graph preserving and graph disrupting CA schemes. We compare \textit{CALM} with three existing interference estimation metrics, and demonstrate that it is consistently more reliable. CALM boasts of accuracy of over 90\% in performance testing, and in stress testing too it achieves an accuracy of 88\%, while the accuracy of other metrics drops to under 75\%. 
It reduces errors in CA 
performance prediction by as much as 75\% when compared to other metrics. Finally, we validate the expected network capacity estimates generated by NETCAP, and show that they are quite accurate, deviating by as low as 6.4\% on an average when compared to experimentally recorded results in performance testing. 

\end{abstract}


\end{frontmatter}

\section{Introduction}
Multi-Radio Multi-Channel (MRMC) Wireless Mesh Networks (WMNS) are poised to serve as the backbone of present day wireless communication due to the proliferation of cheap commodity IEEE 802.11 stock hardware. MRMC WMNs are effortlessly deployed, easily scaled and reconfigured to meet dynamic requirements, and seamlessly integrated with the middle-mile fiber optic networks \cite{11Bruno}. The pervasive presence of WMNs is the direct outcome of an unprecedented surge in data rates that the IEEE 802.11 and IEEE 802.16 protocol standards guarantee. WMNs also enhance the network coverage promising last-mile solutions through multiple-hop transmissions relaying data packets between source-destination pairs across several intermediate nodes. Given its practical and commercial appeal, WMN technology can adequately cater to the vigorous needs of myriad network applications, ranging from institutional and social wireless LANs, last-mile broadband Internet access, to disaster networks. Prominent wireless technologies other than the IEEE 802.11 WLANs that stand to benefit from WMN deployments, are the IEEE 802.16 Wireless Metropolitan Area Networks (WMANs) and the next generation cellular mobile systems, including LTE-Advanced \cite{12Capone}. 

\begin{figure}[htb!] 
                \centering
                \includegraphics[width=9cm, height=8cm]{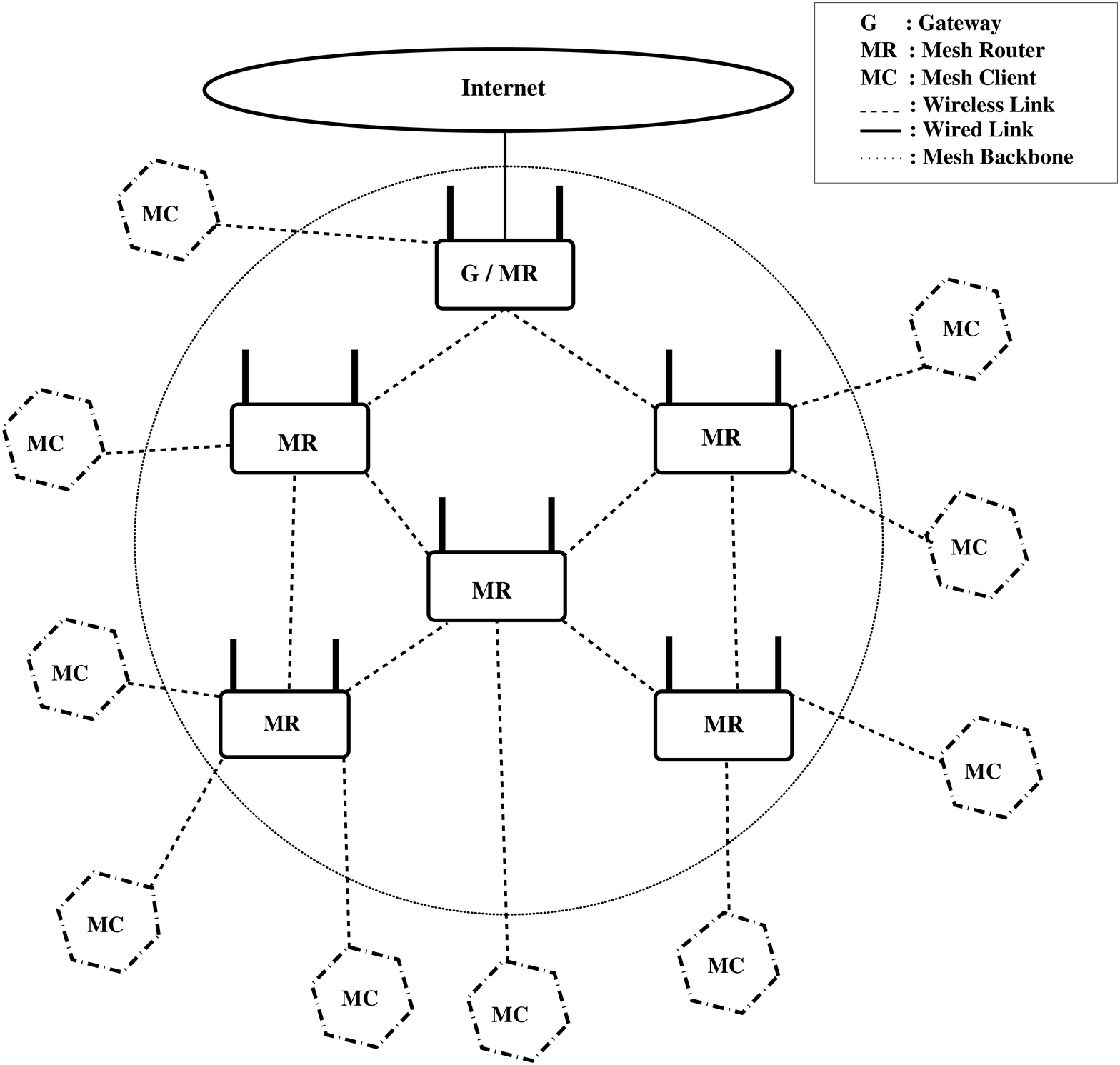}
                \caption{A Simplistic WMN Architecture.}
                \label{WMN}
        \end{figure}

A simple WMN architecture comprises of several mesh-routers (hereafter referred to as  “nodes”), which relay the data packets through multiple-hop transmissions leveraging the mesh topology which is a fundamental characteristic of WMNs. Each node in the \textit{Wireless Mesh Backbone} acts as both, a host and as a router, forwarding packets onto the next hop. The end-user devices constitute the mesh-clients which are serviced by the WMN backbone. Gateways perform the twin operations of functioning as the interface of the WMN with external networks and relaying data across the WMN like other mesh routers. WMN deployments generally employ IEEE 802.11 as the  link layer standard \cite{1IEEE}. A simple lone-gateway WMN is depicted in \mbox{Figure \ref{WMN}}, in which each node is equipped with multiple radios for inter mesh-router communication. This is the WMN model we consider in our study. In real-world applications, several WMN Gateways may be required to establish communication with external networks \cite{
10Wang}. It is for this reason that city-wide WMNs have been deployed in many cities all over the globe such as Taipei, Buffalo, Chaska, Philadelphia, New Delhi, etc.

However, the advent of WMN technology has also spawned a plethora of performance related issues in WMNs, which include the problems of channel allocation to radios, routing, scheduling etc. At the core of these issues lies the impeding factor of interference, which is caused and experienced by concurrently transmitting radios, operating on a conflicting channel and located within each other's interference range. Interference prevalent in a MRMC WMN is generally classified into \textit{External Interference}, \textit{Internal Interference} and \textit{Multipath Fading} \cite{26Koutsonikolas}. The scope of this study is limited to assessing the impact of Internal Interference which is the single most debilitating factor effecting WMN performance, and substantial research effort is focused on mitigating and restraining its adverse impact \cite{6Akyildiz}. We make use of the Protocol Interference Model for its simple mathematical representation of link conflicts. The transmission conflicts determined through the 
Interference Model are represented theoretically through the \textit{Conflict Graph} (CG) \cite{Manas}.

We now define CG and some related concepts. Let $G_{WMN}=(N,L)$ denote an arbitrary WMN consisting of $N$ nodes and $L$ links.
\begin{enumerate}[(i)]
 \item \textbf{\textit{Conflict Link}} :
Let $a\in N$, $b\in N$, such that $(a,b) \in L$, therefore $\forall (c,d) \in L$, if $c$ or $d$ have a transmission range upto or exceeding node $a$ or $b$, it is deemed to be a potential conflict link of link $(i,j)$. Conflict links are often denoted as \textit{Potential Interference Links} or \textit{Contention Links}. 
 \item \textbf{\textit{Interference Degree}}: 
For $a \in N$, $b \in N$, the Interference Degree of link $(a,b) \in L$, is the total number of links in L which are its conflict links.
\item \textbf{\textit{Total Interference Degree}} :  It is a measure of the intensity of interference prevalent in a WMN and is computed by halving the sum of \textit{Interference Degrees} of all links in the WMN.
 \item \textbf{\textit{Conflict Graph}} : $G_c=(N_c,L_c)$ is created from graph $G_{WMN}=(N,L)$ where  
\begin{itemize}
 \item 
   $N_c$ = \{ $(a,b) \in L$  $\lvert$ $(a,b)$ \text { is a transmission link} \}
 \item \{ $((a,b), (c,d)) \in  L_c$  $\lvert$ $(c,d)  $ is a conflict link of $(a,b)$ in $G_{WMN}$ \}.
\end{itemize}
\end{enumerate}

 
The important terms used in our work and the notations used to denote them are presented in Table~\ref{Note}.
In Section~\ref{3}, we elucidate the motivations behind our study and formally define the research problem. In Section~\ref{x}, we carry out a comprehensive literature review of interference estimation metrics and application of sociological concepts in WMNs. Thereafter, we review and critique our earlier works including the two CA performance predication metrics CDAL$_{cost}$ and CXLS$_{wt}$, in Section~\ref{4}. Further, we offer a fresh hypothesis explaining the shortcomings of TID in Section~\ref{5}. Next, we propose and discuss the CALM algorithm in great detail in Section~\ref{6}, followed by the NETCAP model along with a brief evaluation in Section~\ref{7}. In Section~\ref{8}, we present and analyze the results of ns-3 experiments carried out to evaluate the performance of CALM and also validate theoretical NETCAP estimates against recorded network capacity values. Finally, we conclude the findings of our study followed by a discussion on the intended future work in Section~\ref{9}.

 \begin{table} [h!]
\caption{Important notations used in the work.}
 \center 
\begin{tabular}{|M{2.5cm}|M{8.5cm}|}
\hline
 \bfseries Notation&\bfseries Elaboration\\ [0.2ex]
 \hline
\hline
WMN &Wireless Mesh Network  \\
\hline
GWMN & Grid Wireless Mesh Network \\
\hline
MRMC WMN &Multi-Radio Multi-Channel Wireless Mesh Network  \\
\hline
CG& Conflict Graph\\
\hline
MMCG & Multi-Radio Multi-Channel Conflict Graph  \\
\hline
CA& Channel Assignment\\
\hline
GPCA& Graph Preserving Channel Assignment\\
\hline
GDCA& Graph Disrupting Channel Assignment\\
\hline
TPCA& Topology Preserving Channel Assignment\\
\hline
TID&Total Interference Degree  \\
\hline
CDAL$_{cost}$&Channel Distribution Across Links Cost\\
\hline
CXLS$_{wt}$ &Cumulative X-Link-Set Weight\\
\hline
SIBM&Sociological Idea Borrowing Mechanism \\
\hline
CALM& Channel Assignment Link-Weight Metric  \\
\hline
NETCAP & Network Capacity Estimation Framework\\
\hline
MIP&Mixed Integer Programming \\
\hline
MINLP& Mixed Integer Non-Linear Programming \\
\hline
TIM& Trinity Interference Model  \\
\hline
\end{tabular}
\label{Note}
\end{table}  

\section{Research Motivation, Problem Definition and Contribution}\label{3}

\subsection{Motivations Behind the Study}
Interference alleviation is a primary network design consideration often achieved through a prudent \textit{Channel Assignment} (CA) to the radios in the WMN which is an NP-Hard problem. However, the number of CA techniques that have been proposed run into thousands, and even the most comprehensive surveys \cite{14Weisheng, 15Skalli, 2017YingSurvey} can only analyze a few. This makes the task of wading through the research literature to select a suitable CA scheme for a given WMN very tedious. But an even more effort consuming task is to implement and deploy each one of the selected CA schemes on the WMN, or a simulator, to ascertain if its design is compatible with the network architecture and requirements. In order to pick the best CA from a given set of CA schemes, a substantial amount of time, human effort, computational resources and power are consumed. Besides, given the effort intensity of this task, it can be performed only for a limited number of CA schemes, placing 
restrictions on the choice of a WMN administrator. This makes the problem of selecting a suitable CA scheme for a given WMN akin to finding a \textquoteleft needle in a stack of needles\textquoteright {}. The body of research literature that attempts to address this challenge directly is scarce. 

Further, in order to select an appropriate CA scheme for a WMN, the administrator should have at her disposal a model or framework that offers a reliable estimate of the network capacity that can be expected when a given CA is deployed in the WMN. To the best of our knowledge, there is no research work that proposes such a model or framework.

Finally, the emergence of cyber-social paradigms has triggered an increased interplay between the social realm and communication networks. The design, implementation and performance of wireless networks is being increasingly dictated by the demands of the social-world or social-reality. This social-reality consists of objective components which operate at the societal level, and subjective components operating at the level of social-actor or human agency \cite{Giddens}. Owing to these temporal and abstract subjective components, an accurate quantification and mathematical modeling of the social-realm is not possible. Social theory has found limited application in wireless networks as there has been an excessive reliance on quantitative concepts. The abstract sociological ideas are largely overlooked due to the absence of a proper mechanism that could facilitate a methodical and logical borrowing of sociological concepts to the domain of wireless networks. If a versatile and robust method to implement 
sociological concepts in wireless networks is designed, the immense potential of social-theory can be harnessed to augment network performance.

We now formally state the research problem.
\subsection{Problem Definition}
Let $G_{WMN}=(V,E)$ denote an arbitrary MRMC WMN consisting of $n$ nodes, where $V$ represents the set of all nodes and $E$ represents the set of all wireless links present in the WMN. Each node $i$ is equipped with a random number of identical radios $R_i$, which can communicate on channel set $Ch_i$ assigned from the set of available orthogonal channels $Ch$. 

An accurate theoretical estimate of the endemic interference should be devised, to predict with high confidence, the expected performance of CA schemes in $G_{WMN}$. This metric should aid in the selection of suitable CA schemes for $G_{WMN}$ from the given set of CA schemes. The design of this metric should be inspired by social theory and to facilitate the application of sociological ideas in WMNs, an idea borrowing mechanism should be devised. Finally, a theoretical framework based on Mixed Integer Programming (MIP) should be designed using the theoretical values of the metric, to predict the network capacity that a CA scheme can offer upon deployment in $G_{WMN}$.

\subsection{Research Contributions}
Our work constitutes of \textit{four major contributions}. First, we devise a generic \textit{Sociological Idea Borrowing Mechanism} (SIBM) to assist in the application of qualitative sociological concepts in wireless networks. The proposed SIBM aims to draw a functional equivalence between the components of the source-system and target-system, which makes it fairly versatile in application. 
Second, we apply and operationalize the Durkheimian concepts of social-reality in the estimation of network capacity. Third, we design an interference estimation heuristic named \textit{Channel Assignment Link-Weight Metric} (CALM) whose algorithmic design is inspired by concepts borrowed from social theory. CALM is a conflict-graph-free approach and offers both, link-quality estimates and CA performance estimates. In addition, the proposed CALM algorithm is able to accurately indicate whether a CA is suited for the given WMN, regardless of whether the CA is topology preserving or graph preserving. Fourth, we devise a Mixed Integer Non-linear Programming (MINLP) Model to determine maximal network capacity of a WMN. Further, we relax the constraints of the MINLP model to create a MIP optimization heuristic called NETCAP that uses the estimates generated by CALM to predict the network capacity for a given CA.

This work also includes several minor contributions. We propose an inductive hypothesis explaining the failure of TID as a reliable estimate of CA performance. We design and use a Generic Channel Assignment Algorithm (GCAA) which is topology preserving. We highlight the shortcomings of two interference estimation metrics \emph{viz.,} CDAL$_{cost}$ and CXLS$_{wt}$. We evaluate CALM and NETCAP on an IEEE 802.11g WMN environment simulated on ns-3 by considering a large CA test-set of forty CA schemes. We carry out a two-fold testing exercise, which includes performance testing to assess the reliability of CALM and NETCAP and stress testing to evaluate their resilience to an undesirable input of WMN topology disrupting CA schemes. The NETCAP estimates are also validated against results published in prior works. 




\section{Related Research Work} \label{x}

Multi-hop transmission in a WMN are severely impaired by interference resulting in heightened data packet loss and high link-layer latency, thereby degrading the end-to-end throughput \cite{3Saadawi}. Estimation of interference in a wireless network is an NP-Hard problem. Likewise, the various approaches to mitigate the impact of interference, \emph{e.g.,} its alignment and cancellation, routing and scheduling techniques, CA mechanisms etc., are all NP-Hard problems \cite{NPcomplete, Alignment}. These varying methods of interference alleviation are employed at different OSI layers. For example, at the Physical layer interference alignment and cancellation is attempted through techniques such as \textit{soft interference cancellation} \cite{Cancel}. Similarly, routing algorithms are employed at the Network layer, while CA schemes operate in the Link layer. \textit{Channel Assignment} (CA) is a prudent allocation of frequencies from the available IEEE802.11 spectrum to the radios in a WMN, so that the 
transmission conflicts in 
the network are minimized. It is the most popular approach to interference alleviation, due to the simplicity of CA design and implementation through heuristic approaches, smooth deployment on WMN NICs, and easy experimental verifications. 

The focus of interference mitigation studies is almost invariably on the \textit{network capacity} or end-to-end throughput of a WMN. In the classic work \cite{GuptaKumar}, authors show that the transmission conflicts restrict the maximum possible throughput of each node in a WMN of $n$ arbitrarily placed nodes, to $\Theta(1/\sqrt{n \log{} n})$. Similarly, an inverse relationship between the wireless network capacity and the intensity of interference prevalent in the network is depicted through the proposed \textit{signal to interference plus noise ratio} model \cite{Capacity}. The debilitating effect of endemic interference over multi-hop communications is demonstrated in \cite{Impact} through an upper-bound on the maximum achievable throughput, given the specific locations of wireless nodes and a priori information of traffic load. Thus, network capacity becomes a direct measure of the adverse impact of interference as the two are inversely correlated. We focus on CA schemes in our work for two reasons. 
First, it is feasible to design a metric for a CA scheme that generates an estimate of its \textit{interference mitigation potential}. Second, the inverse correlation between the magnitude of the interference estimate and the network capacity of the WMN upon deployment of the CA scheme can be experimentally substantiated, and serves as a yardstick for assessing the accuracy of the estimate. 

\subsection{Interference Estimation in WMNs}

We now discuss a few metrics that have been used in prior research works to generate an estimate of interference.
The conventional measure of intensity of interference in a WMN is TID \cite{TID1}, as it accounts for every \textit{potential} transmission conflict represented in the CG. The magnitude of TID estimate is considered to be an indicator of the severity of endemic interference and is popularly employed to design Interference Aware CAs. The guiding principle of such CA design is to achieve the lowest possible TID value after CA deployment, to arrive at an efficient CA \cite{Arunabha}. CGs for MRMC WMNs are called \textit{multi-radio multi-channel conflict graphs} (MMCG) \cite{Manas}, and are indispensable  in the design of TID based interference-aware CAs. Several CA schemes have been designed using MMCGs with the sole objective to reduce  transmission conflicts in the network \cite{14Weisheng, 15Skalli, 13Jorge, 16Subramanian, 17Xutao, 18Marina, 19Cao, 20Hongkun, 21Hamed, 22Ramachandran, 23Cheng, 24Aizaz, Manas2}. In addition to frequency allocation to radios, TID is used in other aspects of WMN performance 
such as resource allocation \cite{TID1}, scheduling \cite{TID2} etc. However, TID is only an approximate measure of interference and is far from being a reliable interference estimation metric. In addition calculation of TID estimate is computationally expensive. We have experimentally substantiated these arguments in prior studies \cite{Manas, Manas3, Manas4} and offer a novel hypothesis in Section~\ref{5}. 
%
Another metric is the \textit{Link Layer Queue Length} (LLQL) which gives an indirect estimate of interference based on the premise that long link layer queues and network congestion are a result of interference induced packet drops and subsequent retransmissions \cite{LLQ4}. In \cite{LLQ1} authors propose a CA scheme that mitigates interference by considering the average length of the link-layer queue. Likewise, authors propose a joint CA and scheduling scheme in \cite{LLQ3} which accounts for the interfering links through their LLQLs. It is also used to model the impact of interference in implementing a fair rate control mechanism \cite{LLQ2}. Thus, LLQL is a versatile metric for interference estimation in various aspects of WMN design. However, it has certain limitations. The LLQL and network congestion in WMNs are not caused by interference alone, and may be attributed to other factors such as poor link capacity and bandwidth constraints, lack of fairness in channel distribution, routing and mobility 
issues \cite{LLQ5}. Further, the LLQL has a spatial dimension, as WMN Gateway nodes tend to experience intense congestion even though they may not suffer from high link conflicts \cite{LLQ6}. Finally and most importantly, LLQL is a temporal metric as it varies depending on the load and direction of traffic. It needs to be estimated in real-time by deploying the CA in the WMN, and its value varies based on the network traffic. This defeats the objective of our study which aims to design an interference estimation metric that can predict the CA performance, without having to exhaust time and computational resources in running experiments/simulations and which does not require real-time information about network data traffic.
\textit{Traffic Load} has also been used as an interference estimation metric in several studies \cite{4P}. A new WMN architecture named \textit{Hyacinth} and a CA design based on load-aware interference estimation is proposed in \cite{TL1}. Another interference alleviating load-aware CA scheme is proposed by authors in \cite{TL2}. Like LLQL, traffic load too is a temporal metric and can only be observed post CA deployment. Further, its implementation often requires information about the expected traffic beforehand, which renders it unsuitable for our work. The Expected Transmission Count (ETX) is another metric that assesses the link quality based on the probability of successful data packet transmissions. It has been used for maximizing network throughput \cite{ETX1} and for solving routing issues \cite{ETX2}. It also has a limited application as a CA performance prediction metric.

So, interference estimation and its impacts can be modeled in a variety of direct ways (eg. TID) and indirect ways (eg. LLQL). While direct estimation techniques take a theoretical approach to measure interference, indirect techniques rely on certain underlying parameters such as bandwidth, delay, congestion etc., which have an established relationship with interference. Often, more than one parameter is used to optimize WMN performance as demonstrated in \cite{EndRW}. Besides TID, most estimates rely on temporal data and signal characteristics, which makes the task of CA selection and performance prediction without having to deploy and observe CA performance in a WMN an impossible task. 

Prior to our works, little attention was devoted to this research problem, which left TID as the only feasible CA performance prediction metric. Given the complexity of the NP-Hard interference estimation problem, evaluating CA performance without actual deployment requires a novel view of characteristics of interference, and incorporating those characteristics adequately in innovative heuristics. We explore this problem with these considerations, and to the best of our knowledge, propose the first ever CA performance prediction metric called CDAL$_{cost}$ \cite{Manas3}. It is based on a fresh characterization of interference, some aspects of which are inspired by microeconomics theory. We follow up by designing second estimation metric called CXLS$_{wt}$ \cite{Manas4}, which outperforms both TID and CDAL$_{cost}$ in terms of accuracy. These ideas are revisited in great detail in Section~\ref{4}. The twin pioneer CA performance prediction algorithms CDAL and CXLS, sparked greater interest on this problem and 
were employed in several research studies. They were used to design a \textit{Successive Interference Cancellation Algorithm} in \cite{cite1}, but a detailed description of their functional role is not offered. In \cite{cite4} authors propose a framework for network fitness evaluation and test it against CDAL$_{cost}$. However, they employ a relaxed version of CDAL algorithm, without considering the vital \textit{probabilistic channel selection} approach, which we describe later. Both metrics form integral components of frameworks for efficient CA design through \textit{interference mitigation functions} with feedback mechanisms \cite{cite2}. CXLS$_{wt}$ is employed in theoretical assessment of the proposed \textit{Near Optimal Channel Assignment} scheme (NOCAG) which is compared to a brute-force optimal CA in \cite{cite3}. Both metrics along with TID have found use in \cite{cite5} for CA performance evaluation in rural/random WMNs. The two concepts have also contributed towards building a theoretical 
foundation 
in other problems related to interference in WMNs such as intra-flow interference mitigation, fault-tolerant network design etc. \cite{cite6, cite7}. Thus, both CDAL$_{cost}$ and CXLS$_{wt}$ are versatile metrics offering a diverse range of applications, that include interference estimation and cancellation, CA performance prediction, CA design, and network performance evaluation. This functional diversity is a result of their simple algorithmic design and ease of implementation.

However, they suffer from certain limitations which we discuss in detail in upcoming sections. In this work, we aim to predict the expected network capacity, which is a function of link quality which in turn is evaluated by parameters such as link bandwidth, bit error rate, packet drop ratio, received signal strength (RSS), etc. \cite{Link1}. These parameters are determined by three main factors, \emph{viz.,} the ambient environment including background noise, the interference owing to simultaneous transmissions on overlapping channels and the internal noise in transceiver hardware that attenuates the signal \cite{Link2}. The ambient noise and internal noise of communication hardware is beyond the scope of our work. We limit our focus to the inverse relationship between link quality or network capacity, and the internal interference in a WMN. But the primary flaw in the design of both both CDAL$_{cost}$ and CXLS$_{wt}$ is an inability to offer interference estimates at the level of wireless links. Since, 
they fail to assess link quality, they can not serve our objective of theoretically estimating network capacity. For a quantitative assessment of network performance metrics such as theoretical upper-bounds of network Throughput, a link quality estimate is necessary. We take up this problem and devise a prediction estimate based on the individual link quality. 

\subsection{Sociological Theory and Wireless Networks}
In our pursuit, we make use of ideas from sociological perspectives. Challenges in network communications are often resolved through solutions inspired by other disciplines, \emph{e.g.,} genetic theory is used to devise CA schemes \cite{Genetic}, routing algorithms are motivated by ant-colony movements \cite{Ant}, Newtonian laws of gravity inspire CA design in the form of \textit{gravitational search algorithms} \cite{GS}, and optimal resource allocation strategies offered by economics are widely used in wireless networks \cite{LMR}. In comparison, there has been a limited use of sociological theory in solving issues related to WMNs, owing to the abstract nature of ideas and non-material aspects of \textit{social facts}. We contend that among the various strands of sociological theory, the \textit{Positivist} perspective \cite{Giddens}, which emulates methods of natural sciences, is a powerhouse of ideas and concepts that can be harnessed to enhance wireless network performance. \textit{Structural Hole 
Theory} proposed by sociologist Ronald Burt \cite{SH} is the prime example of an idea taken from social theory that 
revolutionized wireless network technology, contributing significantly to paradigms such as \textit{small world networks} and \textit{scale free networks} \cite{socio}. Mobility in Opportunistic Networks benefits from the idea of \textit{social subgraphs} and several other socio-inspired networking solutions \cite{S2}. Concept of \textit{social groups} is likened to a \textit{cluster} in wireless sensor networks, and utilized to enhance energy efficiency through topological modifications \cite{S4}. A radical view of future information and communication systems equates them with social systems, implying that the communication frameworks will be an integral aspect of socialization, constantly shaping human social behavior \cite{S5}. A considerable effort is devoted to enhancing and building social groups and communities through advancements in wireless technologies. For example, the deployment of 4G driven WMNs to enhance community participation through creation of a \textit{Social Layer} which is seamlessly 
integrated with other OSI layers, is proposed in \cite{S6}. Similarly, authors in \cite{S7} envision an integration of the human social network with the \textit{Internet of Things}, creating a platform where the data on \textit{behavior} of people can be collected and analyzed. 

Still, the implementation of sociological concepts in wireless networks is restricted to concepts that are quantitative in nature such as \textit{structural holes} or \textit{swarm intelligence} \cite{SI3}. Likewise in \cite{S3}, the authors propose a mathematical mobility model motivated by social network theory, with the human need to socialize as its driving idea. The overt emphasis on quantitative perspectives is due to a belief in the possibility of mathematical construction of social-reality. But social-construction has a subjective element, and even if we set aside the unceasing debate over the nature of reality (objective or subjective), there is a general agreement that there exist multiple domains of reality \cite{Hacking}. Quantitative analysis relies on the \textit{mathematical or formal reality}, which offers objective and well-defined constructs that are testable and have a logical \textit{truth value}. Social-reality on the other hand involves subjective components and abstract constructs, the 
formalization of which through quantitative or formal methods requires reductionist techniques that may cause aspects of reality to be lost in translation \cite{SocMath}. There exists a trade-off between the ease that formal quantitative techniques offer, and the complexity involved in the comprehensive representation of subjective social-reality. Thus, excessive reliance on quantitative modeling of sociological perspectives is often at the expense of abstract constructs, and offers only a partial view of reality.

Wireless networks stand to benefit significantly more if \textit{qualitative} concepts from social theory are borrowed and applied to network paradigms, as the functions and performance of these networks are shaped by human needs and requirements. Further, communication is a social fact, guided by social norms and conventions, and when this social communication is projected onto technological frameworks, it determines their behavior and performance as well. For example, the quality and quantity of data traffic, and the peak-load in a WMN will be determined by the social groups and communities which are the end users. The data-traffic 
characteristics of a wireless network in a sub-urban residential community will be starkly different from that of a college campus, and the qualitative aspects of this difference are as important as the quantitative ones, if not more. Unfortunately, the over-emphasis on quantitative analysis and mathematical modeling of sociological concepts deprives contemporary network technologies of the opportunity to factor in the sociological aspects that ultimately control their performance. Another example of this selective application is the partial use of \textit{deviancy theory} in the domain of cyber-security, which is vital for understanding the interplay between technological advancement in communication networks and deviant behavior in society. While deviancy models and statistics enhance network security, there is an immense need for a comprehensive study of qualitative \textit{human relationships} to also understand the surge in cyber-crime rates. To meet this need, \textit{Social Network Analysis} (SNA) has 
been proposed as an interdisciplinary tool to deal with cyber-crime and criminal networks with a special focus on human social systems \cite{S8}. However, the adoption of SNA is slow and must be accelerated to meet the emerging socio-technological challenges in the cyber-security domain.

An aversion towards borrowing of qualitative and abstract sociological ideas may stem from a lack of insight and understanding of these concepts. For example, as discussed earlier, authors in \cite{S7} propose the idea of treating data collected on human behavior as data collected for any other object, by processing and analyzing it through a similar clustered platform. But they do not take into consideration that this idea originates from the Durkheimian concept of \textit{Social Fact}, wherein he states that \textquotedblleft The first and most fundamental rule is: Consider social facts as things/objects\textquotedblright \cite{D3}. The primary reason for limited applicability of ideas from social theory is the absence of a straight forward \textit{Borrow-Operationalize-Validate} (BOV) mechanism that is generally applicable to ideas from other disciplines such as genetic algorithms (Biology), resource optimization models (Microeconomics) etc. Although, we agree that a simple BOV approach to benefit from 
sociological concepts is not feasible, we attempt to simplify this meticulous process by suggesting a fresh \textit{idea borrowing mechanism} in Section~\ref{D}. While designing the proposed CALM metric, we employ this mechanism and introduce ideas from social theory to enhance its accuracy and reliability as a CA performance prediction tool. 

Another aspect is the widespread criticism of attempts by pioneers such as Parsons to create a \textit{grand theory of social reality} or \textit{unified social theory} \cite{Parsons}. Today, there is a general consensus in favor of \textit{middle range theories} (MRT) that aim at integrating theoretical foundations with empirically observed data. MRTs are more suited to developing testable hypothesis rather than a grand \textit{theory of everything} that is not falsifiable and thus, not applicable to specific real-world problems. In the potential falsifiability of a theory lies the proof of its testability. Thus, in our work we propose two testable and falsifiable corollaries for a WMN, derived from sociological concepts, that are subjected to the test of empirical evidence through rigorous ns-3 simulations.

In the next section, we review our past work and offer its comprehensive critique that forms the foundation of CALM. 

\section {Revisiting CDAL$_{cost}$, CXLS$_{wt}$ and TID} \label{4}

\subsection {TIM : A Fresh Characterization of Interference} 

We have discussed the popular three-fold classification of interference in Section~\ref{3}. Most theoretical classifications categorize interference based on the \textit{source} of conflicting transmissions, treating it essentially as a spatial notion \cite{26Koutsonikolas, Int1, Int2}. However, this spatial characterization is inadequate to be the conceptual foundation of an interference estimation heuristic. We improve the simplistic classification by proposing a fresh characterization of interference in wireless networks in \cite{Manas3}, which we call the \textit{Trinity Interference Model} (TIM). It reflects the intrinsic \textit{characteristics} central to all types of wireless interference, irrespective of their source or origin, by treating interference as a three dimensional entity, the dimensions being \textit{temporal, spatial and statistical}. The three dimensional classification of characteristics of interference proposed under TIM is elucidated below :

%
\begin{enumerate}
 \item \textbf{Spatial Characteristics} : Proximity of wireless links communicating on overlapping channels leads to link conflicts in a WMN. The spatial dimension necessitates that the transmitting radios must fall within each other's transmission/interference range and operate on non-orthogonal channels.
 \item \textbf{Temporal Characteristics} : They signify the \textit{dynamism} in the intensity of interference. The temporal variation of interference scenarios is caused by the nature, direction and volume of data traffic in a WMN. Further, the multi-hop transmissions add to the temporal dimension of transmission conflicts as they are not synchronized enough to facilitate zero link conflicts. Thus, the interference scenarios in a WMN are as much a function of time as they are of spatial proximity of conflicting links. Interference classifications and models seldom take cognizance of this aspect, and adopt a reductionist spatial view.  
 \item \textbf{Statistical Characteristics} : We introduce this dimension to interference characterization inspired by the concept of distributive efficiency in economic resource allocation theory \cite{Microeco}. It is demonstrated in \cite{Manas2} that the intensity of interference in WMNs has a correlation with fairness in distribution of channels across radios in a WMN. Thus, the statistical dimension introduces the concept of fairness and equitable distribution of channels among radios in interference characterization. 
\end{enumerate}

\subsection {CDAL$_{cost}$ : A Statistical Interference Estimation Metric}

CDAL$_{cost}$ is the interference estimate generated by the \textit{Channel Distribution Across Links} (CDAL) algorithm proposed in \cite{Manas3}. In designing CDAL, we borrow the \textit{Law of Marginal Returns} from microeconomics and apply it to the distribution of channels to radios in a WMN. Simply put, the law states that in a production process, investing more in a particular factor of production while keeping other factors of production constant, leads to a relatively smaller increase in output (marginal utility). A corollary in terms of resource allocation would be that \textit{monopoly causes distributive inefficiency}. The \textquoteleft Law of Marginal Returns\textquoteright {} has found several applications in wireless networks. It is used to design optimal resource allocation frameworks in wireless networks \cite{LMR}, ensure bandwidth allocation in sensor networks \cite{LMR2}, create efficient RFID systems \cite{LMR3}, and offer utility optimization solution to determine time slots in WLAN 
communication \cite{LMR4}. We extend the 
law to CA design in a WMN, and demonstrate that a judicious and equitable distribution of channels among radios will spawn fewer link conflicts as compared to a skewed distribution by proposing a CA scheme called \textit{Optimized Independent Set CA} in \cite{Manas2}. This correlation forms the theoretical foundation for CDAL algorithm and $CDAL_{cost}$, which are briefly explained below.


CDAL algorithm assesses a CA scheme for \textit{statistical evenness} in channel allocation. It links CA performance to a proportionate distribution of available channels among WMN radios. CDAL accepts radio-channel mapping of a CA scheme as input and determines the number of links operating on each available orthogonal channel \emph{i.e.,} \textit{Link-counts}. It then arrives at a quantitative statistical metric, by computing the \textit{standard deviation} of Link-counts. Standard Deviation gives the variation or dispersion of a set of data values from their mean. An equitable distribution of channels among radios would generate a CDAL$_{cost}$ value close to 0. Although CDAL assesses CA design on a purely quantitative basis to arrive at a statistical metric called $CDAL_{cost}$, it accounts for temporal dimension of interference through a mechanism called \textit{Probabilistic Channel Selection}, for links which can be assigned multiple channels. Thus, CDAL$_{cost}$ is a measure of dispersion from the 
desired equitable distribution, and a lower magnitude of CDAL$_{cost}$ would suggest a better expected performance of CA when deployed in a WMN. We compare CDAL$_{cost}$ with TID, and demonstrate that it is a slightly better interference estimation and CA performance prediction metric than TID in \cite{Manas3, Manas4}. Further, CDAL$_{cost}$ incurs lower computational costs when compared to TID, as the size of WMN is scaled up in terms of nodes and the number of radios.


\subsubsection{Inadequacy of CDAL}

While CDAL algorithm offers a more reliable estimation metric than TID, it suffers from certain flaws. In estimating interference, TID relies purely on the spatial dimension, overlooking the other two dimensions. CDAL$_{cost}$ does slightly better by considering the statistical and temporal dimension, but ignores the crucial spatial characteristics of interference. These flaws are rooted in the distributive efficiency theory from which its design is inspired. Ensuring an equitable distribution of resources is not necessarily optimal and may disincentivize production efficiency \cite{Microeco}. Thus, there exists a need to identify bottlenecks and strengths in an economic system, to allocate extra resources to alleviate and incentivize them, respectively. Likewise, a trade-off exists between wireless network performance and absolute distributive fairness \cite{tradeoff}, and it is vital to resolve network bottlenecks to enhance efficiency. CDAL ignores the spatial characteristics of interference, which 
reduces its 
ability to identify interference bottlenecks and account for them in its interference estimate, rendering it only marginally more accurate than TID. 

\begin{figure}[htb!]
    \centering
    \includegraphics[width=8.5cm, height=3cm]{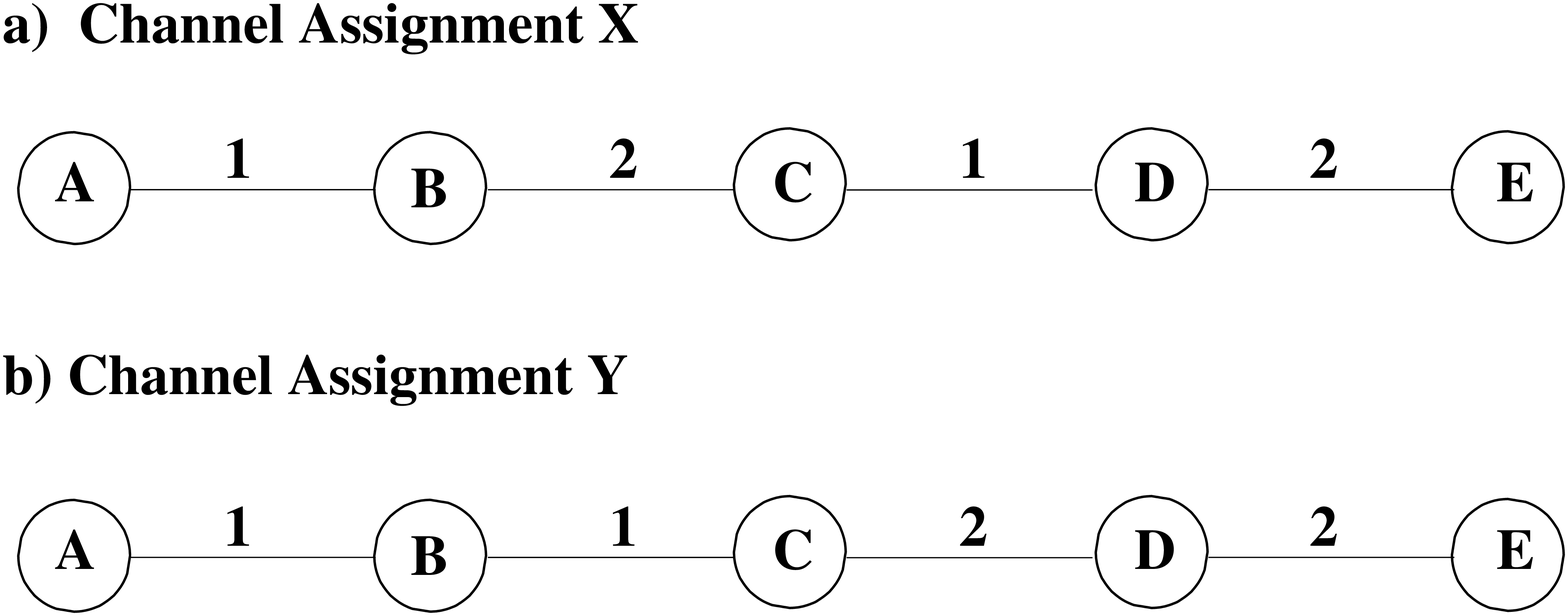}
    \caption{Limitation of CDAL estimation.}
    \label{stat}
\end{figure}

We explain this problem through Figure~\ref{stat}, which exhibits channel allocations to a $5$ node MRMC WMN by two different CA schemes, CA$_X$ and CA$_Y$ from the available set of two orthogonal channels \textit{(1, 2)}. The link quartet \textit{(AB, BC, CD, DE)} is assigned channels {1, 2, 1, 2} by CA$_X$ and {1, 1, 2, 2} by CA$_Y$. It is easily discerned that the two channel allocations are statistically alike \emph{i.e.,} link-counts of Channels 1 and 2 are similar for both CA$_X$ and CA$_Y$. However, the CA schemes differ significantly in terms of spatial distribution channels to the wireless links. Spatial characteristics of CA$_X$ ensure minimal link-conflicts  while CA$_Y$ leads to a high interference scenario marked by transmission conflicts between adjacent links (AB \& BC) and (CD \& DE). The CDAL algorithm is oblivious to the spatial dimension of interference characteristics and assigns the two CA schemes the same CDAL$_{cost}$. This leads to erroneous interference estimation and CA 
performance. 
However, 
lacunae of CDAL$_{cost}$ form the theoretical foundation for a more efficient estimation technique we review next.    

\subsection{CXLS$_{wt}$ : Spatio-Statistical Interference Estimation}
The Cumulative X-Link-Set Weight or CXLS$_{wt}$ algorithm \cite{Manas4} is a remarkable improvement over TID and CDAL approaches. It takes a comprehensive three dimensional \textit{Spatio-Statistical-Temporal} view of endemic interference. CXLS algorithm considers the wireless links in a WMN and assigns a set of links termed \textit{X-Link-Set} (XLS) a certain \textit{weight}, that represents its resilience to interference. An XLS includes $X$  number of links, where $X$ is determined by the ratio of \textit{Transmission Range : Interference Range} (TR:IR), which is considered to be an integer in CXLS design. The interference range usually exceeds the transmission range of a radio by a certain distance, over which the signal transmission is potent enough to attenuate other transmissions on an overlapping frequency, but incapable of successfully delivering data packets.  Thus $X$ determines the \textit{impact radius} of link conflicts, and the CXLS$_{wt}$ algorithm considers an XLS of $X$ consecutive links as 
the fundamental unit for interference estimation. The algorithm treats the set of all XLS in a WMN as the sample space, and processes each XLS to compute its \textit{weight} which depends upon the probabilistic selection of channels for links that can operate on multiple channels. An XLS$_{wt}$ considers all possible permutations of link-channel mappings that are determined by the CA scheme. CXLS$_{wt}$ metric is the sum of all XLS$_{wt}$ in the WMN, and is a measure of CA efficiency. A higher value of the metric signifies a diminished impact of interference. The CXLS$_{wt}$ is a holistic interference estimation metric as it incorporates the three dimensional view of interference in its design, fully capturing the characteristics of wireless interference. The impact radius of $X$ accounts for the spatial dimension, and the temporal dimension is considered by a probabilistic selection of channels for links. Statistical distributive fairness is ascertained as XLS$_{wt}$ is arrived at by mean of 
weights generated by each permutation of XLS. Owing to its excellent design, in \cite{Manas4} we successfully demonstrate its reliability as an interference estimation metric and remarkably high accuracy in CA performance prediction. It significantly outperforms TID and CDAL$_{cost}$ in both aspects.

 Despite being an ideal interference estimation metric with accuracy in CA performance prediction exceeding 90\%, CXLS$_{wt}$ has certain limitations as well. As discussed earlier, both CDAL$_{cost}$ and CXLS$_{wt}$ offer estimates for the whole CA and do not assess individual link quality which is crucial for the objective of this study. In addition, like TID, both algorithms assume a topology preserving graph after CA deployment. They will not be effective if the CA scheme is only graph preserving and employs some measure of topology control to optimize performance. Also, CXLS$_{wt}$ does not provide a measure of fairness of channel allocation as it considers XLS as the primary unit of interference estimation and not individual links. Finally, CXLS algorithm has a high computational overhead. Next, we propose a theory explaining the problems observed with the use of TID as an interference estimation metric. 
 
 \subsection{TID : An Inductive Critique} \label{5}
  \begin{figure*}
  \centering%
    \begin{tabular}{cc}
    
   \subfloat[]{\includegraphics[width=.45\linewidth]{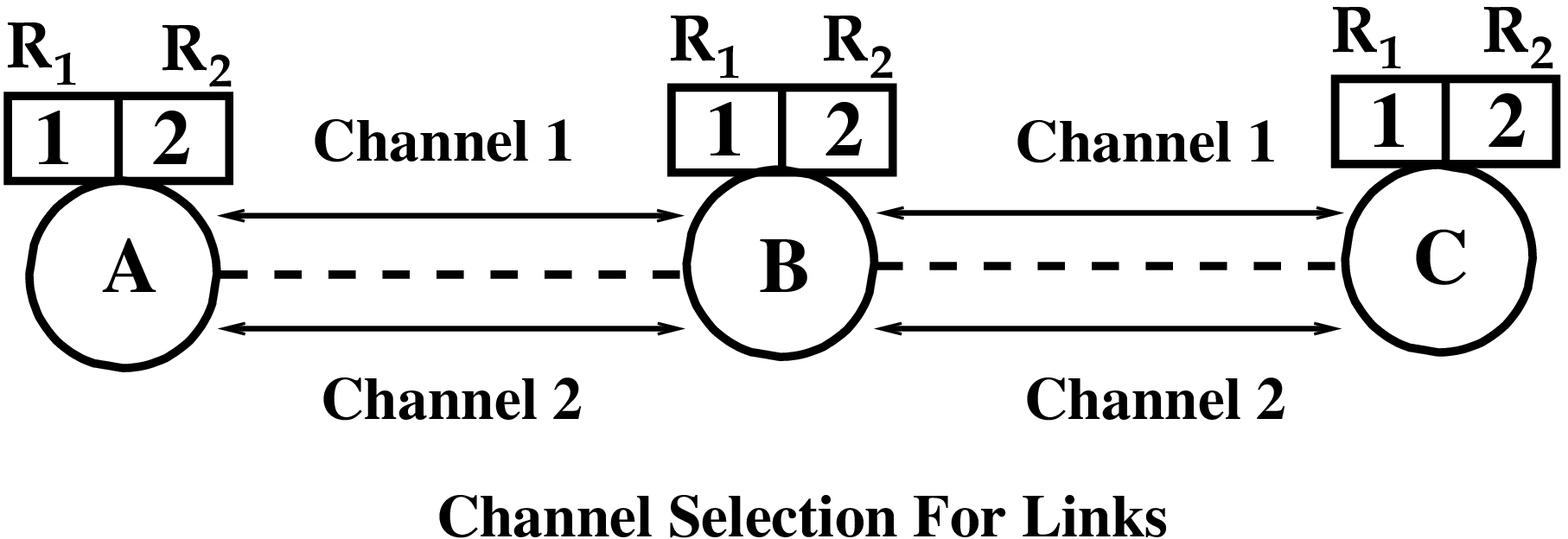}}\hfill%
    \subfloat[]{\includegraphics[width=.45\linewidth]{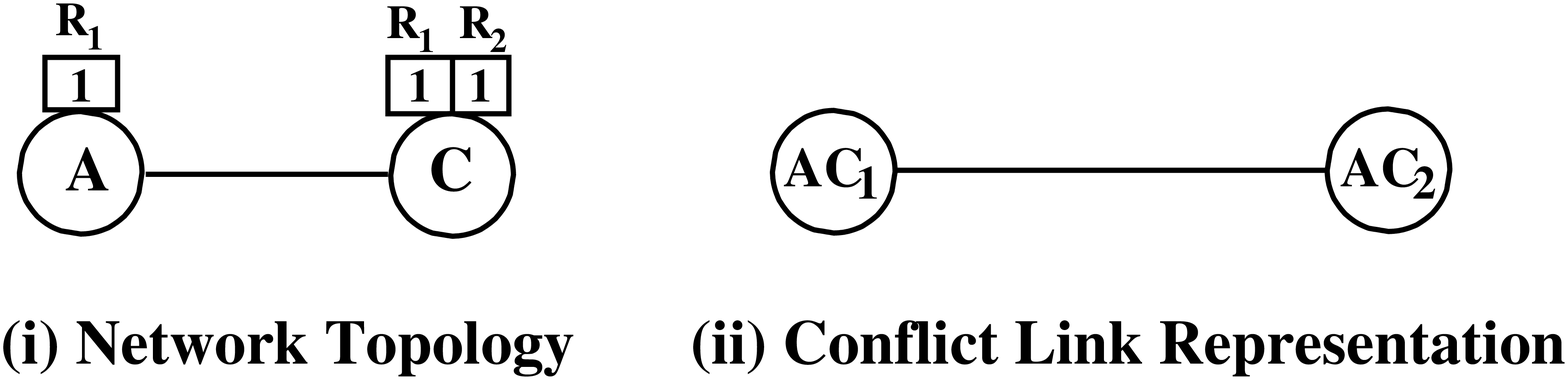}} 
   
    \end{tabular}
   \caption{Potential Conflicts and TID.} 
     \label{BADTID1}
\end{figure*}
 
 TID, as discussed, has conventionally been used as a measure of endemic interference. A strictly inverse relationship between TID and network performance is commonly accepted as a principle. However, during the course of experiments in \cite{Manas}, we arrive at anomalous results that do not conform to this generally accepted view. This being a classic case of serendipity in research \cite{serendipity}, we merely state the findings. Subsequently, in \cite{Manas2, Manas3, cite5} we rigorously compare TID with CDAL$_{cost}$ and CXLS$_{wt}$ demonstrating the new metrics to be better than TID. In \cite{cite2}, CXLS$_{wt}$ is demonstrated to create better interference mitigation function for CA schemes than TID. However, we do not offer a theory substantiating such counter-intuitive observations in these works. Now, we employ inductive research methodology \cite{inductive} to propose a hypothesis explaining this phenomena by analyzing the patterns in observed data. 
  
  \begin{figure}[htb!]
                \centering
                \includegraphics[width=7.5cm, height=6cm]{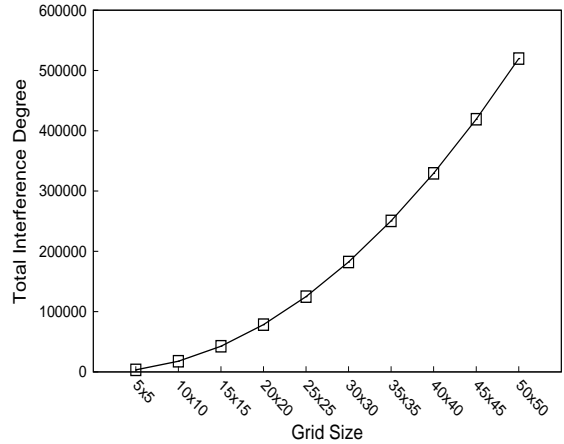}
                \caption{Scalability of TID.}
                \label{G1}
        \end{figure}

\subsubsection{Fictional Potential Conflicts}
TID is a static estimate of interference, \emph{i.e.,} it does not account for temporal interference dynamics in a WMN. It gives a measure of \textit{potential conflict links}, operational word being \textit{potential}, which implies that all possible link conflicts in the WMN ought to be considered, regardless of the likelihood of their actually being  present during active wireless communication. In the 3 node MRMC WMN depicted in Figure~\ref{BADTID1}~(a), the links $A_1B_1, A_2B_2, B_1C_1$ and, $B_2C_2$ exist at the granularity of radios, which can communicate on Channels $1,2,1$ and, $2$ respectively. Assuming parallel transmissions to maximize throughput, all possible conflicts seldom manifest during communication as MAC layer mechanisms employ \textit{multiple rendezvous protocols} to avoid such conflicts \cite{MAC2}. Assuming single channel MAC protocols, at the granularity of nodes, link pair (AB,AC) can be assigned channel pairs (1,1), (1,2), (2,2) or  (2,1) by the channel selection algorithm 
depending upon a variety of real-time 
network parameters. Here too, the probability of the worst case scenario \emph{i.e.,} channel pairs (1,1) or (2,2) being assigned is half. However, the TID estimate inevitably represents the worst case in which all possible conflicts exist concurrently. Secondly, theoretical link-conflicts spawned by \textit{radio co-location interference} \cite{Manas} as shown in Figure~\ref{BADTID1}~(b), further push up the TID estimate. But during active transmissions in wireless networks, such conflicts are successfully avoided by MAC layer. Thus, although TID estimates are scalable, they do not scale up proportionally to the WMN size. We illustrate this by plotting TID estimates of Grid WMNs (GWMN) of size $n \times n$ in Figure~\ref{G1}, where $n \in \{{5,10 \dots50}\}$, each node is equipped with two radios and all radios are assigned an identical channel. For a 100 fold increase in the number of radios, link-conflicts increase by a factor of over 10000. The tendency to always consider the \textit{worst case view} 
makes 
CG and TID a necessary theoretical tool for interference modeling and alleviation, but an inaccurate practical tool for CA performance prediction. Finally, overlooking temporal characteristics of interference also impacts accuracy of TID. We attempt to address this issue up to some extent in CDAL and CXLS algorithms through a \textit{probabilistic selection of channel for links}.  

\subsubsection{Elusive Link Disconnections}

CA design may be topology preserving, ensuring, that for all the edges in the graphical representation of the network there exist corresponding wireless links in the WMN after CA deployment. However, a CA may attempt to minimize link-conflicts by reducing redundancy while maintaining graph connectivity, which may involve some element of topology control embedding within the CA design \cite{18Marina, LLQ1, TopoNPHard, GridWSN, 1P}. There exists a trade-off between topology preservation and interference mitigation \cite{14Weisheng}, but there is also a strong inverse correlation between link disconnections and network performance \cite{Discon}. The reason TID is not a suitable indicator of CA performance prediction is that TID computation does not take into account the fact whether the CA is topology preserving (TPCA) or graph preserving (GPCA). Transmission conflicts in a WMN may reduce in two ways \emph{viz.,}  by assigning different channels to conflicting links while ensuring they are not disrupted or by 
disconnecting 
links in conflict link pairs. In the latter, link-conflicts decrease with removal/disconnection of conflicting wireless links, leading to a lower TID estimate, but this estimate is misleading as it fails to capture the impact of link disconnections on network performance. So, a lesser TID estimate does not necessarily imply an improved CA performance. Therefore, the TID estimates of a GPCA and TPCA can not be compared, as one or more edges in the original network graph may be absent in the WMN after GPCA deployment.

 \renewcommand{\algorithmicrequire}{\textbf{Input:}}
\renewcommand{\algorithmicensure}{\textbf{Output:}}
\begin{algorithm}[htb!] 
\caption{Generic Channel Assignment Algorithm.}
\label{GCAA}
\begin{algorithmic}[1]
{\fontsize{9}{10}
\REQUIRE $G = (V,E)$, $G_c = (V_c,E_c)$, $CS =\{1, 2, 3\}$ \\
$G$ : $5\times5$ Grid WMN Graph, $G_c$: E-MMCG of the WMN, $CS$ : Available Channel Set\\
For $i \in V \rightarrow Adj_{i}$ : Adjacency list, $Ch_{i}$ : Set of channels assigned to radios, $Num_i$ : Sequence number\\
For ${i, j} \in V \rightarrow Ch_{com}$ : Channels common to i \& j, $Ch_{dif}$ : Channels exclusive to either i or j\\

\ENSURE CA : Channel Assignment for $G$ \\
\line(1,0){236}
\STATE $TID_{prev} \leftarrow FindMaxTID(G_c)$. \COMMENT {Initially all radios are set to default channel 1.} 

\FOR {$Node \in G_c$}
\FOR {$Channel \in CS$}
\IF {$G$ is not disconnected}
 \STATE $Node \leftarrow Channel$
  \IF {$((TID_{wmn} < TID_{prev})$)}
  \STATE $TID_{prev} \leftarrow TID_{wmn}$
   \IF {Topology Preserving CA}
    \STATE Run Step 17 to 22
    \ELSE
     \STATE \textit{Output CA}
    \ENDIF
    \ENDIF
\ENDIF
\ENDFOR
\ENDFOR\\

\COMMENT {To Ensure Topology Preservation in $G$.}
\FOR {$i \in V$}
\FOR {$j \in Adj_{i}$} 
\IF {$((Num_i < Num_j)$  $\&\&$ $(\lvert Ch_{i} \cap Ch_{j}\lvert$ $==0))$}
\STATE 	$Ch_{j}  \leftarrow Ch_{j} + \{Ch_{com}\} - \{Ch_{dif}\}$ $\lvert \ \{(Ch_{com}\in Ch_{i})$ $\&\&$ $(Ch_{dif} \in Ch_{j})$  $\&\&$ $(TID_{wmn}$ \textit{satisfies condition in Step 6)}$\}$
\ENDIF
\ENDFOR
\ENDFOR
\STATE \textit{Output CA}
}
\end{algorithmic}
\end{algorithm}
\begin{figure*}
  \centering%
    \begin{tabular}{cc}
    
   \subfloat[Link Disconnections vs TID]{\includegraphics[width=.45\linewidth, height=5cm]{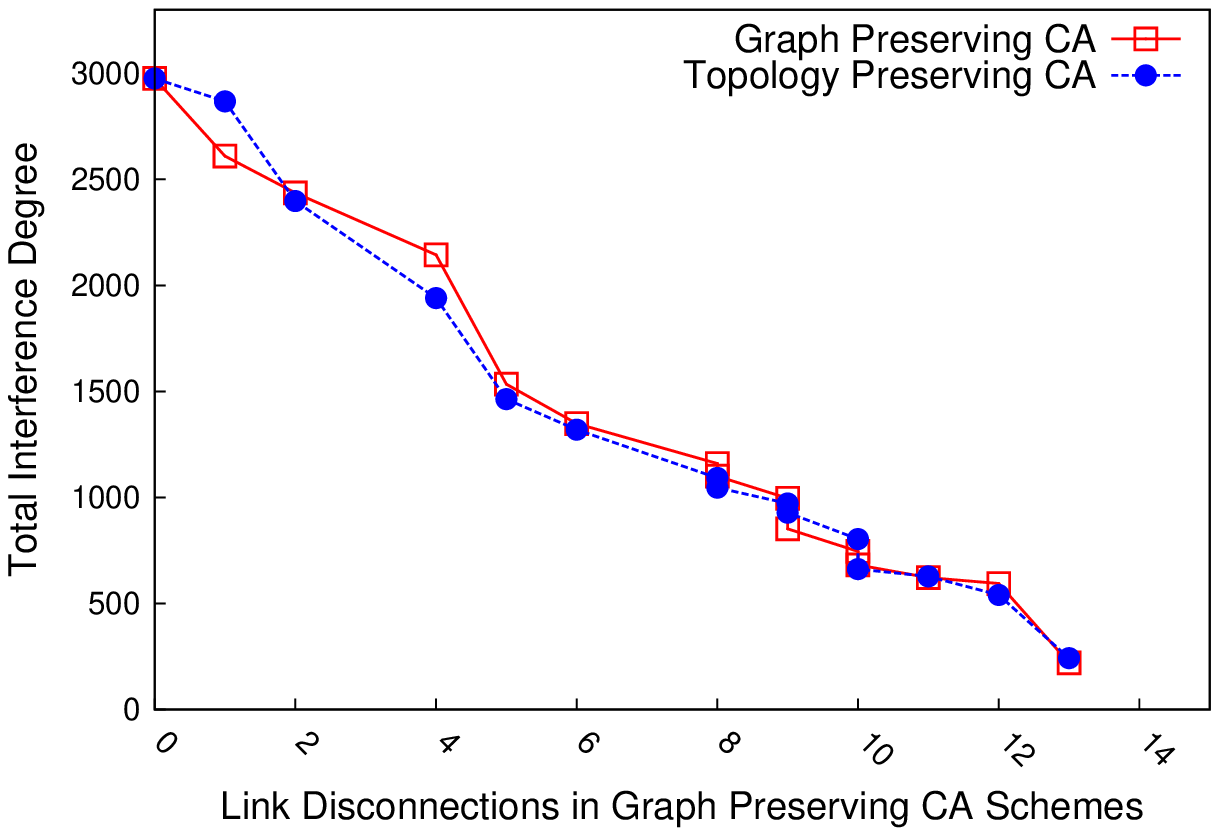}}\hfill%
    \subfloat[$5\times5$ Grid WMN]{\includegraphics[width=.35\linewidth, height=5cm]{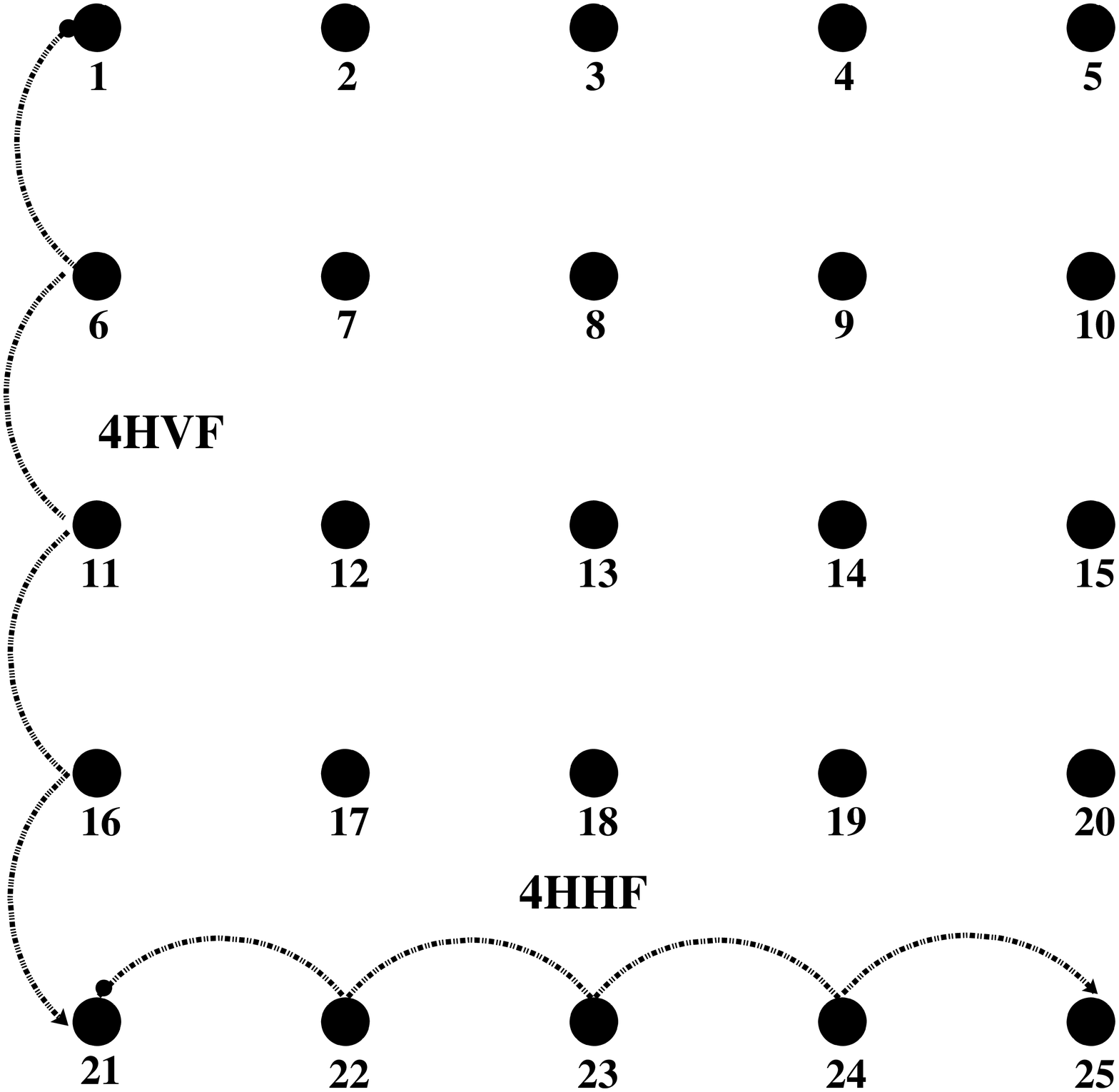}} 
   
    \end{tabular}
   \caption{Link Disconnections and TID.} 
     \label{BADTID2}
\end{figure*}

We demonstrate this by considering a $5 \times 5$ GWMN of $25$ nodes presented in Figure~\ref{BADTID2}~(b), where each node is equipped with two identical radios and there are three IEEE802.11g orthogonal channels to choose from. We generate a string of GPCAs and TPCAs through the \textit{Generic Channel Assignment Algorithm} (GCAA) proposed in Algorithm~\ref{GCAA}, which makes use of an Enhanced-MMCG to compute TID \cite{Manas}. GCAA starts from a maximal interference scenario by assigning all the radios in the GWMN an identical channel. Thereafter, it consistently attempts to mitigate interference by intelligently assigning channels to radios and outputs CAs graded in terms of improved TID values at regular intervals. For TPCAs, we ensure topology preservation through a linear time \textit{Forward Correction Algorithm}. We find that several TPCAs and GPCAs have comparable TID estimates, and while TPCAs preserve all the $40$ links in the WMN, the number of disrupted links in GPCAs generally increases with 
lowering TID 
values. These two observations are shown in Figure~\ref{BADTID2}~(a). 

Thus, TID is unable to evaluate a CA if it is not topology preserving, which is the reason why TID is not accurate. Both CDAL$_{cost}$ and CXLS$_{wt}$ also give best results with TPCAs, although CXLS$_{wt}$ is not impacted to the same extent if it encounters a disrupted link owing to its design. CDAL$_{cost}$ considers channels allocated to radios, without considering the wireless links so its algorithmic design can be said to be oblivious to any topological modification by the CA scheme. In comparison, magnitude of CXLS$_{wt}$ estimate is directly correlated to the expected CA performance and as link disconnections lower CXLS$_{wt}$ value, it indirectly accounts for their occurrence. Though GPCAs and GDCAs are not an ideal input for either CDAL$_{cost}$ or CXLS$_{wt}$, neither of them erroneously generates a misleading estimate for a CA upon encountering link disruptions. In sharp contrast, TID estimates \textquoteleft benefit\textquoteright {} from link disconnections making the CA look far more efficient 
than it actually is. Further, if the CA design is faulty and leaves one or more nodes disconnected in the WMN, it will also result in a smaller TID estimate, while the network performance will be poor. This causes an inconsistency in correlation with the experimentally observed network capacity and TID estimate is unable to predict the CA performance with high confidence as demonstrated in our prior studies. 


Having discussed and critiqued the three prominent interference estimation and CA performance prediction techniques, we now propose the CALM algorithm that aims to address the lacunae of its predecessors.

 \section {CALM : Channel Assignment Link-Weight Metric} \label{6}
From our discussion, we identify two primary challenges in designing a reliable interference estimation metric which are elucidated below  
\begin{enumerate}
\item Winnowing relevant link conflicts from the enormous set of all potential link conflicts, by taking temporal selection of channels into consideration.
\item Accounting for the interplay between CA scheme and WMN topology. 
\end{enumerate}
To address the twin challenges, we propose the \textit{Channel Assignment Link-Weight Metric} or \textit{CALM}. Unlike other theoretical metrics which assess the impact of interference on a WMN node/radio or the entire WMN, \textit{CALM} takes a \textit{link centric} view of interference. It attempts to translate the adverse impact of interference on an individual wireless link into its \textit{Link-Cost}, which in turn is used to obtain a measure of link quality we call \textit{Link-Weight}. In addition to offering an interference aware CA quality estimate, its link-centric design enables \textit{CALM} to be used in the $NETCAP$ model proposed later to predict the \textit{expected network capacity}. \textit{CALM} succeeds where the three metrics discussed in earlier sections fail. It takes a \textit{Spatio-Statistical-Temporal} view of link conflicts guided by the Trinity Interference Model. It successfully incorporates concepts from social theory in its design, which enhances its efficiency. It is capable 
of considering link-disruptions while comparing CA performance regardless of the impact of CA scheme on the topology of the WMN based on the type of CA scheme \emph{viz.,} topology preserving, graph preserving or graph disconnecting (GDCA). The design of \textit{CALM} is presented in Algorithms~\ref{CALM1}, \ref{CALM2}, and \ref{CALM3}. Before delving into the algorithm, we describe how \textit{CALM} resolves the two challenges. 

\subsection{Probabilistic Selection of Channels for Links}
      
A theoretical interference estimation metric like TID is inherently static \emph{i.e.,} it is unable to acknowledge the \textit{temporal} dimension of interference which is dynamic in nature. Selecting channels for link is a Media Access Control (MAC) function and the generally adopted approach of \textit{one flow transmission per radio} makes use of the channel that experiences minimum impact of interference and has the highest \textit{signal to interference plus noise ratio} (SINR), among the available channels \cite{PSL1}. Further, multi-channel MAC protocols facilitate \textit{parallel transmissions} which involve transmitting data over multiple channels concurrently through multiple radios installed on a node \cite{Parallel}. Communicating radios may monitor or sense the transmission channel through a control channel or make use of efficient techniques such as synchronized channel hopping \cite{MAC2}. This makes the process of channel selection as much a function of MAC protocol, as it is of the CA 
scheme. While the CA scheme assigns channels to 
radios on the link layer, MAC layer determines which channels among those will be operationalized. Thus, frequency switching may involve a dynamic interference-aware CA scheme, or a channel sensing and sequencing MAC protocol, or both \cite{MAC1}. However in MRMC WMNs, where nodes have multiple radios and several channels at their disposal to transmit and receive data, the channel that may be assigned to a link for communication is limited to the set of channels common to the communicating radios. An accurate estimation of interference is therefore predicated upon the ability of the estimation algorithm to  account for temporal variations in channel selection. As stated earlier, TID fails to account for this temporal variation.
 \begin{figure}[htb!]
                \centering 
                \includegraphics[width=7.5cm, height=1.7cm]{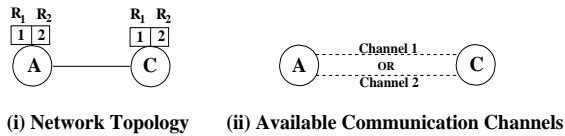}
                \caption{Link Selection for Transmission.}
                \label{links}
        \end{figure}
 
To factor in the temporal dimension of interference described in TIM, we employ the \textit{probabilistic channel selection for links}. For an easy discourse, we first consider wireless links at the granularity of radios which make use of multichannel MAC protocols. Each wireless link is assigned a channel which is determined by the CA scheme. For example, in Figure~\ref{BADTID1}~(a), the links $A_1B_1, A_2B_2, B_1C_1$ and, $B_2C_2$ are mapped to Channels $1,2,1$ and, $2$ respectively. The role of MAC layer mechanisms is limited to the selection of best radio-level link for transmission, which introduces an element of temporal randomness. We tackle this by introducing randomness in transmission link determination which is explained through the trivial 2 node MRMC WMN illustrated in Figure~\ref{links}~(i), where nodes $A$ and $C$ are equipped with two identical radios. Let $Channel_1$ and $Channel_2$ be two non-overlapping channels that are assigned to one radio each of both the nodes, and we consider links 
at the 
granularity of nodes ($Link_{AC}$). When the two nodes intend to communicate, MAC layer makes a temporal selection between one of the two  orthogonal channels for $Link_{AC}$. Since there is no way to obtain a priori knowledge of this selection, we take a probabilistic view that if the WMN in context were to actively transmit data for an infinite period, $Channel_1$ and $Channel_2$ would be equally likely to be selected for transmission on $Link_{AC}$, as the probability of being chosen would converge to $1/2$ for either of the two channels. We invoke the \textit{central limit theorem} to state that in case of availability of multiple links for transmission, each is equally likely to be chosen \cite{CLT}. This approach offers the novel concept of \textit{probabilistic link conflicts}, which represent the real-time interference dynamics in a WMN more accurately than the conventional approach of considering \textit{potential link conflicts}. \textit{CALM} incorporates temporal characteristics of interference 
by accounting for \textit{probabilistic link conflicts} instead of considering all \textit{potential link conflicts} in Algorithm-\ref{CALM3}.
 \begin{figure*}[ht!]
  \centering%
  \begin{tabular}{cc}
   \subfloat[SIBM's View of a System]{\includegraphics[width=.70\linewidth]{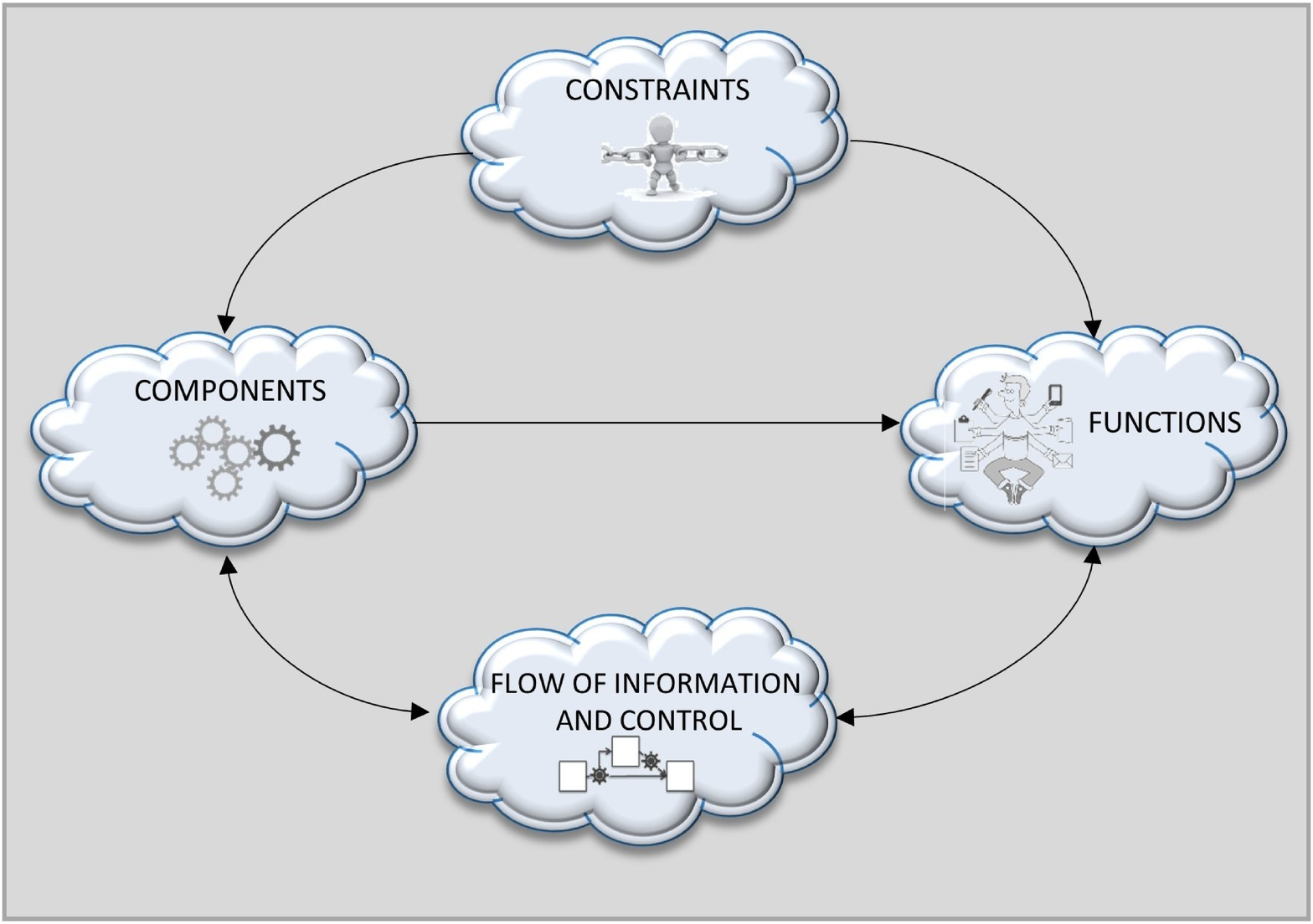}}\hfill%
    \subfloat[SIBM Flowchart]{\includegraphics[width=.295\linewidth]{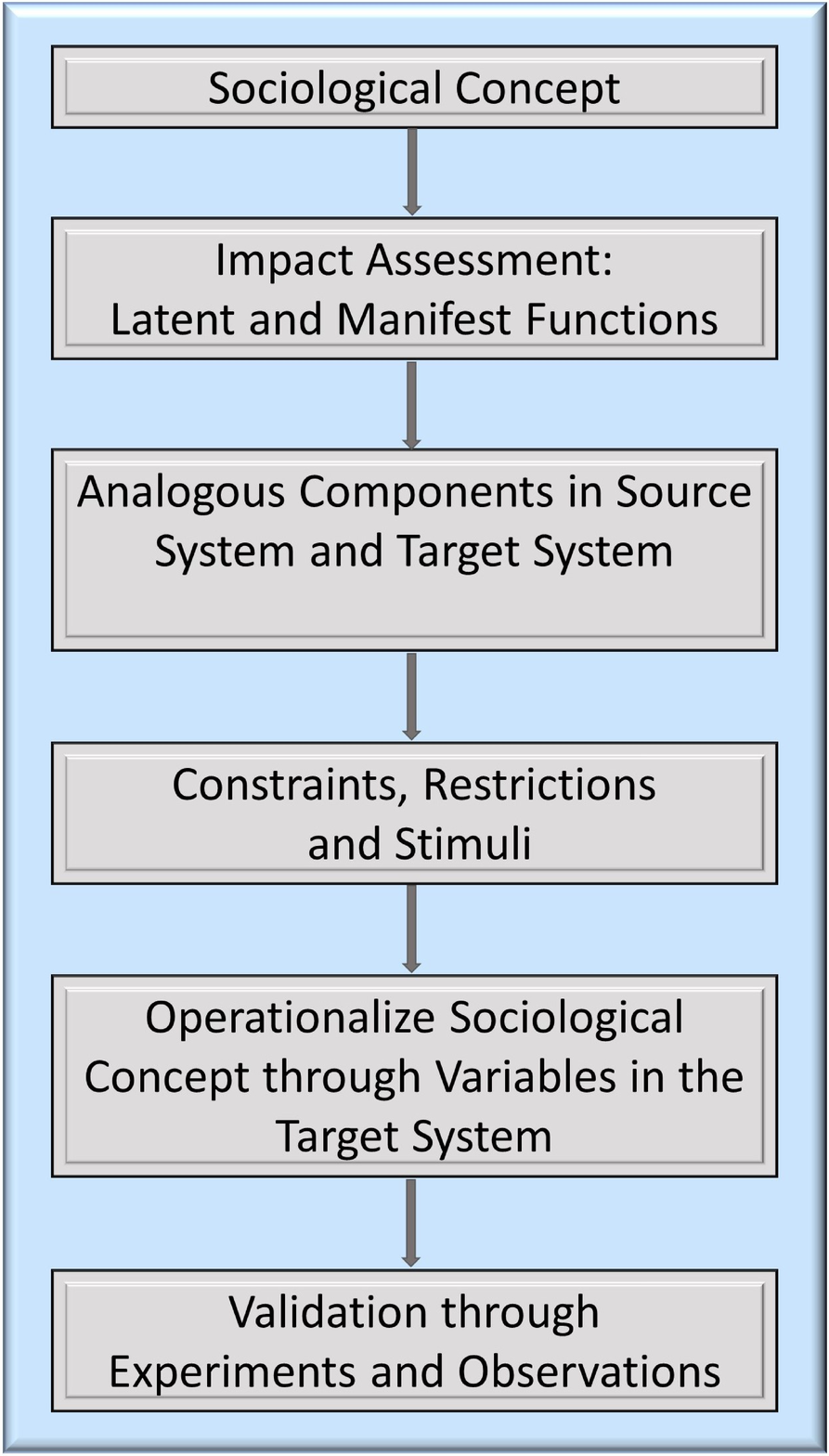}}\hfill%
      \end{tabular}
      \caption{Sociological Idea Borrowing Mechanism (SIBM)} 
     \label{SIBM}
\end{figure*}
\subsection{The Sociological Foundation of CALM} \label{D}
 The transfer of ideas from the sociological realm to the domain of wireless-networks involves a complex process, which we attempt to simplify by proposing a novel Sociological Idea Borrowing Mechanism (SIBM). 

 \subsubsection{Sociological Idea Borrowing Mechanism (SIBM)}
 A sociological idea operates in a social-construct or a social-system, which is the \textit{source-system}. The system where these ideas are to be applied becomes the \textit{target-system}. The application of ideas may be in their original form or they may be \textit{contextualized} to meet the target-system requirements. The proposed design of SIBM is inspired from \textit{Cybernetics}, a trans-disciplinary view of systems in diverse domains and the flow of information and control within them. It pays a special attention to how flow of information exercises control over the system functions and operation \cite{Cybernetics}. For example, DNA is coded information, which controls the development and functioning of the muscular system. The famous \textit{AGIL} schema of Parsons is based upon a cybernetic 
hierarchy and is a prime example of application of Cybernetics in Sociology \cite{Parsons}.
 
 The SIBM considers both, the source-system and the target-system, as a \textit{closed system} consisting of components that serve specific functions and offer particular services to the system. It assumes an interdependence between the components and takes a \textit{Functionalist} view that all components of a system are vital for its existence and continued operation. Further, every system is subject to regulatory provisions, restrictions and constraints within which it operates. SIBM recognizes two flows in the system, an \textit{informational flow} and a \textit{control flow}. SIBM's view of a source or target system is depicted in Figure~\ref{SIBM}~(a). 
 
 Further, in line with Cybernetics, SIBM operates on the principle that a system (or sub-system) high on information will invariably control a system (or sub-system) high on energy, \emph{i.e.,} it will be at a higher level on the \textit{Cybernetic Hierarchy}. For example, the electronic panel (sub-system) of a washing machine which is high on information controls the rotating drum, which is high on energy. This principle is crucial, as SIBM will be applicable to a target-system only if the following conditions are fulfilled. 
 \begin{itemize}
  \item The source-system and target-system are at comparable or similar levels in their respective cybernetic hierarchy.
  \item The source-system is at a higher cybernetic level and there exists a sub-system within the target-system that is situated at a level similar to the source-system and regulates the target-subsystem. 
 \end{itemize}
Also, for every sociological paradigm, the \textit{manifest} (intended) functions must be evaluated for desirability, while the impact and consequence of \textit{latent} (unintended) functions must be investigated and assessed. A manifest function has desirable and predictable consequences, while a latent function is inherently unpredictable and may have positive or negative consequences \cite{Merton}.

Now, we describe the stepwise procedure of SIBM which is also illustrated in Figure~\ref{SIBM}~(b).
\begin{enumerate}
 \item Develop a thorough understanding of the sociological concept(s) : Before attempting to operationalize a concept, a researcher must ensure a deep insight of its meaning, its social context and its limitations. 
 \item Impact assessment of latent and manifest functions of the sociological paradigm in the target-system.
 \item Draw an analogy between the social system and the target system : A systems view is similar in many respects to the Functionalist paradigm of Sociology, as both visualize a system (natural, social or technological) as made of components. The parts or components of a system are integrated and interdependent, with each component serving a particular and indispensable function. 
 \item Identify analogous functional components in both the systems : Determine analogous components either in terms of their functional role in the system, or in terms of their relationship with other components or the system as a whole. Cybernetic Hierarchy comes into play here.  
  \item Identify sociological factors that influence the components of a social system (constraints, restrictions, stimuli etc.), and carefully implement them in the target-system through appropriate variables. This step involves operationalization of sociological concepts by identification of target-system variables.
 \item Validate the successful implementation through direct or indirect experimentation and observation. 
\end{enumerate}
We follow the steps prescribed by SIBM while applying social theory in the design of CALM.

\subsubsection {Wireless Mesh Network : An Emergent Reality} \label{calmsoc}
To make \textit{CALM} agnostic to the CA type and the impact of CA scheme on WMN topology, we borrow concepts from social theory. One of the founding fathers of Sociology, Emile Durkheim states that \textit{social reality is more than the sum of its individual components} and that \textit{the individual is subject to an external social reality that transcends her} \cite{Durkheim1,Durkheim2}. Within the functional perspective of sociology, society is considered to be an \textit{emergent reality}. The social system has individuals as its components, but it is not merely a collection of individuals. Social reality is \textquoteleft sui generis\textquoteright, \emph{i.e.,} it has an independent existence of its own, which lies at a higher plane of reality than that of its bio-psychic components. Further, this social reality is external to the individual, it is diffused in the collectivity, and it exerts control over the individual. An individual as a social actor responds to the external stimuli, acting in 
accordance with social norms, and therefore human behavior is governed by this overarching social reality \cite{Giddens}. The Positivist strand of Sociology takes an objective view of human behavior. Although it does not dismiss \textit{subjective disposition} or the subjective aspects of human agency that include meanings and motives, it gives priority to \textit{objective consequences} that are observable and verifiable. This facilitates a simpler borrowing and operationalization of ideas from social theory for application in wireless networks, as individual network components too simply respond to the ambient conditions in the wireless network, and lack the consciousness to attach any subjective meanings to their roles in the network. 
We follow the proposed SIBM and draw an analogy between a social system and a wireless network. This allows us to treat a WMN as an \textit{emergent reality}, external to, and independent of the individual network components which it regulates, and upon which it places constraints. Equating a WMN system with a distinct social-reality has two implications for its components, which we now propose as corollaries to Durkheim's hypothesis. We propose these corollaries with regards to a WMN, but they can be extended to any wireless network.
\begin{enumerate}
 \item \textbf{Corollary 1 : } \textquotedblleft A network is more than the sum of the individual nodes that it comprises of\textquotedblright, \emph{i.e.,} when nodes in a wireless network communicate, the capacity of the WMN is not equal to the sum of individual capacities of radios or links. When several nodes in a network transmit and receive data in tandem, the interaction of these concurrent transmissions spawns a variety of factors such as link-conflicts, congestion, latency, data packet corruption etc. Thus, a WMN is more than just a collectivity of mesh routers and operates at a plane above its individual components. 
 \item \textbf{Corollary 2 : } \textquotedblleft An individual node/link is subject to constraints created by the ambient network conditions\textquotedblright, \emph{i.e.,} the capacity of a node/link is determined not by the capacity of that node/link and its adjacent nodes/links alone, but also by the prevailing network-wide factors. This implies that there are network-wide constraints that have a bearing upon the performance and functioning of individual nodes/links, and with a change in ambient network conditions such as topology, these external constraints will change too. 
\end{enumerate}
     \begin{figure}[h!]
                \center 
                \includegraphics[width=1\linewidth, height=15cm]{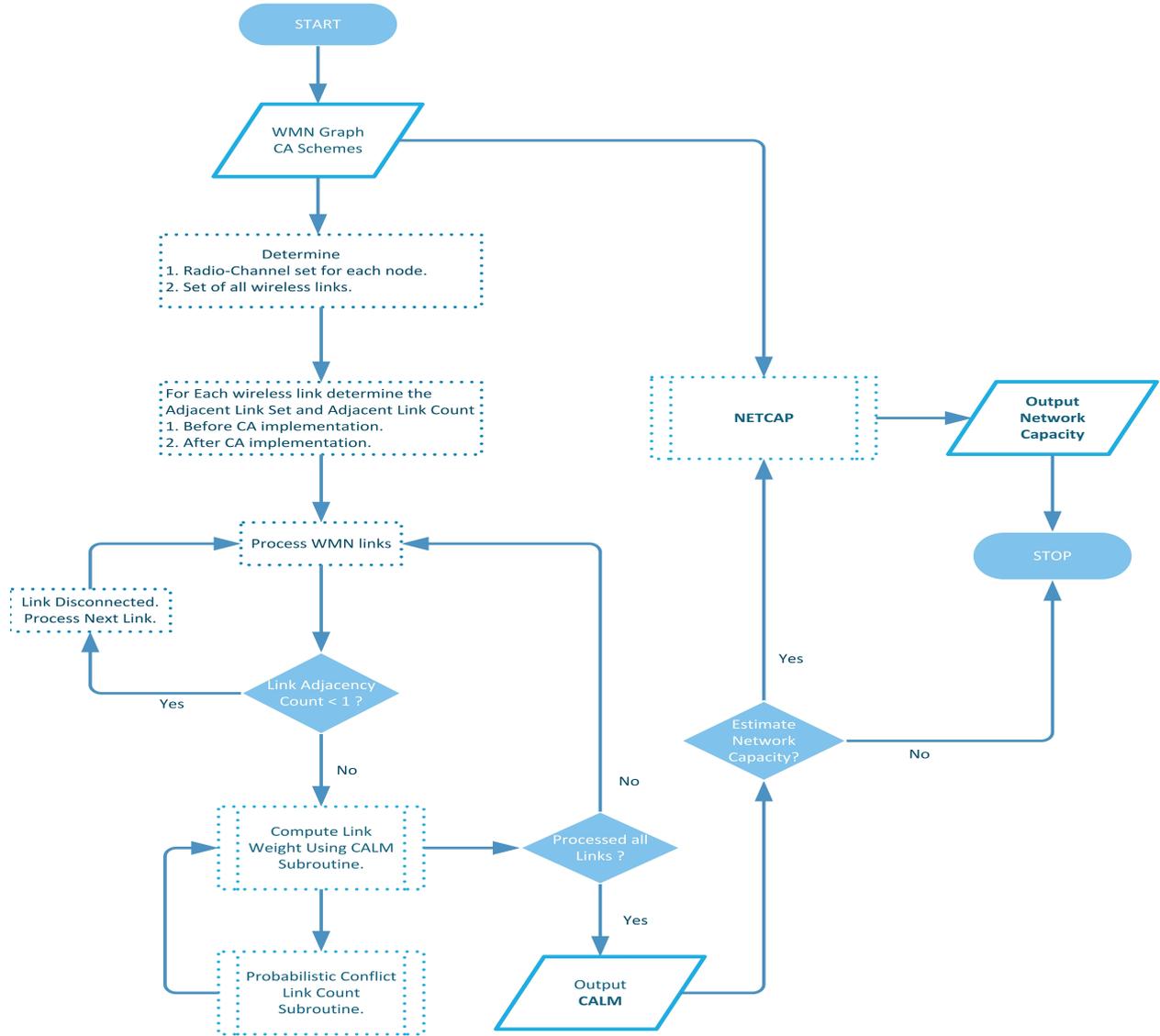}
                \caption{The High-level Schema of CALM-NETCAP Framework.}
                \label{FC}
        \end{figure}
Continuing with the steps listed in SIBM, we now use these corollaries to determine two variables related to wireless links in a WMN \emph{viz.,} the Link-Cost (LC) and the Link-Weight (LW). LC represents the adverse impact of interference on the link quality, while LW represents a link's resilience to this detrimental impact. The sociological concepts are incorporated in \textit{CALM} through two pan-network variables, both of which are dependent on network topology and are used in the computation of LCs and LWs for individual links. The first variable is the \textit{Maximum Link Adjacency Count} ($MaxAdjCount$), which denotes the maximum number of adjacent links among all links in the WMN, before a CA is deployed. Second variable is the \textit{Average Number of Adjacent Links} ($AvgAdjLink$), which represents the mean link adjacency for all links in the WMN after CA is deployed. For TPCAs, \textit{CALM} makes use of $MaxAdjCount$ to compute LC and LW as it depends on original WMN graph and the CA scheme 
does not alter the WMN layout. For the other two CA categories \emph{viz.,} GPCA and GDCA, LC and LW are computed using the $AvgAdjLink$, which depends on the WMN topology after CA deployment. Both variables are determined by the network topology as a whole, and are system variables external to, and above the individual nodes/links, which is in line with \mbox{Corollary 1}. Further, the use of these pan-network variables to determine individual link quality gives shape to \mbox{Corollary 2}, as they determine link-quality estimates. Variable $AvgAdjLink$ takes into account the change in network topology after CA deployment, ensuring that these changes are reflected in the individual link quality assessment, and the overall interference estimate. This further reinforces the idea behind \mbox{Corollary 2}. Finally, the efficacy of SIBM is validated through the high accuracy of CALM demonstrated in Section~\ref{C}.

Thus, in the computation of individual link quality \textit{CALM} takes a network-wide view, while also ensuring that a change in network topology is reflected in interference estimation. These features are absent in TID, CDAL and CXLS algorithms, which are oblivious to changes in WMN topology.


\renewcommand{\algorithmicrequire}{\textbf{Input:}}
\renewcommand{\algorithmicensure}{\textbf{Output:}}
\begin{algorithm}[htb!] 
\caption{Channel Assignment Link-weight Metric.}
\label{CALM1}
\begin{algorithmic}[1]
{\fontsize{9}{10}
\REQUIRE $G_{WMN} = (V_{WMN},E_{WMN})$, $R_i (i \in V_{WMN})$, $CA = \{(R_i,CS), i \in V_{WMN}\}$, $CS =\{1, 2,...M\}$\\
\textbf{Notations} $:$ $G_{WMN}$ $\leftarrow$  Original WMN Graph, $R_i$ $\leftarrow$ Radio-Set, $CA$ $\leftarrow$ Channel Assignment, $CS$~$\leftarrow$~Available Channel Set, $Adj_i$ $\leftarrow$ Set of nodes adjacent to node $i$ in $G_{WMN}$, $Ch_i$  $\leftarrow$ Set of channels allocated to the radios at node $i$ in $G$ by $CA$, $LinkSet$ $\leftarrow$ Set of all wireless links in $G_{WMN}$, $_{G}LnAdjCnt$~$\leftarrow$~Number of adjacent links link $Ln$ has in $G_{WMN}$ before $CA$ implementation, $_{CA}LnAdjCnt$ $\leftarrow$ No of adjacent links link $Ln$ has in $G_{WMN}$ after $CA$ implementation, $LinkChSet$ $\leftarrow$ Set of channels link $Ln$ can transmit on, $LinkAdjSet$~$\leftarrow$~Set of adjacent links with channels common to link $Ln$ after $CA$ implementation, $LCMap$~$\leftarrow$~Contains entire link related information for all operational links. \\
\ENSURE \textit{CALM} \\
\line(1,0){236}
\FOR {$i \in V_{WMN}$}
\STATE Determine $Adj_i$ \& $Ch_i$
\FOR {$j \in Adj_i$}
\STATE $LinkSet \leftarrow InsertLn(i,j)$. 
\ENDFOR
\ENDFOR
\FOR {$Ln \in LinkSet$}
\STATE Determine $enode_1,$ $enode_2,$ $_{G}LnAdjCnt$ \COMMENT {$enode_1$ $\in$ $V_{WMN}$, $enode_2$ $\in$ $V_{WMN}$}

\STATE $LinkChSet\leftarrow Ch_{enode_1} \cap Ch_{enode_2}$

\IF {$LinkChSet \neq \emptyset$} 
\STATE $_{CA}LnAdjCnt \leftarrow GetNumAdjacentLinks(enode_1,enode_2, LinkChSet)$
\STATE $LinkAdjSet \leftarrow GetAdjacentLinks(enode_1,enode_2, LinkChSet)$  
\ENDIF 
\IF {($_{CA}LnAdjCnt < 1$)}
\STATE Mark $Ln$ disconnected/non-operational. 
\ELSE
\STATE $LCMap \leftarrow InsertLC(LCMap, enode_1, enode_2, LinkChSet, LinkAdjSet, _{CA}LnAdjCnt,_{G}LnAdjCnt)$
\ENDIF
\ENDFOR
\STATE  $CALM \leftarrow CompCALM(LCMap, LinkSet)$. \COMMENT {Function $CompCALM()$ implements Algorithm~\ref{CALM2}}
\STATE{Output the \textit{CALM}}

}\end{algorithmic}
\end{algorithm}

\renewcommand{\algorithmicrequire}{\textbf{Input:}}
\renewcommand{\algorithmicensure}{\textbf{Output:}}
\begin{algorithm}[htb!] 
\caption{Computation of CALM.}
\label{CALM2}
\begin{algorithmic}[1]
{\fontsize{9}{10}
\REQUIRE$LCMap,  LinkSet$\\

\textbf{Notations} $:$ $TotLinks$ $\leftarrow$ Total number of links in $LinkSet$, $AvgAdjLink$  $\leftarrow$ Average number of adjacent links for a link in $G_{WMN}$ after CA implementation, $MaxAdjCount$ $\leftarrow$ Maximum number of adjacent links for any link in $G_{WMN}$ before CA implementation, $NumProbConfLn$ $\leftarrow$ Number of probabilistic link conflicts for link $Ln$ in $G_{WMN}$ after CA implementation, $LinkWeight$ $\leftarrow$ Performance weight of link $Ln$, $LinkCost$ $\leftarrow$ Interference cost of link $Ln$ \\
\ENSURE Return \textit{CALM} \emph{i.e}, Channel Assignment Link-weight Metric\\
\line(1,0){236}
\STATE $CALM \leftarrow 0, NumProbConfLn \leftarrow 0, AvgAdjLink \leftarrow 0, MaxAdjCount \leftarrow 0$
\FOR {$Ln \in  LinkSet$}
\STATE $LinkWeight_{Ln} \leftarrow 1.0$
\IF {$_{CA}LnAdjCnt_{Ln} > 0$}
\STATE $AvgAdjLink \leftarrow AvgAdjLink + _{CA}LnAdjCnt_{Ln}$
\STATE $MaxAdjCount \leftarrow GetMaximum(_{G}LnAdjCnt_{Ln})$
\ENDIF
\ENDFOR
\STATE $AvgAdjLink \leftarrow AvgAdjLink/TotLinks$ 
\FOR {$Ln \in  LinkSet$} 
\IF {$_{CA}LnAdjCnt < 1$}
  \IF {$(_{G}LnAdjCnt/AvgAdjLink > 1)$}
    \STATE $LinkCost \leftarrow 1$
  \ELSE
    \STATE $LinkCost \leftarrow (_{G}LnAdjCnt/AvgAdjLink)$
  \ENDIF
\ELSE
  \STATE $NumProbConfLn=ProbAdjConfLink(Ln,LCMap)$
  \IF{$(_{CA}LnAdjCnt < _{G}LnAdjCnt)$}
    \STATE $NumProbConfLn\leftarrow NumProbConfLn + _{G}LnAdjCnt - _{CA}LnAdjCnt$
  \ENDIF
  \STATE $LinkCost \leftarrow NumProbConfLn/(MaxAdjCount + 1)$
\ENDIF
\STATE $LinkWeight  \leftarrow LinkWeight - LinkCost$
\STATE $CALM  \leftarrow CALM + LinkWeight$ 
\ENDFOR
\STATE Return \textit{CALM} 
}
\end{algorithmic}
\end{algorithm}

\renewcommand{\algorithmicrequire}{\textbf{Input:}}
\renewcommand{\algorithmicensure}{\textbf{Output:}}
\begin{algorithm}[htb!] 
\caption{Function ProbAdjConfLink(Ln, LCMap) : Probabilistic Computation of Conflict Links.}
\label{CALM3}
\begin{algorithmic}[1]
{\fontsize{9}{10}
\REQUIRE$Ln,  LinkSet$\\
\textbf{Notation} $:$ $CommCh_{i,j}$ $\leftarrow$ Set of common channels shared by links $(i \ \& \ j)$ \\
\ENSURE Return $NumProbConfLn$ \emph{i.e.}, No of probabilistic link conflicts of each link $Ln$ \\
\STATE $NumProbConfLn \leftarrow 0$

\FOR {$Adj \in LinkAdjSet_{Ln}$}
\STATE $CommCh_{Ln,Adj} \leftarrow LinkChSet_{Ln} \cap LinkChSet_{Adj}$ 
\STATE $X \leftarrow |LinkChSet_{Ln}| $
\FOR {$ channel \in CommCh_{Ln,Adj}$}
\STATE $NumProbConfLn \leftarrow NumProbConfLn + (1/X)$
\ENDFOR
\ENDFOR
}
\end{algorithmic}
\end{algorithm}
   \begin{figure}[h!]
                \center 
                \includegraphics[width=1\linewidth, height=20cm]{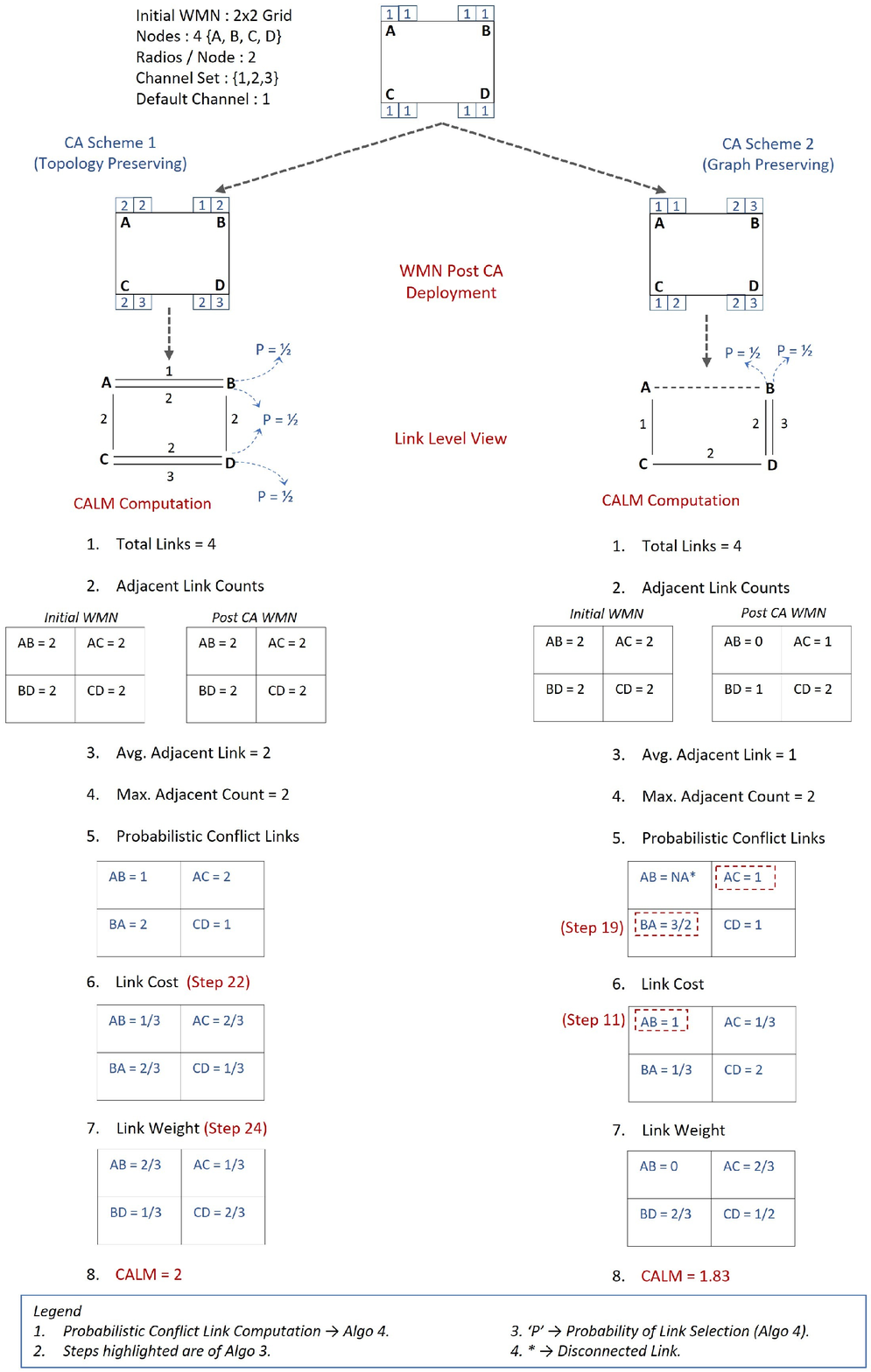}
                \caption{Computation of CALM : A Detailed Illustration.}
                \label{wex}
        \end{figure}
\subsection{CALM : A Functional Description}
\textit{CALM} takes a link view of a wireless network at the granularity of nodes which may be equipped with multiple radios operating on multiple channels. In Algorithm-\ref{CALM1}, all links between nodes in the WMN prior to CA implementation are listed in $LinkSet$. For each link \textit{CALM} ascertains all the relevant information such as the two WMN nodes sharing the link ($enode_1, enode_2$), number of adjacent links ($_{G}LnAdjCnt$) before CA implementation, and post CA implementation information including set of channels on which the link can transmit ($LinkChSet$) and number of adjacent links ($_{CA}LnAdjCnt$) of the link. \textit{CALM} identifies the CA type by checking for disconnected links and labels them. Thereafter, information for all connected/operational links is stored in a structure ($LCMap$) for further processing. 
\textit{CALM} has a single design for all types of CAs and is capable of evaluating all CA schemes regardless of whether they preserve WMN topology or not. In Algorithm~\ref{CALM2}, \textit{CALM} computes $MaxAdjCount$ and $AvgAdjLink$ described earlier, and uses them to compute Link-Cost ($LinkCost$). If CA alters WMN topology, it assigns every disconnected link $Ln$ a Link-Cost of $1$ or $(_{G}LnAdjCnt/AvgAdjLink)$, whichever is less, where  $_{G}LnAdjCnt$ is the number of adjacent links $Ln$ had in the original WMN topology. For operational links, \textit{CALM} determines the probabilistic link-conflicts ($NumProbConfLn$) by invoking Algorithm~\ref{CALM3}. It accounts for the impact of CA scheme on link adjacency count of an operational link by computing $_{CA}LnAdjCnt$ and determines the number of adjacent links missing after CA implementation. \textit{CALM} treats a missing adjacent link also as a link-conflict to account for the adverse impact on performance caused by it, and increments 
the number of probabilistic 
link-conflicts. Next, the Link-Cost for an operational link is computed by dividing the total number of probabilistic link-conflicts by ($MaxAdjCount$ + 1). Even if the adjacency of a link equals $MaxAdjCount$, and all adjacent links probabilistically conflict with the given link, the link is still operational and capable of transmitting and receiving data. Thus, the reason for incrementing the Maximum Link Adjacency by $1$ is to ensure that that the Link-Cost is less than $1$, which can only be the Link-Cost of a disconnected link. The Link-Weight ($LinkWeight$) is initially set to $1.0$, signifying an ideal link which can transmit and receive data at its full capacity, without any obstruction from the endemic interference. \textit{CALM} evaluates and assigns the Link-Weight by deducting Link-Cost from the initial value. Finally, the \textit{CALM} estimate is generated by summing up the Link-Weights of all links in the WMN. Thus, it offers both, a link quality estimate for all links in the WMN and an 
overall CA interference estimate. While the CA interference estimate indicates the suitability of a CA for a given WMN, the Link-Weights are used to generate a measure of network capacity through the NETCAP model proposed in the next section.

 A high-level schema of CALM is presented in Figure~\ref{FC}. The Link-Weight for each link and the CALM for the CA scheme is computed by the CALM subroutine. The detailed operation of the CALM subroutine is demonstrated through the illustration in Figure~\ref{wex}, in which both, a \textit{graph preserving} and a \textit{topology preserving} CA scheme are considered. The two examples help us present all possible scenarios and conditions involved in the computation of CALM.
\subsection{Time Complexity of CALM}
Consider an arbitrary MRMC WMN graph $G=(V,E)$, comprising of $n$ nodes, with $r$ identical radios installed on every node and $c$ orthogonal channels for allocation. The upper-bound on algorithmic complexity of TID, CDAL$_{cost}$ and CXLS$_{wt}$ algorithms is O($n\textsuperscript2 r\textsuperscript3$) and O($n\textsuperscript2 r\textsuperscript2$), and O($n\textsuperscript3 r\textsuperscript2$) respectively \cite{Manas4}. While Considering CALM, we assume that \textit{c $>$ r} and \textit{n $>>$ r}. Also, the number of radios installed on each node is subject to the number of available orthogonal channels. In this context, we can determine that the worst-case complexity of Algorithms~\ref{CALM1}, ~\ref{CALM2}, and ~\ref{CALM3} is O($c n\textsuperscript3$), O($c n\textsuperscript2$) and O($cn$), respectively. Thus, the overall complexity of CALM is O($c n\textsuperscript3$). 
In comparison, while CXLS$_{wt}$ has a higher computational overhead, TID and CDAL$_{cost}$ have a better worst-case algorithmic complexity than CALM. Results will demonstrate that this slight increase in complexity is a marginal cost to incur for significantly improved accuracy in CA performance prediction.

\section{NETCAP : Framework for Network Capacity Estimation} \label{7}
WMN deployments involve a variety of challenges for which solutions are sought so as to maximize network performance, minimize fixed costs, and ensure optimal resource allocation. Linear Programming and Mixed Integer Linear Programming (MILP) are often used to provide optimal or near-optimal solutions to challenges in WMNs. Authors in~\cite{amaldi2008optimization,amaldi2007optimization} propose an MILP planning model which aims to optimize the deployment of WMN nodes in a large scale terrain. The work offers optimal solutions to the problems of network installation costs, greater network span, and deeper penetration of WMN. WMN optimization problems are difficult to solve in real-time due to their NP hard nature, and in order to execute them in polynomial time, a relaxed version of problems is considered with fewer constraints in the model. For example, in \cite{amaldi2008optimization}, authors propose a polynomial time heuristic approach with relaxed constraints to address issues related to both, 
WMN planning and channel assignment. In~\cite{valois2013energy}, authors propose a multi-objective optimization model by considering the \textit{throughput} and \textit{energy consumption} of a WMN. This model considers a physical interference scenario wherein the nodes exercise continuous power control, and make use of a discrete set of data rates. The authors make use of several network parameters and network engineering insights. For example, they show that power control and multi-rate functionalities enable a WMN to attain optimal throughput with lower energy consumption, by using a mix of single hop and multi-hop routes. Authors in~\cite{li2009joint} suggest an efficient algorithm for throughput optimization through optimal flow routing, scheduling of transmissions, and dynamic channel assignment. This model visualizes a situation wherein each node levies a charge for relaying data to a neighboring node and every packet flow has a budget constraint. In~\cite{kuperman2016fast}, authors propose a 
scalable algorithm which efficiently determines the maximum achievable unicast and multicast flows under the influence of interference, and also provides feasible routes and schedules to support those flows. Hence, the proposed algorithm aims to maximize the capacity for a given wireless mesh network. In~\cite{cherif2017joint}, an MILP model is formulated with the goal of maximizing the throughput while minimizing energy consumption. Authors demonstrate that using directional antenna in wireless mesh network will increase the beam-width, and decrease the power levels, through effective power control. 

\par Research works focusing on WMN capacity optimization generally aim to solve optimal network capacity problem in one of two ways \emph{viz.,} by devising optimal routing, CA and scheduling schemes or by considering considering these factors as input along with real-time network information. Studies that fall in the first category try to achieve maximal capacity by devising MIP/MILP/MINLP based optimal or near optimal CA schemes \cite{MILP1} or may try to devise formulations combining routing or scheduling with optimal CA scheme design \cite{MILP2, MILP3}. Studies that lie in the latter category optimize network capacity based on information about factors such as traffic routing, rate adaptation etc. \cite{amaldi2008optimization}. Almost invariably, these solutions consider a polynomial time relaxed heuristic of the original MIP/MILP formulation due to the NP-hard nature of finding optimal solutions to the network capacity problems. To the best of our knowledge, we haven't come across an optimization 
study 
that accepts a 
CA scheme as an input, and estimates the optimal network capacity without requiring any additional real-time information such as scheduling, data routing, SINR, congestion, latency etc.

\par We now propose a framework to estimate the network capacity offered by a CA scheme, which unlike prior studies on WMN throughput optimization, does not require any additional real-time network information to offer a capacity estimate. Proposed NETCAP model only requires three inputs \emph{viz.,} original WMN topology, the CA scheme and the Link-Weights generated by the CALM metric, which serve as the constraints on Wi-Fi link quality. With these three inputs alone, the proposed framework proposes an expected network capacity which is remarkably close to the experimentally recorded throughput. We begin by proposing a generic comprehensive MINLP formulation with an objective to maximize the overall system capacity, followed by a simple MIP optimization model called NETCAP, which converges faster and incurs lesser computational cost. 
\begin{table}[htb!]
\caption{List of notations used in the problem formulations.}
\centering
\begin{tabular}{|p{2cm}| p{13cm}|}
\hline\bfseries
Notation&\bfseries \hspace{5cm} Definition \\
\hline\hline
$S$& Set of WMN nodes.\\
\hline
$d_i$& Data demand on each WMN node $i$. \\
\hline
$\alpha$& Number of radios each node support. \\
\hline
$\lambda$& For each link, the SINR should be maintained above the threshold value. \\
\hline
$g_{ij}$& Channel gain on $link_{ij}$ between WMN node $i$ and node $j$. \\
\hline
$P_{max}$& Maximum transmission power of WMN node.\\
\hline
$W_n$& If WMN node $n$ is present then its treated as 1 else 0.\\
\hline
$f_s (i,j)$& The incoming flow of data packets on a WMN node. \\
\hline
$f_s (j,k)$& The outgoing flow of packets on a WMN node. \\
\hline
$Rate_{ij}$ & The maximum rate at which a WMN link from node $i$ to node $j$ can transfer data. \\
\hline
\end{tabular}
\label{tab1}
\end{table}

\subsection{WMN Network Capacity Optimization Problem}
\textbf{Objective Function:}\\
\par The objective of the optimization model\footnote{An alternate optimization goal can be the maximization of the minimum power (max (min($P_{S}$)) consumed by a WMN node where all the constraints are identical to the proposed MINLP model} is to maximize the overall network throughput of the given WMN between every source-sink pair.
Our objective is to maximize capacity despite the adverse impact of interference. Hence the objective function is given by, 
\begin{equation}
Maxmize(B\times log(1+SINR))
\end{equation}
$B$ denotes the \textit{Capacity or bandwidth} of the network. SINR (Signal to Noise plus Interference Ratio) represents the total attenuation suffered by data transmissions, due to ambient noise and concurrent transmissions on identical or overlapping channels, from within the network or from external sources. The notations used in the formulation are presented in Table~\ref{tab1}. \\ \\
\textbf{Constraints:}\\ \\
The sum of all the incoming packet flows in the WMN must be equal to the sum of all outgoing packet flows at every node, whether it is a gateway node or an intermediate node.
\begin{equation}\label{e2}
\sum_i f_s(i,j) = \sum_k f_s(j,k) \hspace{0.3cm} \forall j
\end{equation}
 We also assume that every link is bidirectional \emph{i.e.,} data traffic can travel in both directions within the WMN.
\begin{equation}\label{ex}
\sum_{network} (i,k) = \sum_{network} (j,k)
\end{equation}
For a gateway node, the source of incoming data may be an external network or an ISP (denoted by $Q_1$), while for an intermediate node the source of data is usually a neighboring node. Equations~\eqref{e3} and ~\eqref{e4}, represent the source of the data of a WMN node.
\begin{equation}\label{e3}
Q_1 = \sum_k f_s (j,k) \hspace{0.3cm} \forall j^*
\end{equation}
\begin{equation}\label{e4}
\sum_i f_s (i,j) = Q_1 \hspace{0.3cm} \forall j^*
\end{equation}
The incoming and outgoing flow on any link is non negative, which leads to Equation\eqref{e5}.
\begin{equation}\label{e5}
f_s(i,j) > 0
\end{equation}
Let $S$ represent the set of WMN nodes. In order to ensure optimal link quality for uninterrupted data transmission by each WMN node, the SINR constraints for each channel should be greater than the threshold value. Here, $N_o$ represents the noise in the system, $P_{max}$ represents the maximum transmission power of the WMN node, and $g_{ij}$ represents the channel gain on $link_{ij}$, between WMN node $i$ and WMN node $j$. $W_n$ is a binary variable that denotes the presence ($W_n \leftarrow 1$), or absence ($W_n \leftarrow 0$), of a WMN node $n$. It helps in the consideration of interfering nodes in proximity to a given node, and disconnected nodes in case of non-topology preserving CA schemes. It also holds relevance if the nodes are mobile, but in our work we assume a static planned deployment \emph{i.e.,} topology of the WMN and the location of WMN nodes is known and fixed. Further, if the SINR constraint is not greater than the threshold value then a different channel will be assigned to the WMN link.
\begin{equation}\label{e6}
\frac{Inf \times  (1-x_{ij})+ g_{ij}P_{max}W_{a}}{N_{o}+ \displaystyle\sum\limits_{{b \in S \setminus i}}g_{bj}P_{max}W_{b}} \geq \lambda  \hspace{0.3cm}\forall i, j \in S
\end{equation}
The numerator in Equation~\eqref{e6} represents the signal strength of the transmissions from node $a$, while the denominator represents the sum total of interference offered by all nodes in the proximity of $a$ which are transmitting on conflicting channels, plus the ambient noise. Variable $x_{ij}$ represents the SINR status of $link_{ij}$, where ($x_{ij} \leftarrow 1$) implies that the SINR value above the the threshold is guaranteed by the link. In contrast, ($x_{ij} \leftarrow 0$) signifies a poor SINR, and a \textit{Virtual Infinite} value $Inf$ ($\approx 10^6$) is used, so that the link can be ignored. Thus, the Virtual Infinite value ensures that all links provide a minimum SINR threshold value.
Equation~\eqref{e6} is represented in linear form as,
\begin{equation}\label{e7}
Inf \times (1-x_{ij})+ g_{ij}P_{max}W_{a}  \geq \lambda \times  (N_{o}+ \displaystyle\sum\limits_{{b \in S \setminus i}}g_{bj}P_{max}W_{b})  \hspace{0.3cm}\forall i, j \in S
\end{equation}

Further, each MRMC WMN node ($i$) may be equipped with multiple radios ($r_i$) which leads to the equations below, where $\alpha_{max}$ and $\alpha_{min}$ represent the minimum and maximum number of radios installed on the nodes in a WMN, respectively.
\begin{equation}\label{e8}
\sum r_{i} \geq \alpha_{min} 
\end{equation}
\begin{equation}\label{e9}
\sum r_{i} \leq \alpha_{max} 
\end{equation}

Next, depending on the load/demand in the network, appropriate links will be selected for transmission. If the traffic demand is high, a link that experiences low levels of interference may be selected and vice versa.
\begin{equation}\label{e10}
\displaystyle\sum\limits_{\alpha}\bigg\lceil{\frac{d_i (f_s(i,j))}{mcs(sinr_{i})}}\bigg\rceil \leq C
\end{equation}
Further, the binary variable $H_{i}^k$ is $1$ when WMN node $i$ is communicating using channel $k$, and $0$ when no link emanating from $i$ is operating on channel $k$. Also, $\theta$ represents the maximum number of channels assigned to the radio-set of each WMN node.
\begin{equation}\label{eca}
\sum_{k \epsilon Ch} H_{i}^k \leq \theta
\end{equation}

The mixed integer non-linear problem (MINLP) optimization model proposed above cannot run in polynomial time owing to the inclusion of non-linear constraints. 
Secondly, some of the constraints require a priori information such as status of channel, link quality degradation due to the impact of interference, SINR threshold, ambient noise and data traffic demand. The objective of our work is to theoretically assess a CA scheme and predict with high confidence, the expected network capacity that a CA scheme can offer, without actually deploying it in the WMN and running simulations/experiments. Thus, the need for real-time network information renders the model unsuitable in its current form. To overcome this challenge, we propose a MIP heuristic optimization model called NETCAP, that is not only faster due to simpler constraints, but also does away the need for real-time network data by making use of theoretical Link-Weight values generated by CALM. The twin benefits offered by NETCAP are faster convergence of optimization process, and use of simple theoretical link quality estimates.

\subsection{NETCAP : A Network Capacity Framework}
\subsubsection{Role of CALM}
CALM generates estimates of link quality for each individual link in the WMN. Instead of removing some of the constraints in the optimization model above to reduce the convergence time, we relax them by making use of these theoretically generated link quality estimates. This helps us maintain the efficacy of the proposed optimization model in determining maximal WMN capacity. The LWs generated by CALM are able to relax the following constraints :
\begin {enumerate}
 \item Equations~\eqref{e6}~and~\eqref{e10} : These constraints represent the impact of noise and interference on a transmission, and channel selection. CALM is a theoretical estimate of interference and the LWs offer a measure of probabilistic link conflicts, taking into account the spatial aspects of Equation~\eqref{e6} and temporal aspects of Equation~\eqref{e10}. Thus, CALM LWs can be used to relax these constraints, reducing execution time and doing away with the need for real-time SINR information. 
 \item Equations~\eqref{e8}~and~\eqref{e9} : CALM operates at the granularity of links, incorporating temporal selection of channels in the LW, thereby doing away with the need to consider links at the granularity of radios and channels.
 \end {enumerate}
In addition, CALM makes it easier for NETCAP to take into account the topological modifications brought about by CA implementation, without having to consider the presence of every node through the variable $W_n$ in Equation~\eqref{e6}. As described earlier the use of sociological concepts enables CALM to reflect topographical changes and account for their impact on WMN performance on every individual link through LWs, which lie in the range of 0 to 1. 
Now we propose the CALM aided simplified MIP optimization model NETCAP, listing its constraints, followed by a brief evaluation. 
\subsubsection{NETCAP Constraints}
NETCAP is a simpler WMN optimization model subject to the constraints represented by Equations~\eqref{e14}, \eqref{e15}, \eqref{e16}, \eqref{e17}, and \eqref{e18}. It makes use of link quality estimates generated by CALM to capture the impact of interference on WMN performance. These weighted link-quality estimates are used to calculate the throughput of the entire WMN system. Link-Weights are used to relax several constraints, substantially reducing computational and time costs, while maintaining accuracy of estimates. NETCAP aims to estimate maximal network throughput by maximizing the packet flow between all source-sink pairs, considering all multi-hop paths between them. Thus, the objective function of the NETCAP MIP is :


\begin{equation}
Maxmize \sum_{z} Y_z
\end{equation}
Where $Y_z$  is the packet flow on the ultimate link in a source-sink pair, and $z$ is the number of source-sink pairs in the network.\\
\begin{equation}\label{es}
Y_z = \sum_i \gamma_i^k
\end{equation}
Where $\gamma_i^k$ denotes all possible multi-hop routes between the sink-source pair $k$. Hence, maximizing $Y_z$ leads to the maximizing flow in $\gamma_i^k$.\\
Further, for every WMN node, the sum of all incoming packet flows must be equal to the sum of all outgoing packets flows. 
\begin{equation}\label{e14}
\sum_i f_s(i,j) = \sum_k f_s(j,k) \hspace{0.3cm} \forall j
\end{equation}
We maintain the constraint that the data traffic flows are bi-directional. 
\begin{equation}\label{e15}
\sum_{network} (i,k) = \sum_{network} (j,k)
\end{equation}
The incoming and outgoing flows on a WMN node are considered to be non-negative.
\begin{equation}\label{e16}
f_s(i,j) > o
\end{equation}
Link Weight values generated by CALM for each link lie in the 0 to 1 range. A value of $0$ represents a non-operational link, however not all disrupted links are assigned a LW of $0$ by CALM. Thus, topological constraints are simplified by the use of LWs and there is no longer the need for NETCAP to separately account for topology modifications done by a CA scheme. This makes NETCAP agnostic to CA type and enhances its versatility.
\begin{equation}\label{e17}
0 \leq lw \leq 1
\end{equation}
The final constraint that we consider in NETCAP concerns the maximum rate at which a WMN node $i$ can transfer data to an adjacent node $j$. We assume that multichannel MAC protocols are being used, enabling parallel transmissions between nodes. The number of channels to be assigned will depend upon the capacity ($C$) of a link, and the demand or load on a WMN node. Further, the overall data rate will be constrained by the Link-Weight of each link, as that gives us the effective capacity of the link under the impact of interference.
\begin{equation}\label{e18}
Rate_{ij} \leq C \times lw
\end{equation}
 The NECTAP  optimization model with CALM LW input is solved using the GAMS CPLEX solver \cite{lima2014ibm}, developed by IBM in the year 1988. The solver is a high-level modeling system for optimization and utilizes branch and bound algorithm framework for solving MINLP/MIP based optimization problems \cite{clausen1999branch}. Our choice of optimizer was guided by its capability to solve large and numerically difficult models with features such as priorities of integer variables, choice of different branching, and node selection strategies. 

\subsection{NETCAP : An evaluation}
We carry out a layered evaluation of NETCAP by considering both, a holistic view and a sectional/directional view of the WMN. Here, we evaluate NETCAP against previously published data and later in Section~\ref{8}, we validate it against fresh experimental results. We select three recently proposed CA schemes, which are at three different levels in the spectrum of observed network capacity. A grid specific near optimal CA (GSCA/NOCAG) \cite{Manas3, cite3}, followed by a high performance radio co-location aware CA (EIZM) \cite{Manas2} and a sub-par centralized CA scheme (CCA) \cite{23Cheng, Manas3}. NETCAP estimates were generated by considering identical WMN layout, network parameters and data traffic scenario for which the simulation results are generated in the corresponding work. The $5\times5$  802.11g GWMN considered in these works is depicted in Figure~\ref{BADTID2}~(b), in which a $10$ MB data file is sent from source to destination and the \textit{Net Aggregate Throughput} (NAT) of the network is 
observed. The source nodes are the first nodes of every row and column, while the destination nodes are the last nodes of every row and column. Thus, a 4-Hop-Flow (4HF) is established between every source-destination pair. We distinguish between horizontal and vertical flows by naming them 4-Hop-Horizontal-Flows (4HHF) and 4-Hop-Vertical-Flows (4HVF), respectively. Following test-scenarios were considered in these works which we evaluate on NETCAP.
\begin{enumerate}
 \item \textit{R$_5$C$_5$} : All ten 4HFs concurrently.
 \item \textit{R$_5$} : All five 4HHFs concurrently.
 \item \textit{C$_5$} : All five 4HVFs concurrently.
\end{enumerate}
 \begin{table} [h!]
\caption{NETCAP Evaluation: Comparing NETCAP Estimates with Experimental Results.}
\tabcolsep=0.11cm
\begin{tabular}{|M{1.4cm}|M{1.4cm}|M{1.5cm}|M{1.5cm}|M{1.4cm}|M{1.4cm}|M{1.5cm}|M{1.3cm}|M{1.3cm}|M{1.3cm}|}
\hline 
    \multicolumn{1}{|c|}{\textbf{CA}}&\multicolumn{3}{|c|}{\textbf{NETCAP Estimates (Mbps)}}&\multicolumn{3}{|c|}{\textbf{Observed Values (Mbps)}}&\multicolumn{3}{|c|}{\textbf{Margin of Error (\%)}}\\ \cline{2-10} 
     \multicolumn{1}{|c|}{\textbf{Scheme}}&\textbf{R$_5$}&\textbf{C$_5$}&\textbf{R$_5$C$_5$}&\textbf{R$_5$}&\textbf{C$_5$}&\textbf{R$_5$C$_5$}&\textbf{R$_5$}&\textbf{C$_5$}&\textbf{R$_5$C$_5$}\\
\hline  

CCA	&	6.4	&	3.2	&	9.9	&	7.1	&	3.0	&	8.5	&	11.8	&	6.6	&	14.2	\\
\hline
EIZM	&	17.5	&	12.3	&	27.7	&	17.7	&	13.5	&	26.3	&	1.4	&	9.7	&	4.8	\\
\hline
NOCAG	&	28.5	&	22.0	&	38.9	&	26.7	&	21.0	&	38.8	&	6.4	&	4.5	&	0.4	\\

\hline  
\end{tabular} 
\label{NE}
\end{table}

R$_5$C$_5$ is a maximal interference scenario which takes a comprehensive view of the WMN, while R$_5$ and C$_5$ test-scenarios offer a sectional/directional view of the WMN. Evaluation of NETCAP estimates against observed simulation results is presented in Table~\ref{NE}. Estimates of both, the sectional view along rows (R$_5$) and columns (R$_5$), and a comprehensive view of overall WMN performance (R$_5$C$_5$), exhibit low \textit{Margin of Errors} (MoE), leading to an accuracy of over 90\% in most cases, and as high as 99\% for NOCAG. Also, NETCAP estimates demonstrate a direct relationship with the efficiency of CA scheme \emph{i.e.,} higher the network capacity offered by CA scheme, more accurate are the NETCAP estimates. Upon arranging CAs in terms of observed NAT, we get the sequence CCA $<$ EIZM $<$ NOCAG which is also the sequence of increasing accuracy of NETCAP predictions. Thus, NETCAP gives a fairly reliable estimate of expected NAT whether we take a sectional/directional view of the WMN or 
consider the overall WMN performance.

\section{Simulations, Results and Analysis} \label{8}
\begin{table} [h!]
\caption{ns-3 Simulation Parameters.}
   \center 
\begin{tabular}{|p{6cm}|p{5cm}|}
\hline
\bfseries
 Parameter&\bfseries Value \\ [0.2ex]
 \hline
\hline
WMN Grid Size&$5\times5$\\
\hline
IEEE Standard & 802.11g (2.4 GHz)\\
\hline
No. of Radios/Node&2   \\
\hline
Range of Radios&250 mts   \\
\hline
Available Orthogonal Channels&3  \\
\hline
Maximum 802.11g PHY Datarate &9 Mbps \& 54 Mbps  \\
\hline
Datafile size &10 MB  \\
\hline
Maximum Segment Size (TCP)&1 KB   \\
\hline
Packet Size (UDP)&1 KB\\
\hline
MAC Fragmentation Threshold&2200 Bytes  \\
\hline
RTS/CTS &Enabled  \\
\hline
Routing Protocol &OLSR    \\
\hline
Propagation Delay Model&Constant Speed \\
\hline
Propagation Loss Model&Range Propagation\\
\hline
Rate Control&Constant Rate \\
\hline
Transmission Power&16 dBm\\
\hline
\end{tabular}
\label{sim}
\end{table}   

We now validate the proposed CALM and NETCAP techniques by subjecting them to exhaustive experimental evaluation on ns-3 \cite{NS-3}. We compare CALM against three existing \textit{CA performance prediction} or \textit{Interference Estimation Metrics} \emph{viz.}, TID, CDAL$_{cost}$ and CXLS$_{wt}$. For each CA scheme, TID is obtained through the \textit{E-MMCG} algorithm proposed in \cite{Manas}, while CDAL$_{cost}$ and CXLS$_{wt}$ metrics are generated through implementations suggested in \cite{Manas3} and \cite{Manas4}, respectively. Further, we use LW values generated by CALM in the NETCAP framework to predict the expected WMN capacity and assess it against observed results.
\subsection{WMN Topology Selection}
WMNs are conventionally designed in grid-like topologies, most common being the square, triangular and hexagonal grids \cite{Grid2}. There are several benefits of grid placement of wireless mesh routers over random or arbitrary topologies, such as, enhanced network capacity, reduced latency, improved network span, greater access-tier coverage area, back-haul connectivity and fairness in channel allocation. Authors in \cite{Grid} demonstrate that the network capacity of a grid WMN is almost twice that of a random WMN consisting of an identical number of nodes and an identical set of orthogonal channels. Gateway placement techniques also make use of GWMNs to ensure optimal network aggregate throughput \cite{Grid3}. GWMN deployments are preferred in research studies as they closely resemble real-world networks in terms of topological characteristics \cite{Complex}. For example, the Network Density of a GWMN ($\delta(G_{wmn})$), which is the ratio of active wireless links in a network and the total number of 
possible links in the network, is very close to its real-world network values \cite{TOPO}. We select a $5\times5$ GWMN configuration depicted in Figure~\ref{BADTID2}~(b) with a $\delta(G_{wmn})$ value of $0.13$, which is quite close to a real-world network value that generally falls in the range $(0.09 - 0.1)$ . 

%
\subsection{Test Case and CA Selection}
\subsubsection{Test Case}
In our discussion earlier, we classified CA schemes as TPCA, GPCA and GDCA. With that classification in mind, we carry out a twofold evaluation of CALM which is explained below.
\begin{enumerate}
 \item \textit{White Box Testing (WBT)} : The purpose of this \
 \textit{performance testing} is to evaluate the structural design of \textit{CALM} for accuracy and reliability, by verifying \textit{CALM} estimates against experimental results of TPCAs. We run two set of experiments on the IEEE802.11g standard. First, on a high interference and low capacity 9 Mbps data-rate (WBT$_{9Mbps}$), and thereafter on a low interference and high capacity 54 Mbps data-rate (WBT$_{54Mbps}$) \cite{80211g}. This exercise ensures exhaustive verification of the reliability of CA performance predications by \textit{CALM}.
  \item \textit{Black Box Testing (BBT)} : We carry out a \textit{stress-test} to ascertain the functional robustness and resilience of \textit{CALM}, by validating its output against an \textit{undesirable input} of a mix of GPCA and GDCA schemes. We run simulations on the 802.11g 54 Mbps environment (BBT$_{54Mbps}$).
\end{enumerate}

\subsubsection{Selection of CA Schemes\\}

\begin{enumerate}
 \item \textit{WBT CA Scheme Selection} : We consider a heterogeneous mix of well known TPCA schemes from the existing research literature, and implement them on an IEEE 802.11g $5\times5$ GWMN. Six of these TPCAs are based on an MMCG. A graph-theoretical centralized CA scheme (BFSCA) is proposed in \cite{22Ramachandran}, in which the channels are assigned to the nodes after subjecting the WMN MMCG to a Breadth First Traversal with the gateway node as the starting point. In \cite{17Xutao}, authors suggest a static maximal clique based CA algorithm (CLICA). A CA scheme (MISCA) in which channel allocation is done by partitioning the set of WMN radios into independent sets is proposed in \cite{24Aizaz}. Authors devise a centralized static CA heuristic (CCA) in \cite{23Cheng}, wherein the channel allocation is done to achieve minimal link conflicts. Two \textit{radio co-location aware} CA schemes are considered \emph{viz.}, an optimized independent set based CA scheme (OISCA), and a spatio-statistically designed,
 Elevated Interference Zone Mitigation approach (EIZMCA) \cite{Manas2}. We implement the six MMCG based CAs using two different MMCG models namely, the conventional MMCG and \textit{radio co-location aware} Enhanced-MMCG \cite{Manas, 22Ramachandran}, to generate twelve different CA schemes.
 We also include a grid-specific linear-time near-optimal CA scheme (NOCAG), which is a non-conflict graph based approach suggested in \cite{cite3}. We employ the three interference estimates being considered, \emph{viz.,} TID, CDAL$_{cost}$ and CXLS$_{wt}$, to implement three optimal brute force CA schemes (BFCA$_{TID}$, BFCA$_{CDAL}$ and BFCA$_{CXLS}$). Further, we implement three CA schemes (KOCA$_{TID}$, KOCA$_{CDAL}$ and KOCA$_{CXLS}$) based on the \textit{kernel optimization} CA classification proposed in \cite{cite2}, with the three metrics as the \textit{Interference Mitigation Functions}. Finally, we also include a \textit{hybrid optimized} CA scheme suggested in \cite{cite2} which employs CXLS$_{wt}$ to minimize interference. Thus, we have a CA test-set of 20 CA schemes to conduct a thorough evaluation of CALM and NETCAP.
 
 For WBT$_{9Mbps}$, we make use of 11 CA schemes, and scale up the CA test-set to include all 20 CAs for WBT$_{54Mbps}$. Our objective is to assess the theoretical estimation metrics, against experimental values of network performance parameters for the entire CA test-set, and we are not concerned with individual CA performances. Therefore, for ease of reference and presentation, we denote the CA schemes as $CA_n$, where $n \in \{{1\dots11}\}$ for WBT$_{9Mbps}$, and $n \in \{{1\dots20}\}$ for WBT$_{54Mbps}$.\\
 
 \item \textit{BBT CA Scheme Selection} : We create a CA test-set comprising of 20 GPCA and GDCA schemes, with the help of the Generic CA Algorithm~\ref{GCAA}, for the $5\times5$ IEEE 802.11g GWMN. We ensure that despite graph disruption, a path exists between every source-sink pair in the WMN, so that the data traffic flows can be established. Here too, we represent the CA schemes as $CA_n$, where $n \in \{{1\dots20}\}$. 
\end{enumerate}

  \begin{figure}[htb!]
                \centering
                \includegraphics[width=12cm, height=2.5cm]{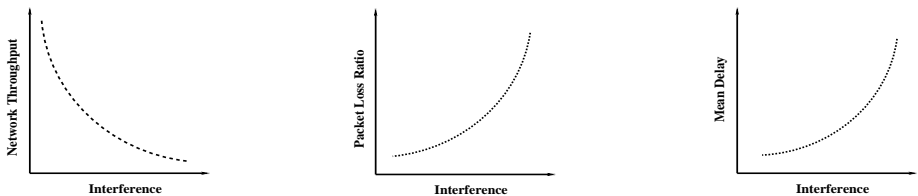}
                \caption{Expected correlation of CA performance parameters with interference.}
                \label{cor}
        \end{figure}
\subsection{Simulation Parameters and Test Scenario}

\subsubsection{Simulation Parameters}

We perform exhaustive simulations by implementing IEEE 802.11g $5\times5$ GWMN in ns-3, using the simulation parameters presented in Table~\ref{sim}. A datafile is transmitted from the source to the destination, over multi-hop flows. We use TCP and UDP as the underlying transport layer protocols, implemented through \textit{BulkSendApplication} and \textit{UdpClientServer} frameworks inbuilt in ns-3. The TCP application offers us the \textit{Network Aggregate Throughput} (NAT) in Mbps, while through UDP simulations we determine the \textit{Packet Loss Ratio} (PLR) as a $\%$, and the \textit{mean delay} (MD) in $\mu$seconds. For WBT$_{9Mbps}$ and WBT$_{54Mbps}$, we consider all three network performance parameters, while for BBT$_{54Mbps}$ we consider only NAT, as it is a stress-test case.
\subsubsection{Test Scenario}
Multi-hop data transmissions are the characteristic feature of WMNs. To create maximal adverse impact of interference, we design a test-scenario in which all nodes of the network participate in the communication of data traffic. Nodes assume one or more of the three network roles \emph{viz.}, a source, a destination and an intermediate data relaying node. In the $5\times5$ GWMN depicted inFigure~\ref{BADTID2}~(b), we consider the nodes of first row and first column as source nodes, and nodes of the last row and last column serve as their corresponding destinations. A 10 MB datafile is sent across every source-sink pair via multi-hop transmissions or $n$-Hop-Flows (nHFs). Thus, we create the test-scenario R$_5$C$_5$ consisting of ten 4HFs, which forms the benchmark on which the overall performance of a CA can be assessed. Due to a large number of concurrent multi-hop flows it is an ideal scenario to observe CA efficiency in transmitting high traffic loads which reflects CA resilience to  elevated interference 
levels. 
         

   \begin{figure*}[ht!]
  \centering%
  \begin{tabular}{cc}
   \subfloat[TID vs Throughput]{\includegraphics[width=.25\linewidth]{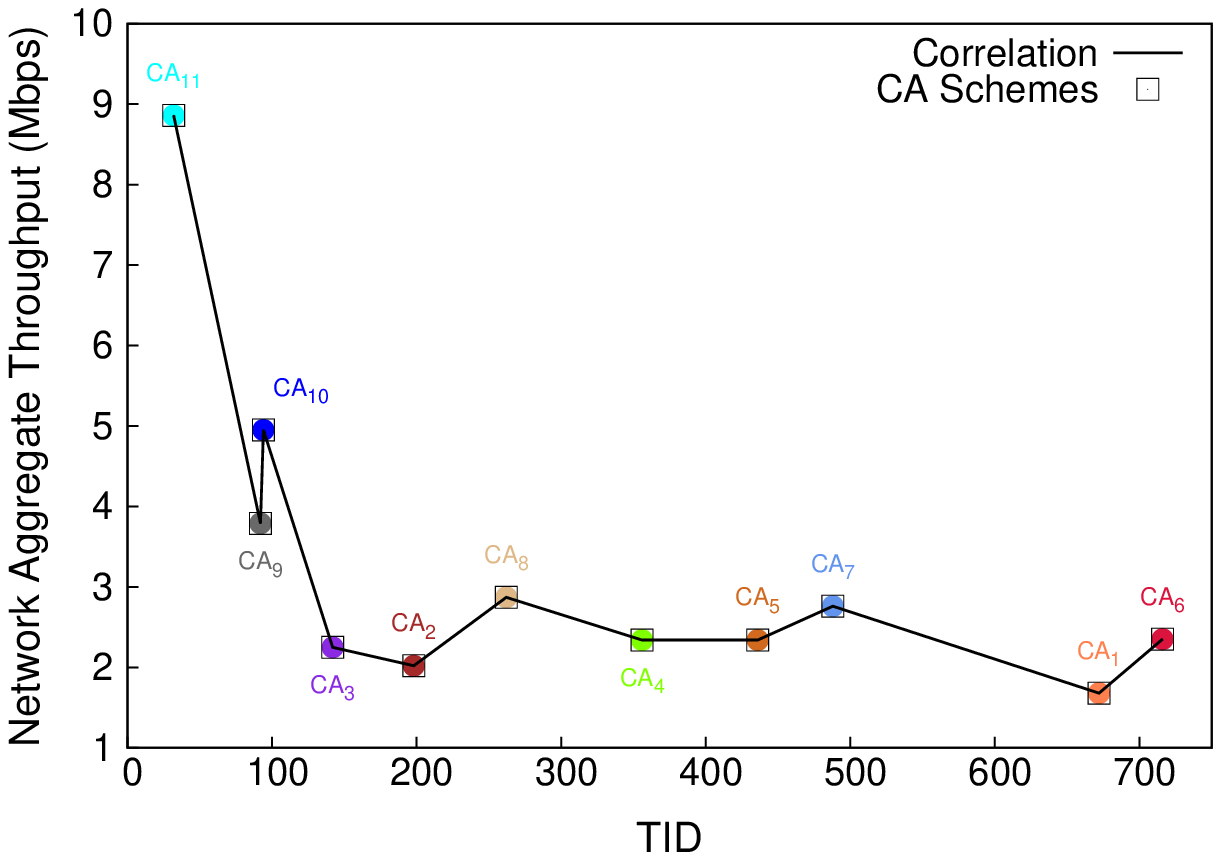}}\hfill%
    \subfloat[CDAL$_{cost}$ vs Throughput]{\includegraphics[width=.25\linewidth]{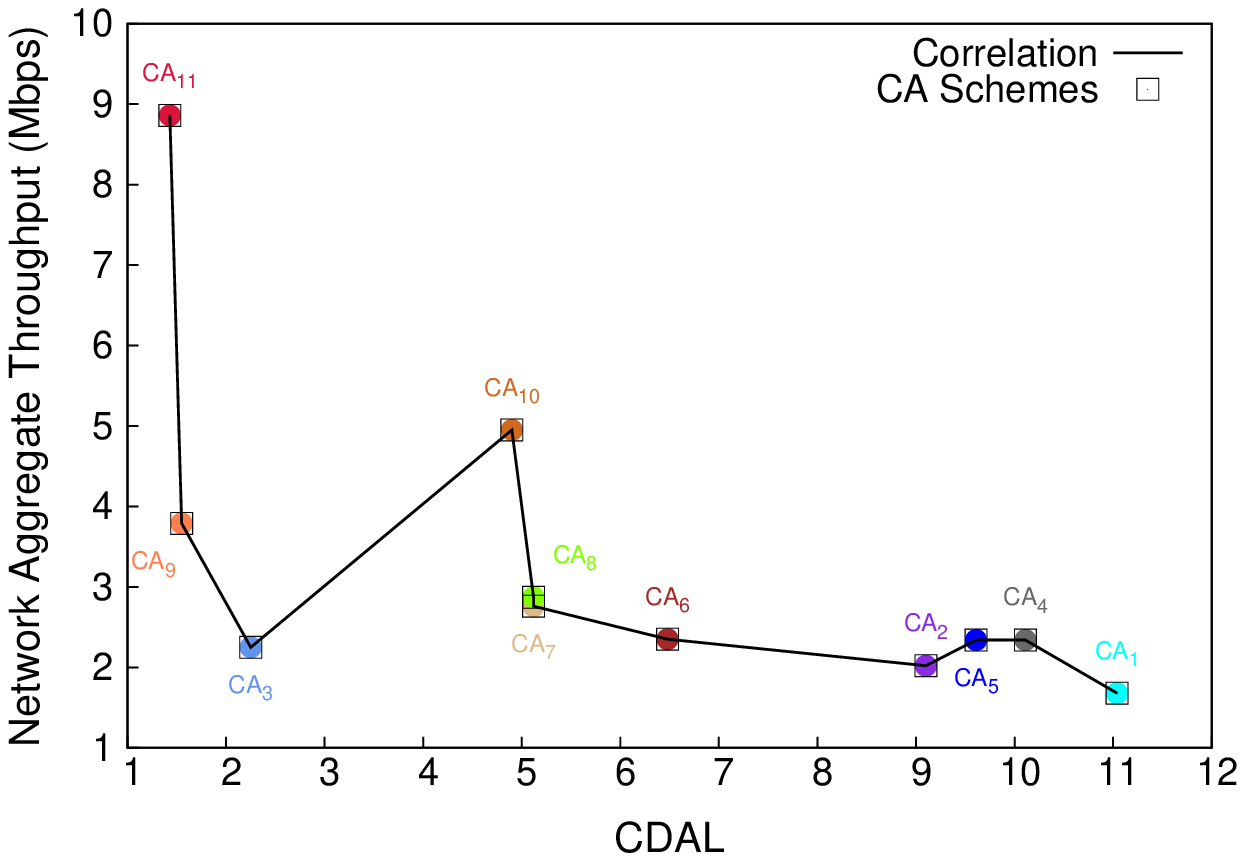}}\hfill%
   \subfloat[CXLS$_{wt}$ vs Throughput] {\includegraphics[width=.25\linewidth]{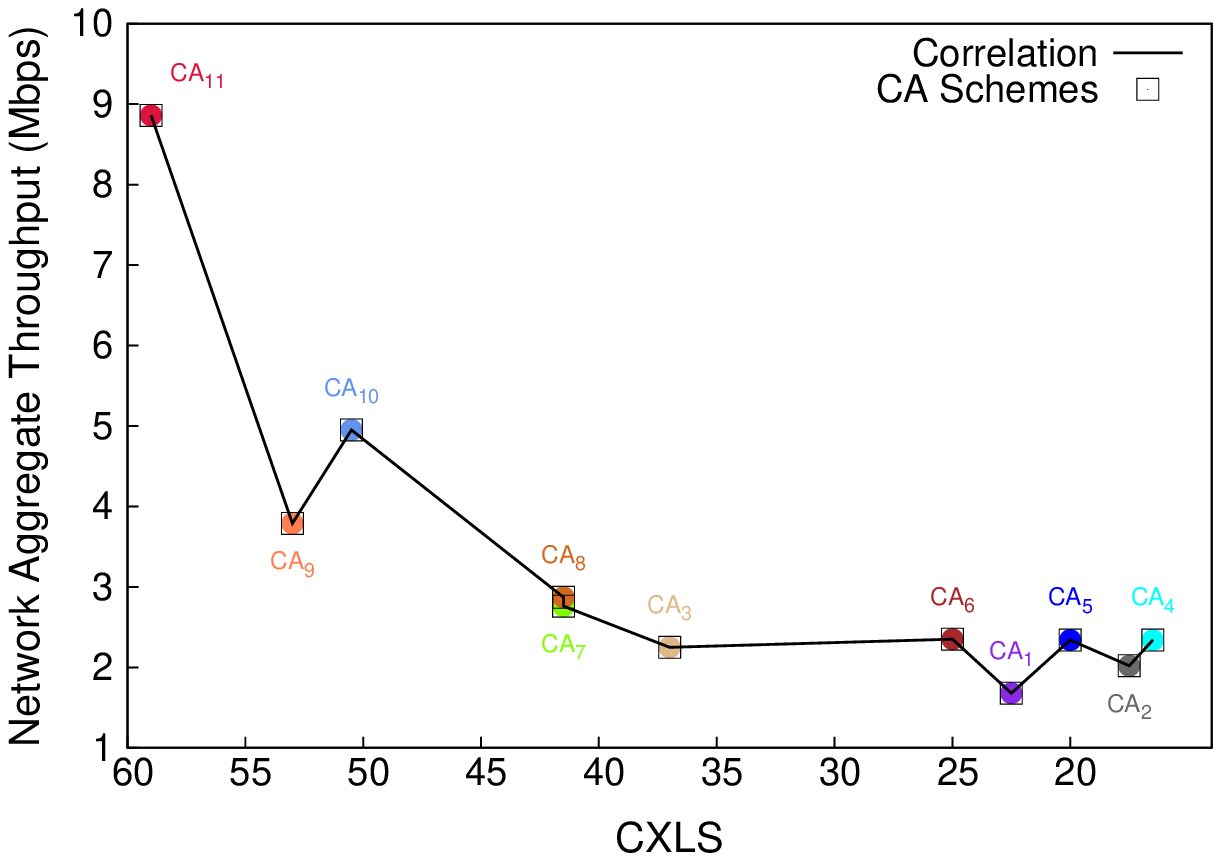}}\hfill%
   \subfloat[CALM vs Throughput] {\includegraphics[width=.25\linewidth]{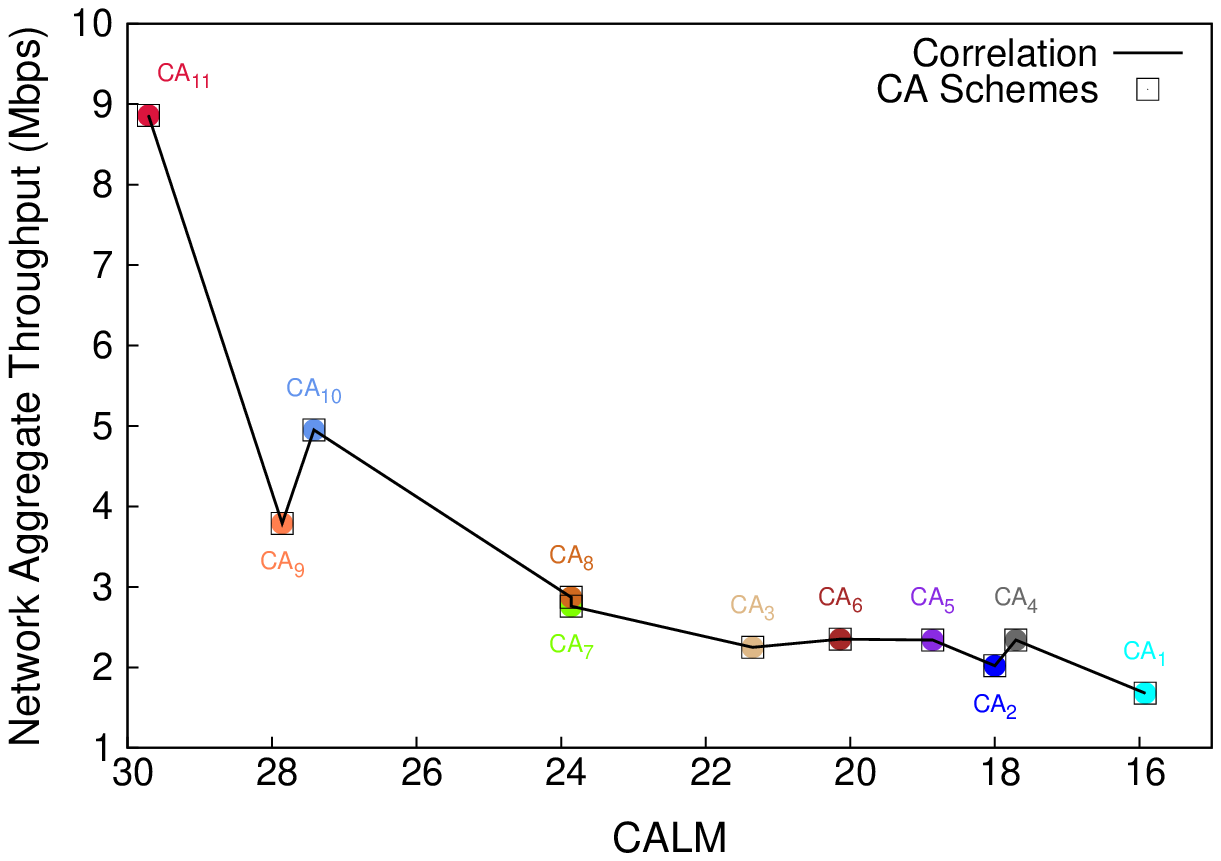}}%
    \end{tabular}
    \caption{WBT (9 Mbps) : Observed correlation of estimation metrics \& Throughput.} 
     \label{9Th}
\end{figure*}

   \begin{figure*}[ht!]
 
  \centering%
  \begin{tabular}{cc}
   \subfloat[TID vs PLR]{\includegraphics[width=.25\linewidth ]{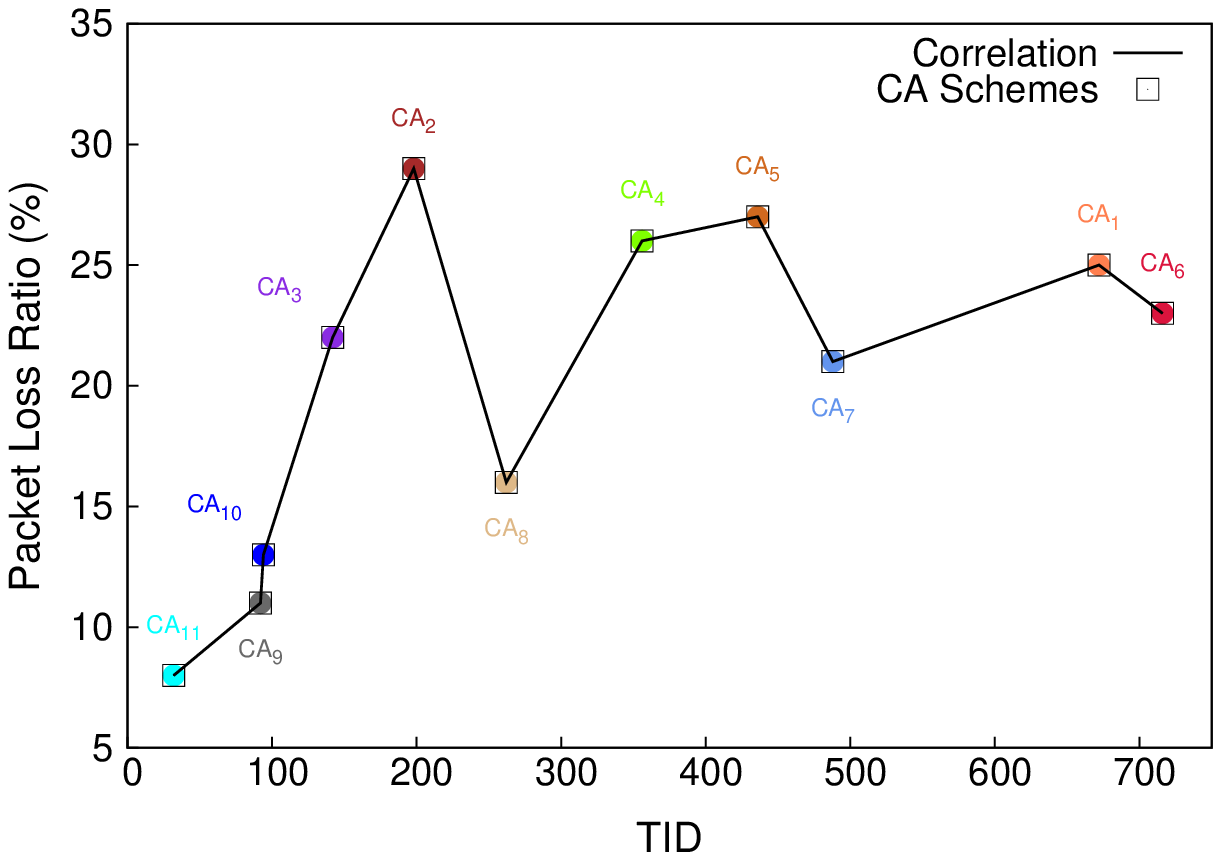}}\hfill%
    \subfloat[CDAL$_{cost}$ vs PLR]{\includegraphics[width=.25\linewidth ]{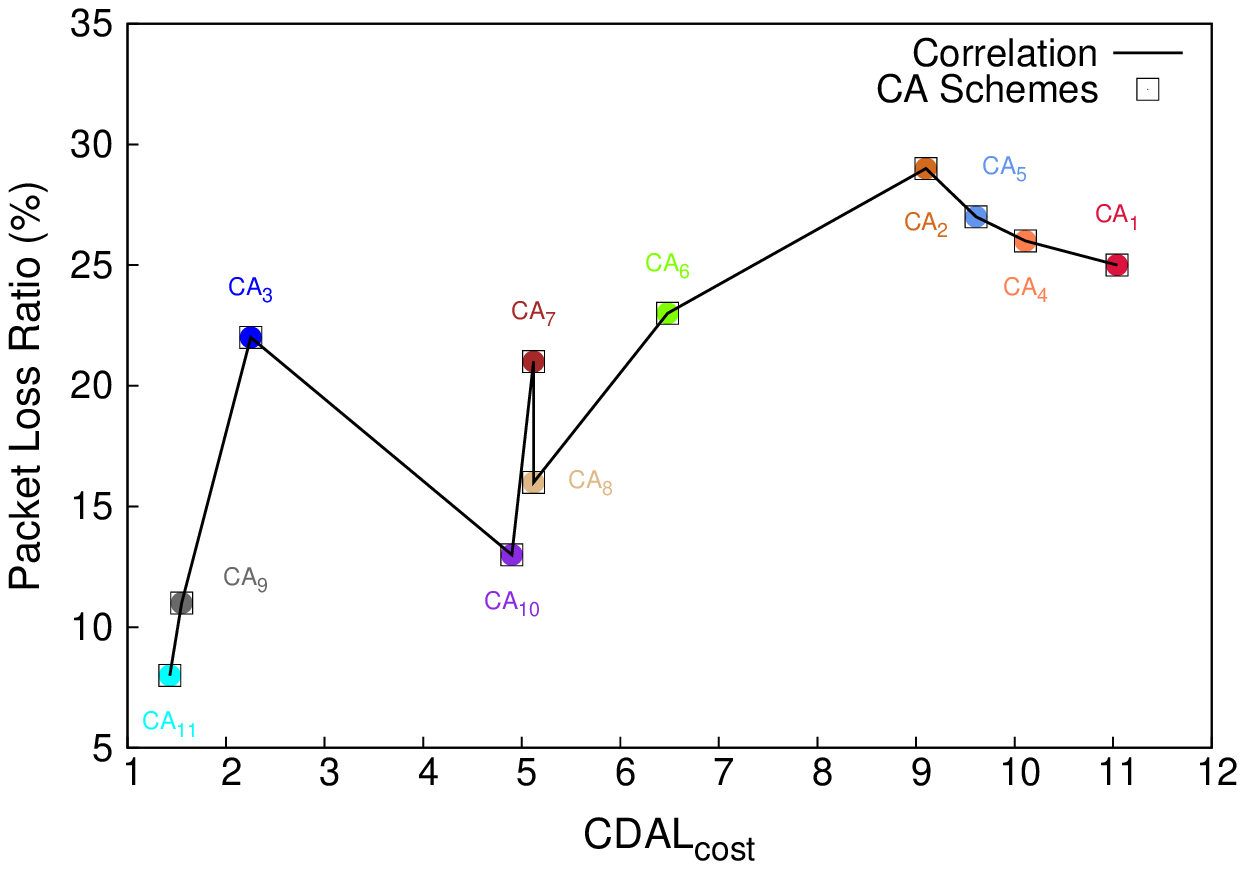}}\hfill%
   \subfloat[CXLS$_{wt}$ vs PLR] {\includegraphics[width=.25\linewidth ]{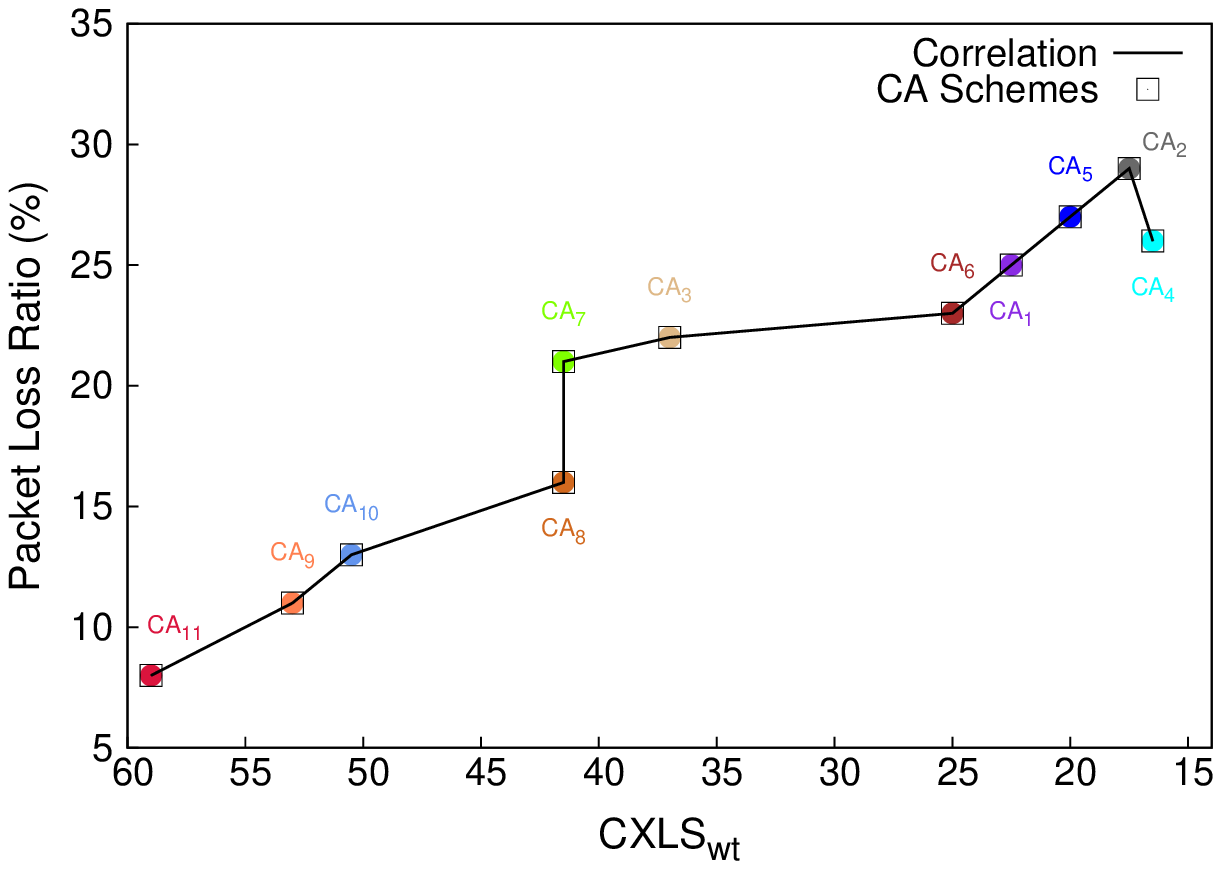}}\hfill%
   \subfloat[CALM vs PLR] {\includegraphics[width=.25\linewidth ]{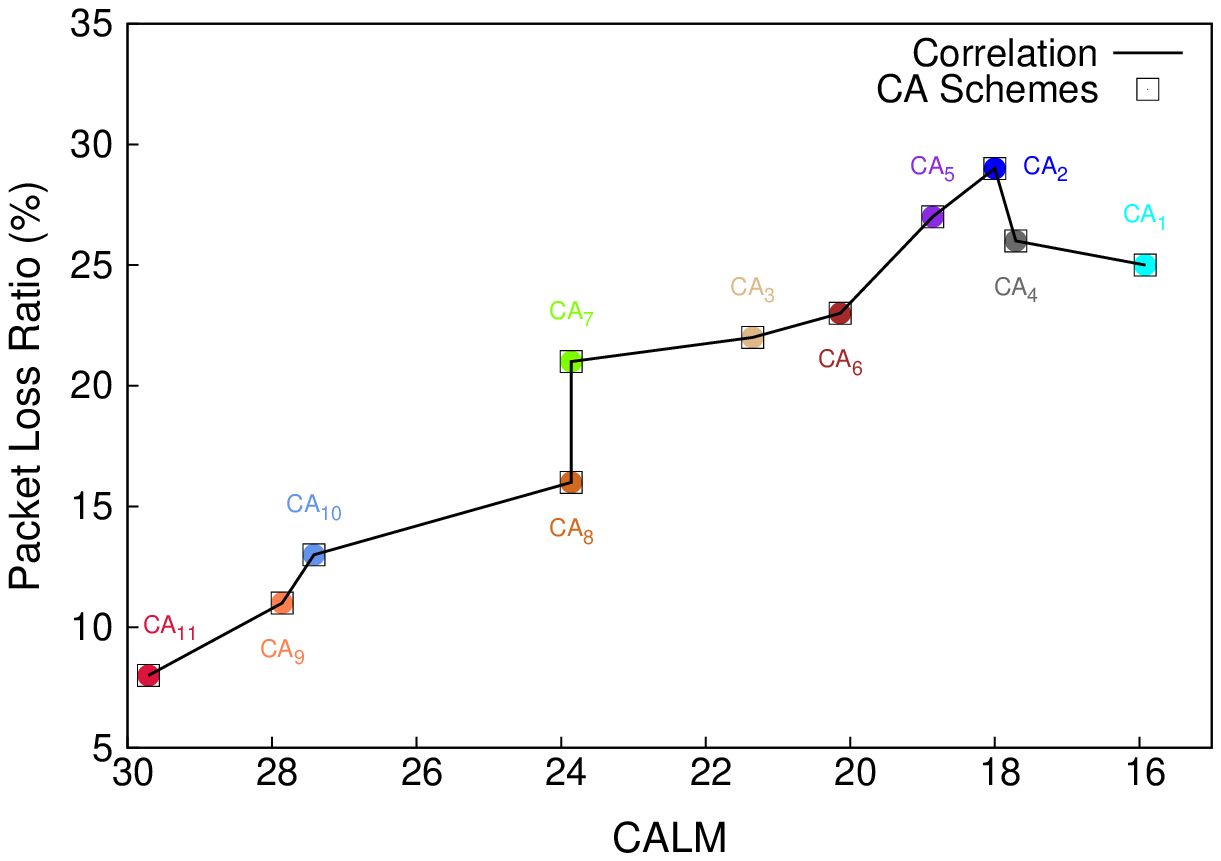}}%
    \end{tabular}
    \caption{WBT (9 Mbps) : Observed correlation of estimation metrics \& Packet Loss Ratio.} 
     \label{9P}
\end{figure*}

   \begin{figure*}[ht!]
 
  \centering%
  \begin{tabular}{cc}
   \subfloat[TID vs MD]{\includegraphics[width=.25\linewidth ]{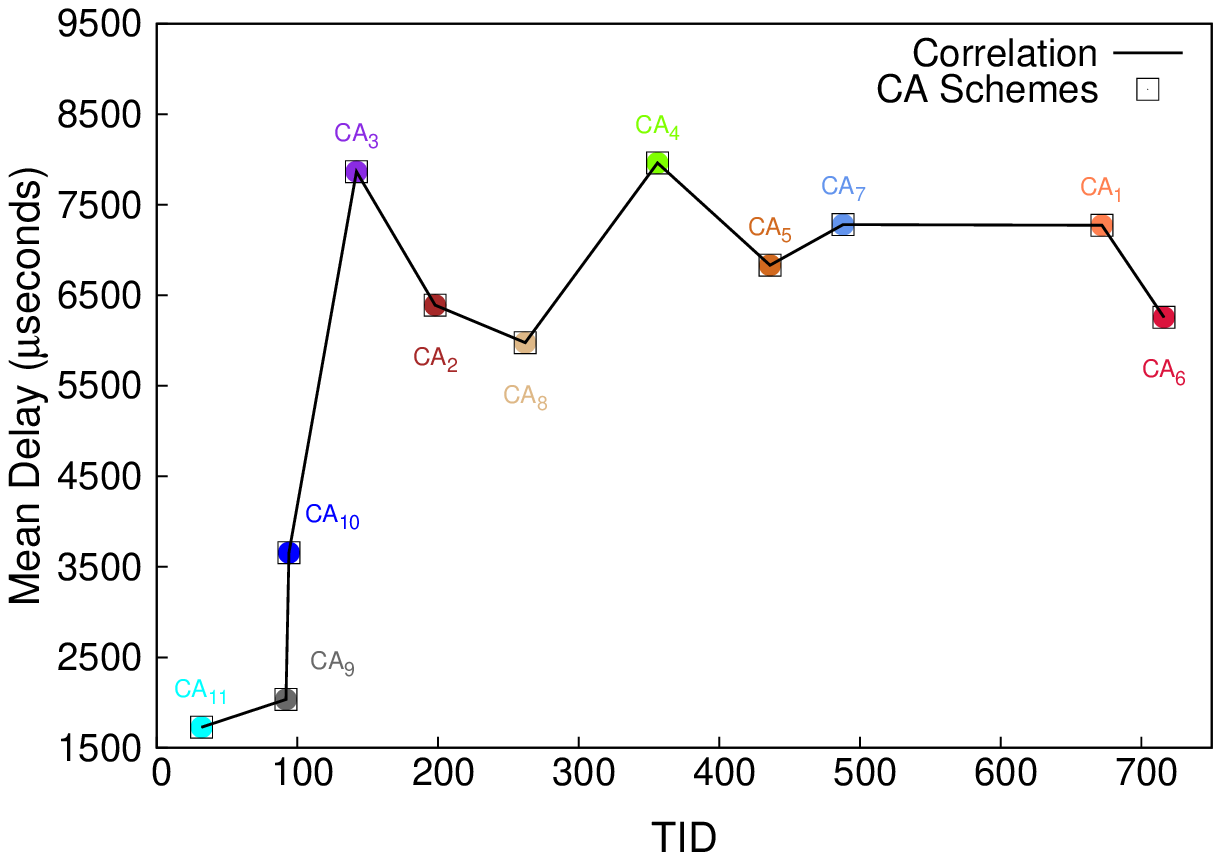}}\hfill%
    \subfloat[CDAL$_{cost}$ vs MD]{\includegraphics[width=.25\linewidth ]{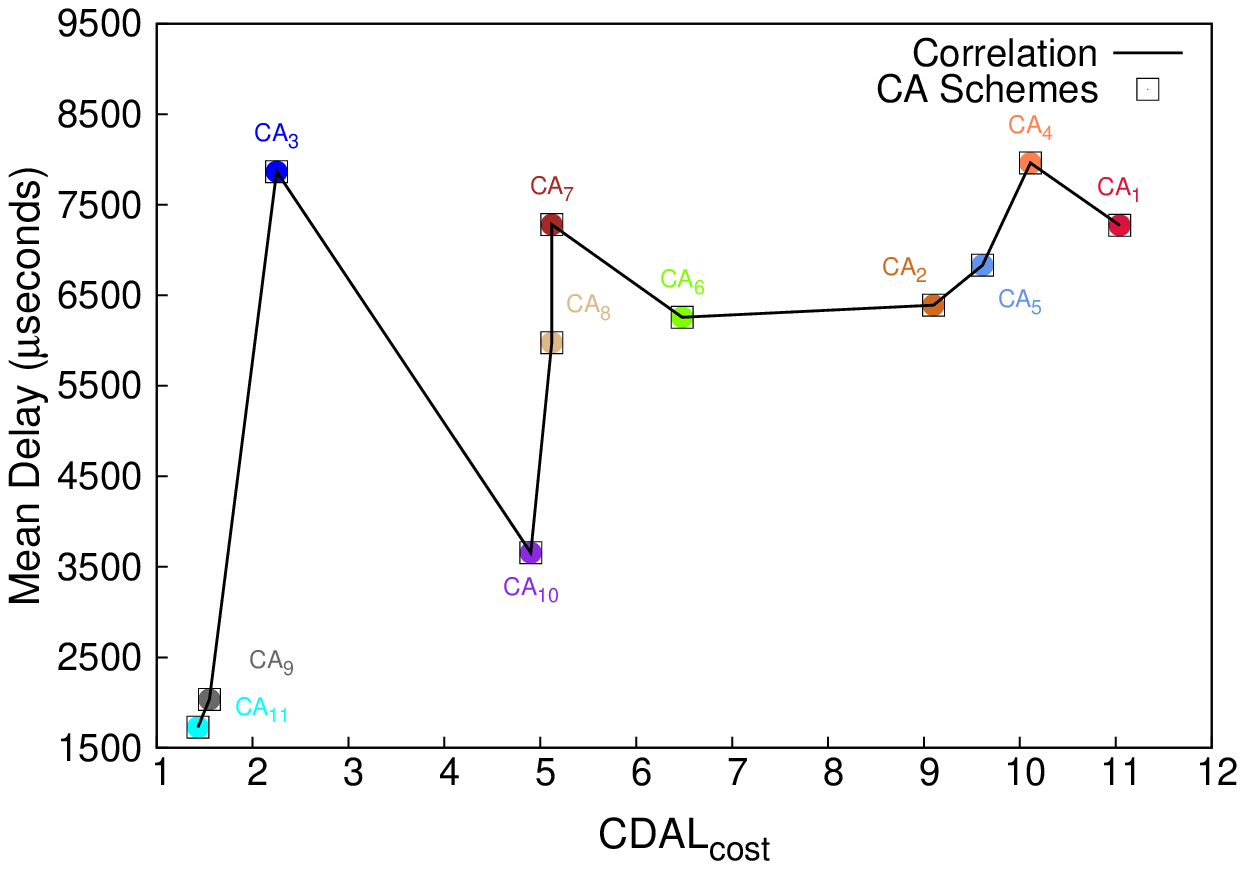}}\hfill%
   \subfloat[CXLS$_{wt}$ vs MD] {\includegraphics[width=.25\linewidth ]{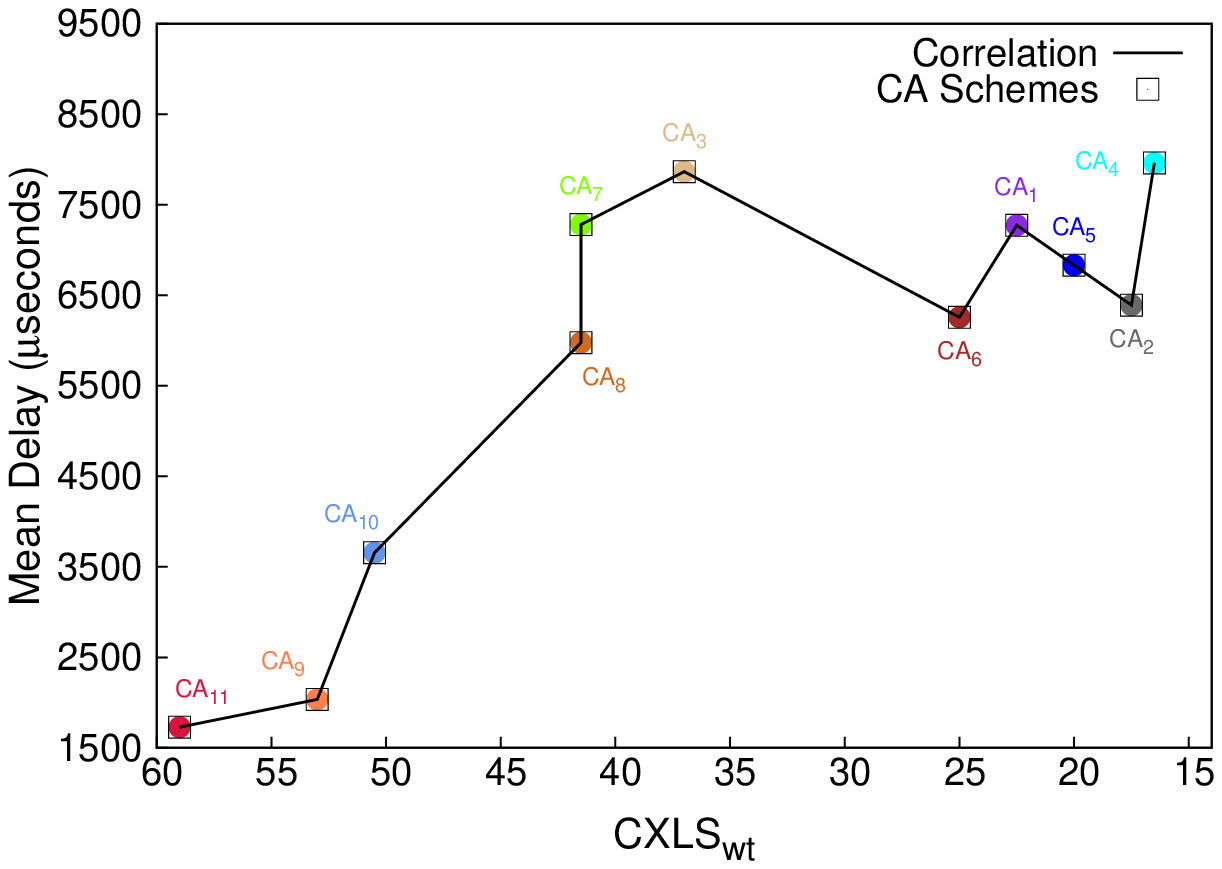}}\hfill%
   \subfloat[CALM vs MD] {\includegraphics[width=.25\linewidth ]{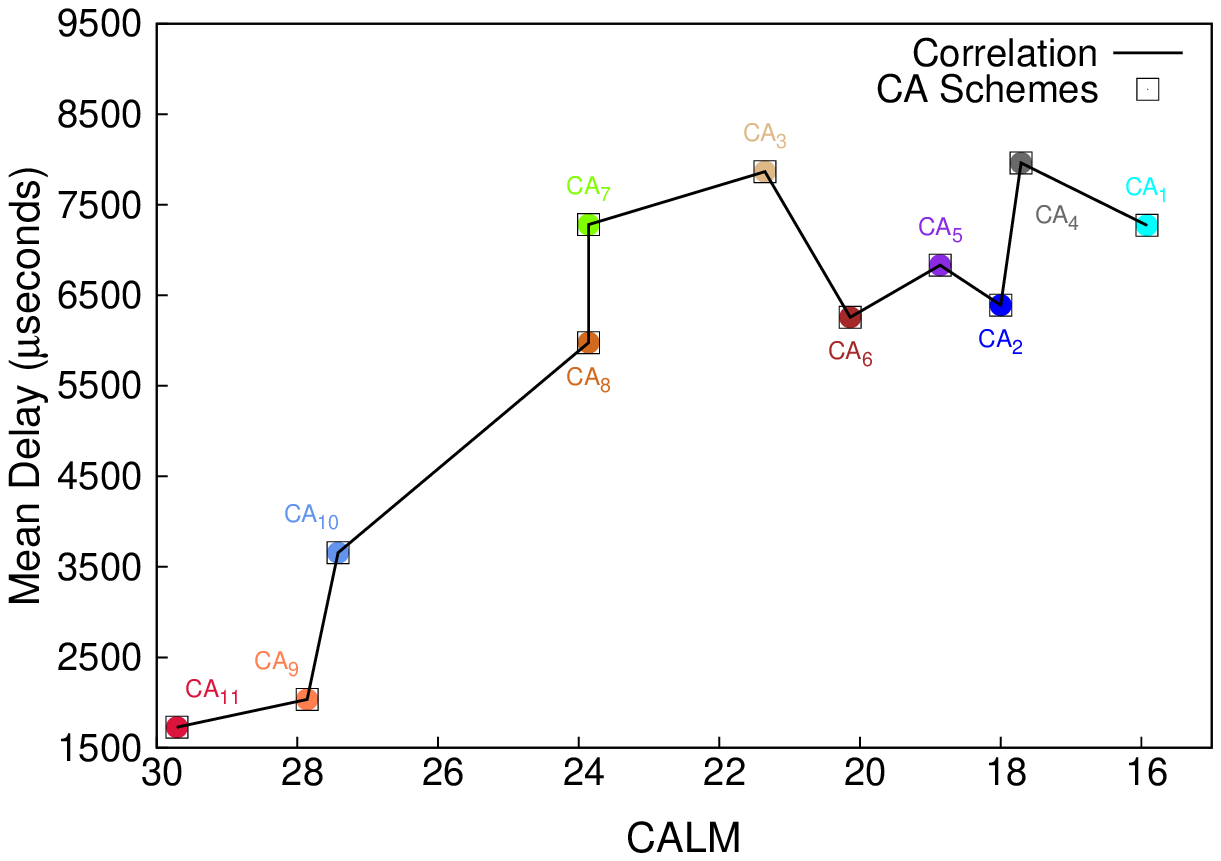}}%
    \end{tabular}
    \caption{WBT (9 Mbps) : Observed correlation of estimation metrics \& Mean Delay.} 
     \label{9M}
\end{figure*}

\subsection{Evaluation Methodology}
Our evaluation methodology is a multi-stage stage process involving the following steps :
\begin{enumerate}
 \item Consider the interference estimation metrics \emph{i.e.,} TID, CDAL$_{cost}$, CXLS$_{wt}$ and CALM, and compute the theoretical interference estimates for each CA scheme.
 \item Identify the expected correlation between interference and network performance indicators \emph{viz.,} NAT, PLR and MD. 
 \item Run simulations to observe the real-time values of network performance indicators.
 \item Determine the actual relationship between the theoretical estimates of interference and the recorded network performance indicators.
 \item Compare the actual relationship to the expected correlation and determine the accuracy of the theoretical interference estimation metric.
\end{enumerate}

The expected correlation of wireless network performance and the endemic interference is illustrated in Figure~\ref{cor}. It can be discerned that the capacity of a WMN deteriorates with the increase in the intensity of interference. WMNs also experience a rise in data packet corruption and delivery failures and an increase in end-to-end latency in packet delivery. A theoretical estimate of interference is considered to be reliable and accurate if it exhibits similar relationships with the experimentally observed network performance parameters. We evaluate the various interference estimates on their ability to conform to the expected patterns. Further, for each observed performance parameter, we arrange the CAs in a \textit{sequence} of increasing performance, which serves as the \textit{reference sequence}. Similarly, we arrange the CAs in the increasing order of expected performance, based on the estimates of the CA performance prediction metrics. Thereafter, we compare the reference CA sequence based on 
experimental data with CA sequence derived from theoretical estimates, to determine the \textit{Errors In Sequence} (EIS). In a sequence of $n$ CAs, there exist a total of  $\textsuperscript n C_2$ \textit{value comparisons} of observed parameters or theoretical estimates, between individual CA schemes. The pairwise comparisons of theoretically expected CA performances are verified against the experimentally determined reference sequence, and the total number of comparison mismatches are considered as \textit{Errors In Sequence} of that theoretical metric. An error in pairwise comparison signifies that the theoretically expected performance relationship between two CAs, as predicted by the estimation metric, is contrary to the actual relationship observed through experimental results. Since EIS for a particular interference estimation metric is the sum of all erroneous pairwise comparisons in its CA sequence, it represents the level of accuracy in the predictions of that estimation metric. 
Finally, we evaluate the \textit{Measure of Accuracy} (MoA) of an interference estimation scheme, which is given by the expression $MoA = (1-(EIS/\textsuperscript n C_2))\times100$, where $n$ is the number of CAs in the sequence.
\subsection{CALM : Results and Analysis} \label{C}
%
 
 \begin{figure*}[ht!]
   \centering%
  \begin{tabular}{cc}
   \subfloat[TID vs Throughput]{\includegraphics[width=.25\linewidth]{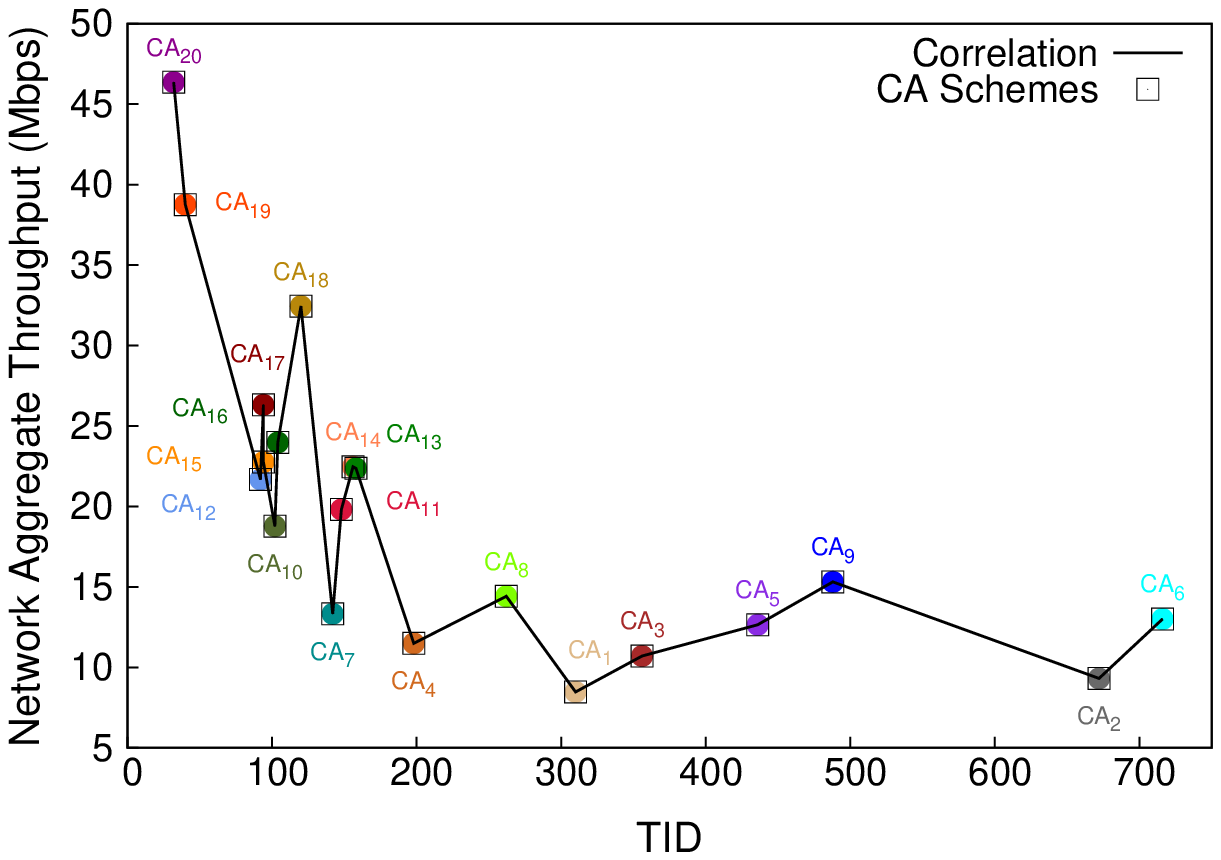}}\hfill%
    \subfloat[CDAL$_{cost}$ vs Throughput]{\includegraphics[width=.25\linewidth]{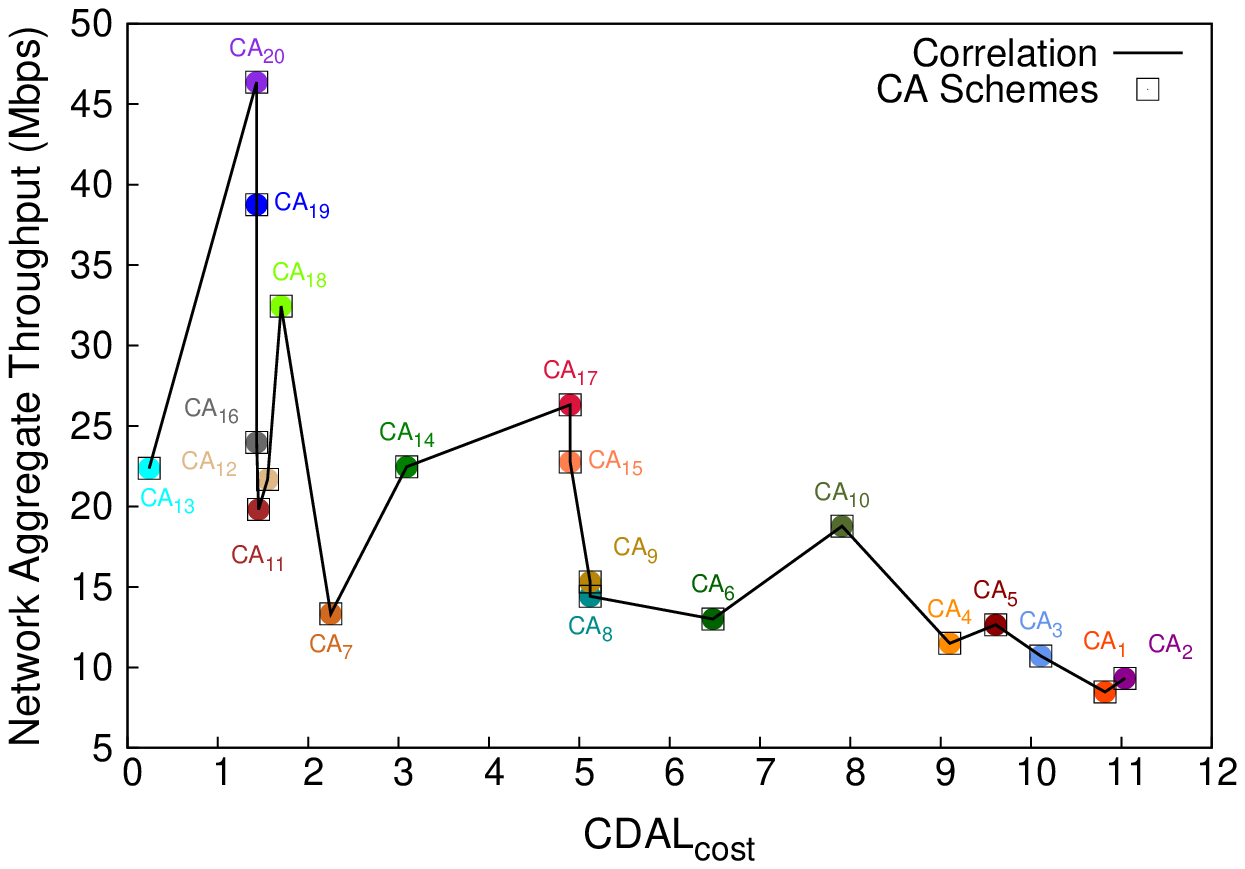}}\hfill%
    
   \subfloat[CXLS$_{wt}$ vs Throughput] {\includegraphics[width=.25\linewidth]{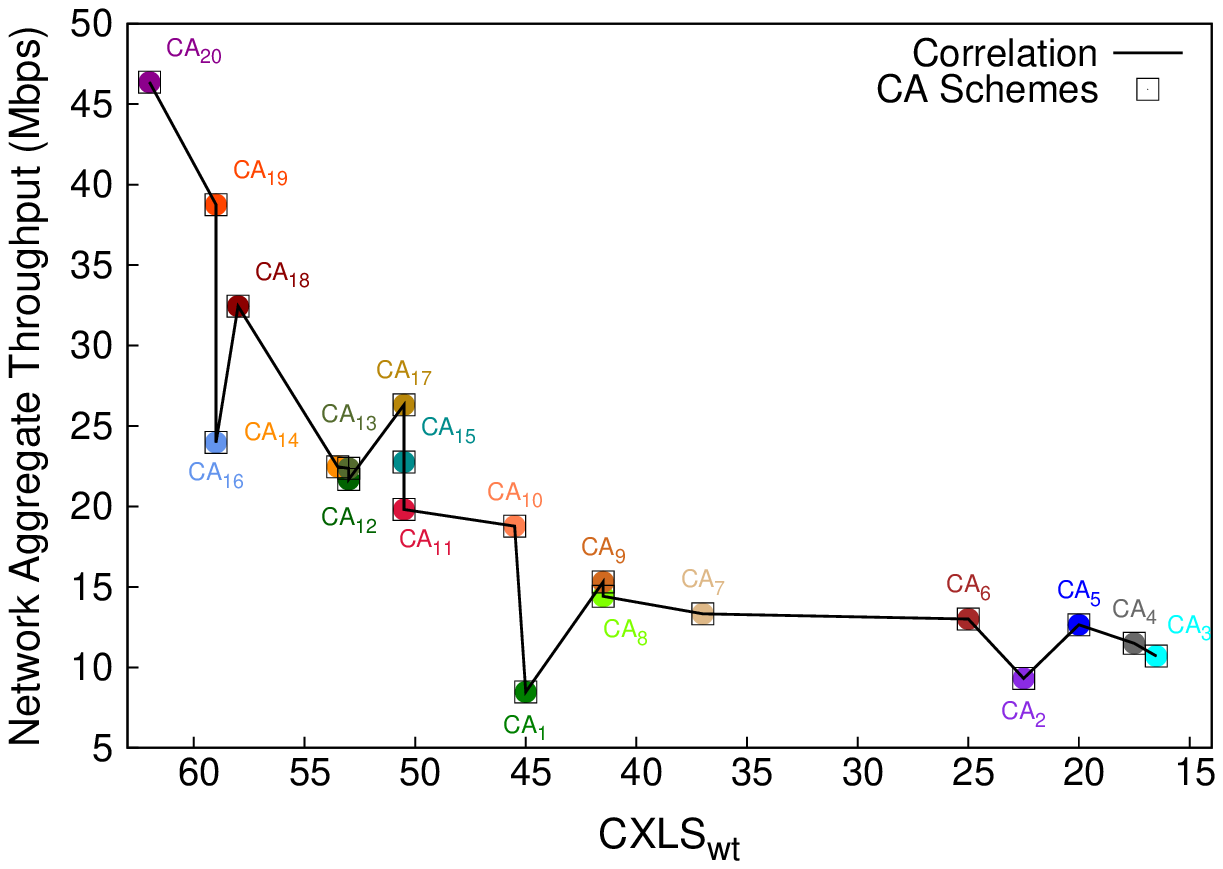}}\hfill%
   \subfloat[CALM vs Throughput] {\includegraphics[width=.25\linewidth]{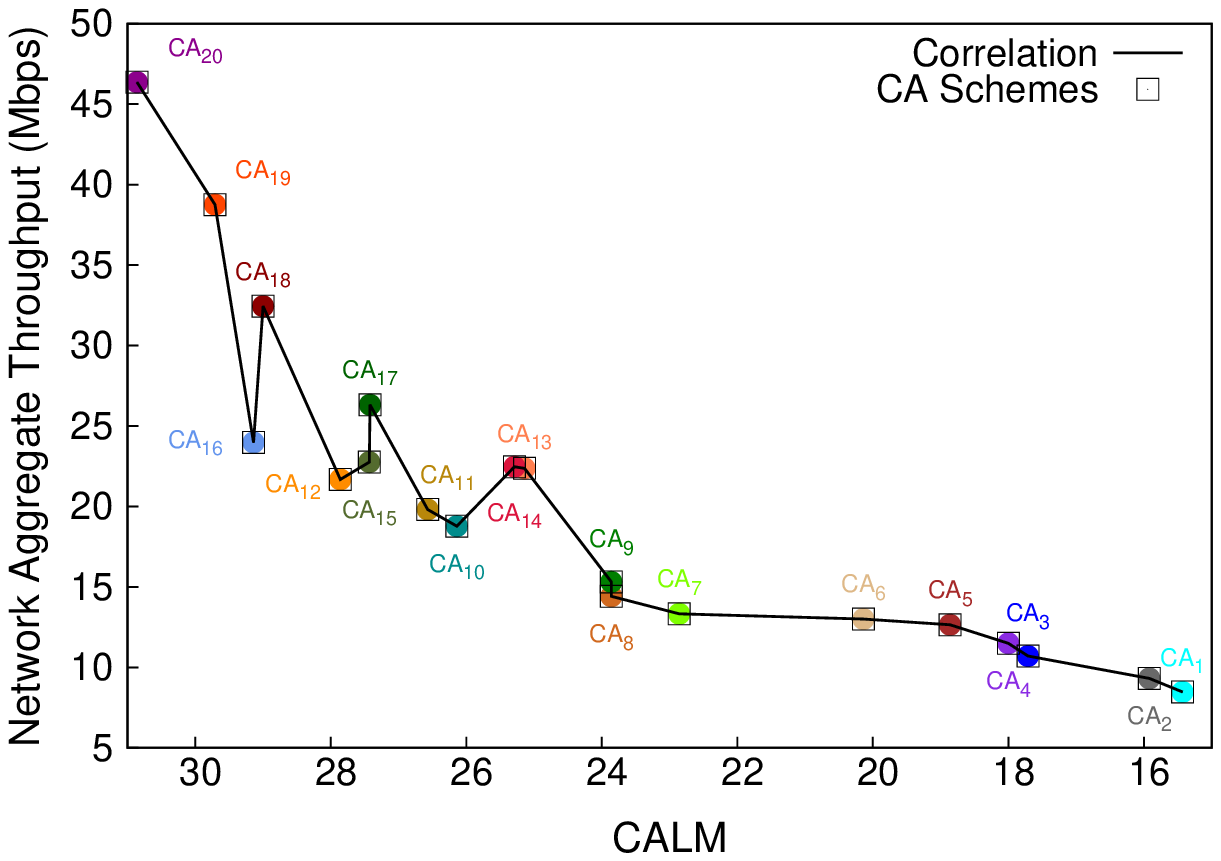}}%
    \end{tabular}
    \caption{WBT (54 Mbps) : Observed correlation of estimation metrics \& Throughput.} 
     \label{54Th}
\end{figure*}
   \begin{figure*}[ht!]
  \centering%
  \begin{tabular}{cc}
   \subfloat[TID vs PLR]{\includegraphics[width=.25\linewidth ]{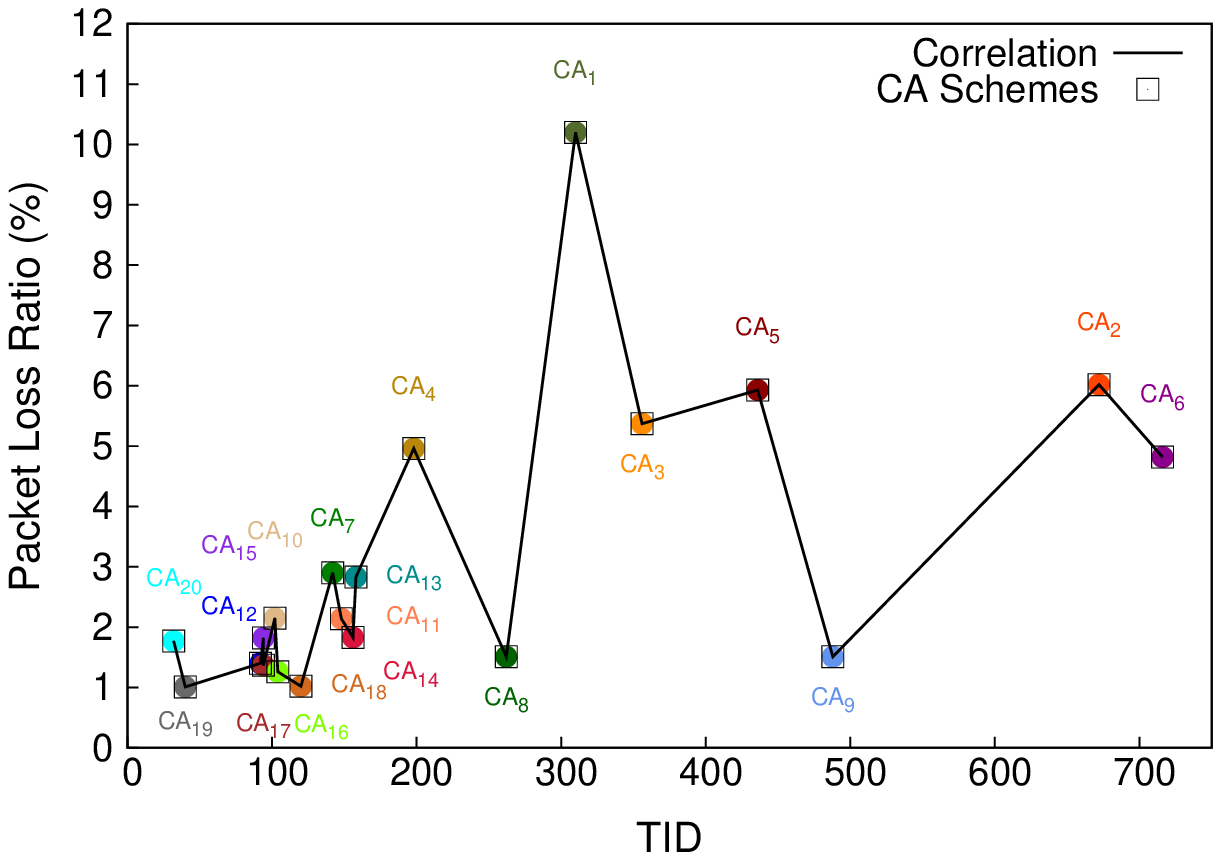}}\hfill%
    \subfloat[CDAL$_{cost}$ vs PLR]{\includegraphics[width=.25\linewidth ]{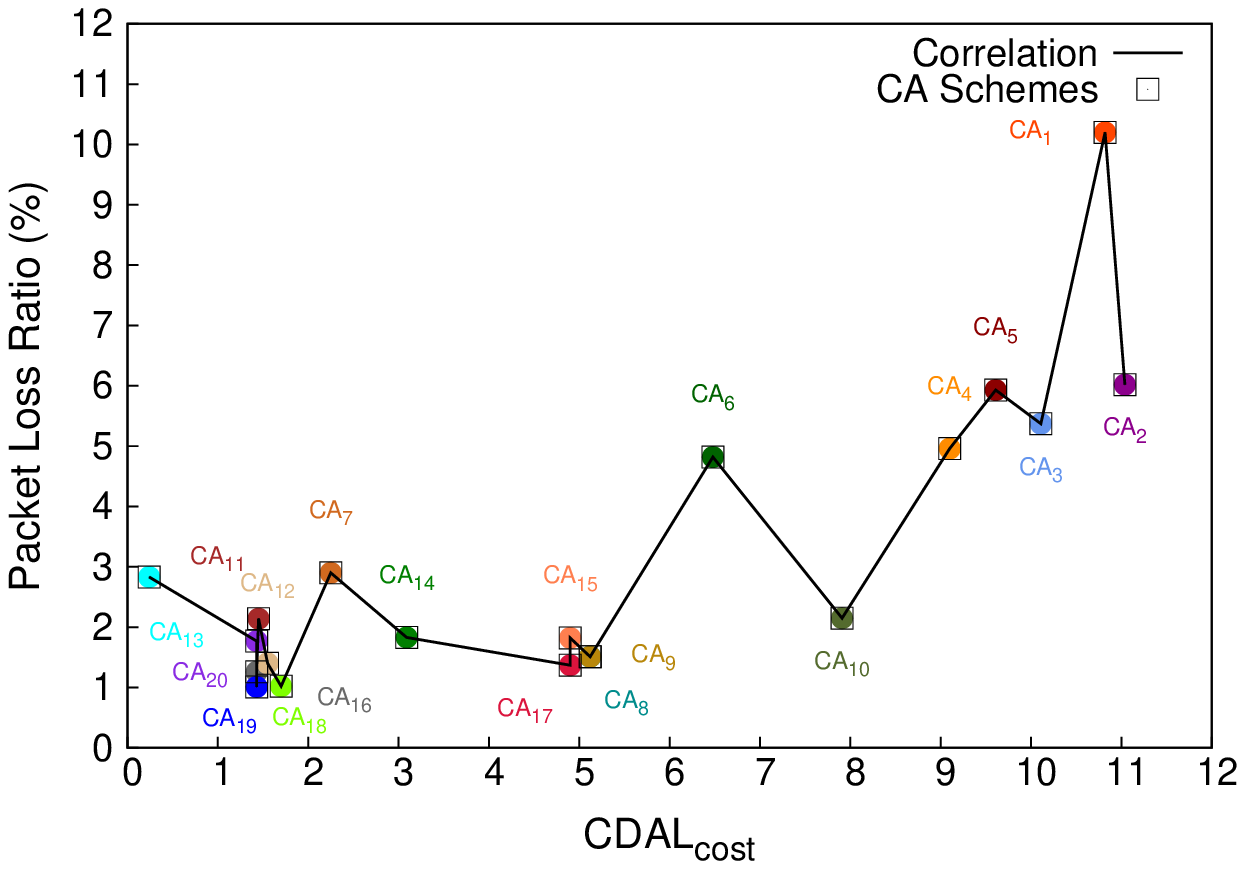}}\hfill%
   \subfloat[CXLS$_{wt}$ vs PLR] {\includegraphics[width=.25\linewidth ]{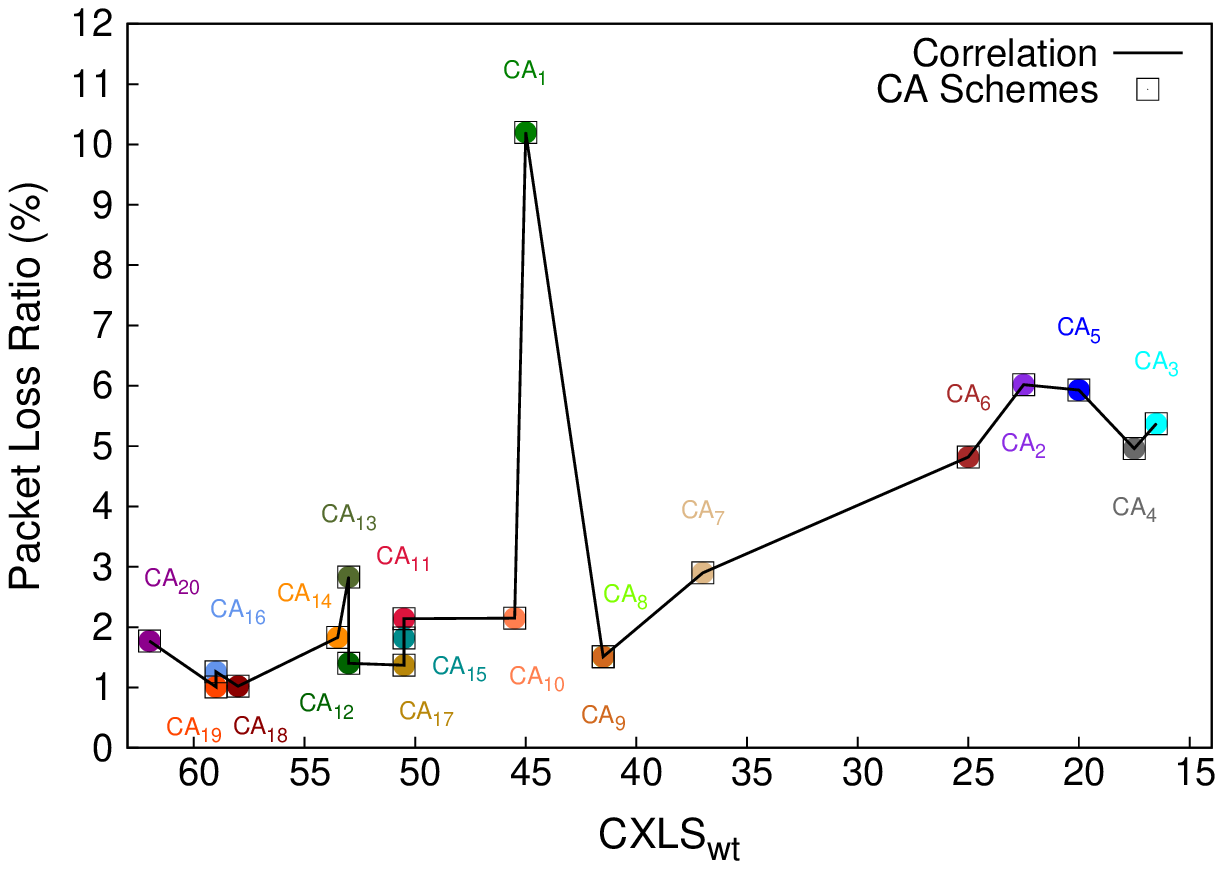}}\hfill%
   \subfloat[CALM vs PLR] {\includegraphics[width=.25\linewidth ]{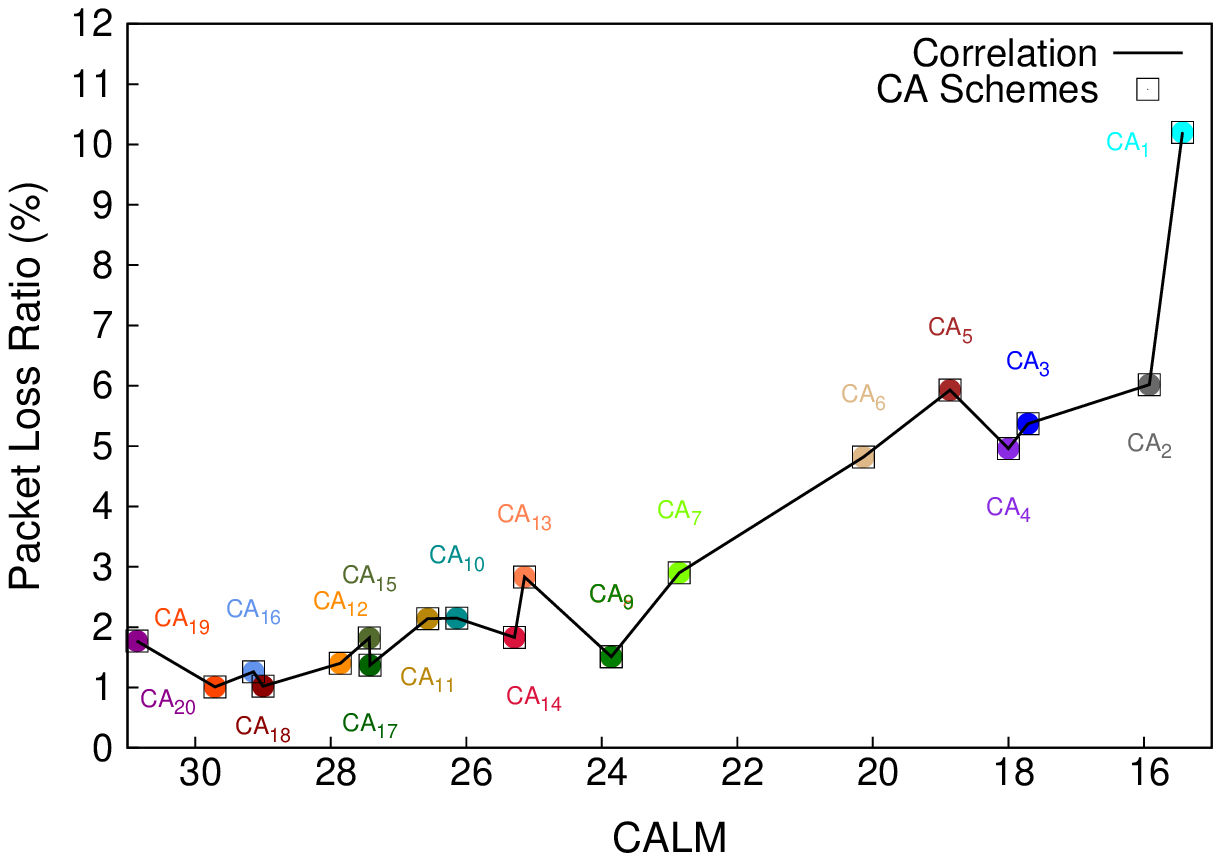}}%
    \end{tabular}
    \caption{WBT (54 Mbps) : Observed correlation of estimation metrics \& Packet Loss Ratio} 
     \label{54P}
\end{figure*}

   \begin{figure*}[ht!]
  \centering%
  \begin{tabular}{cc}
   \subfloat[TID vs MD]{\includegraphics[width=.25\linewidth ]{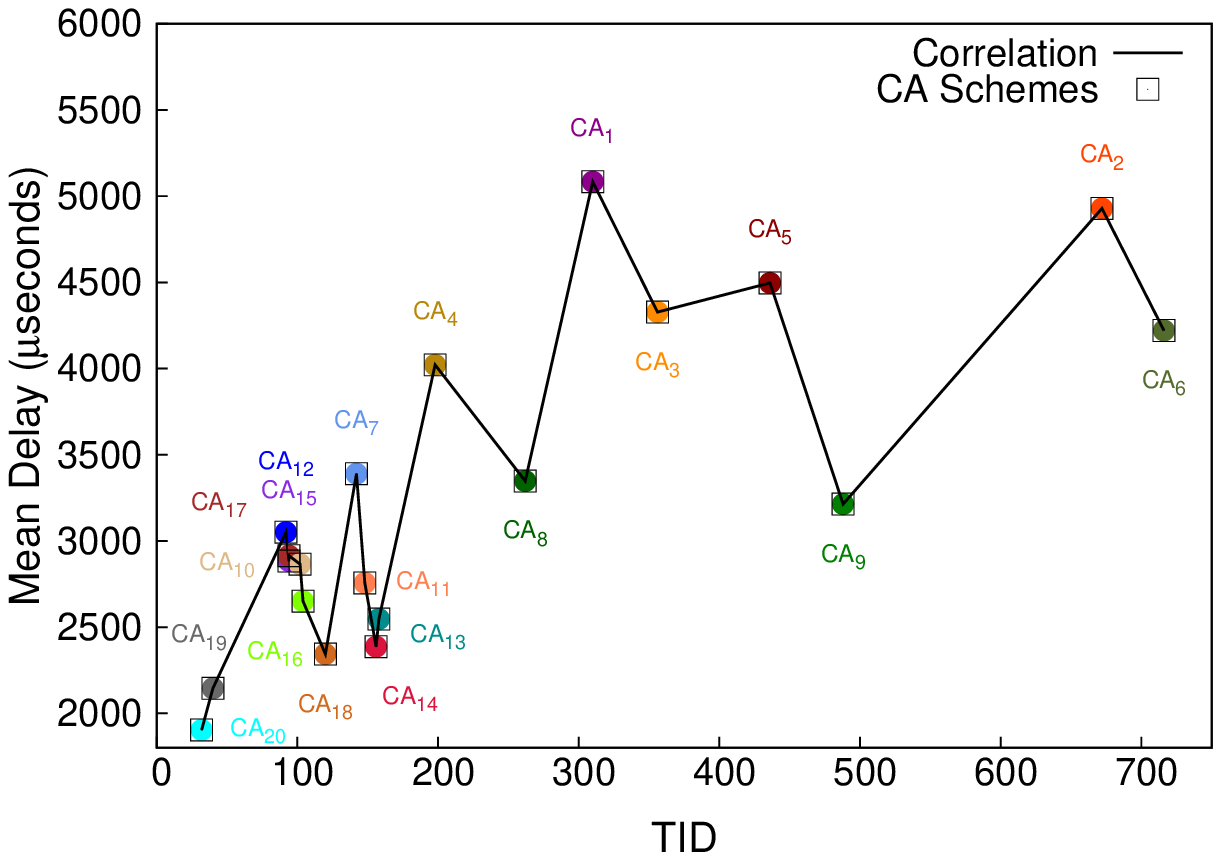}}\hfill%
    \subfloat[CDAL$_{cost}$ vs MD]{\includegraphics[width=.25\linewidth ]{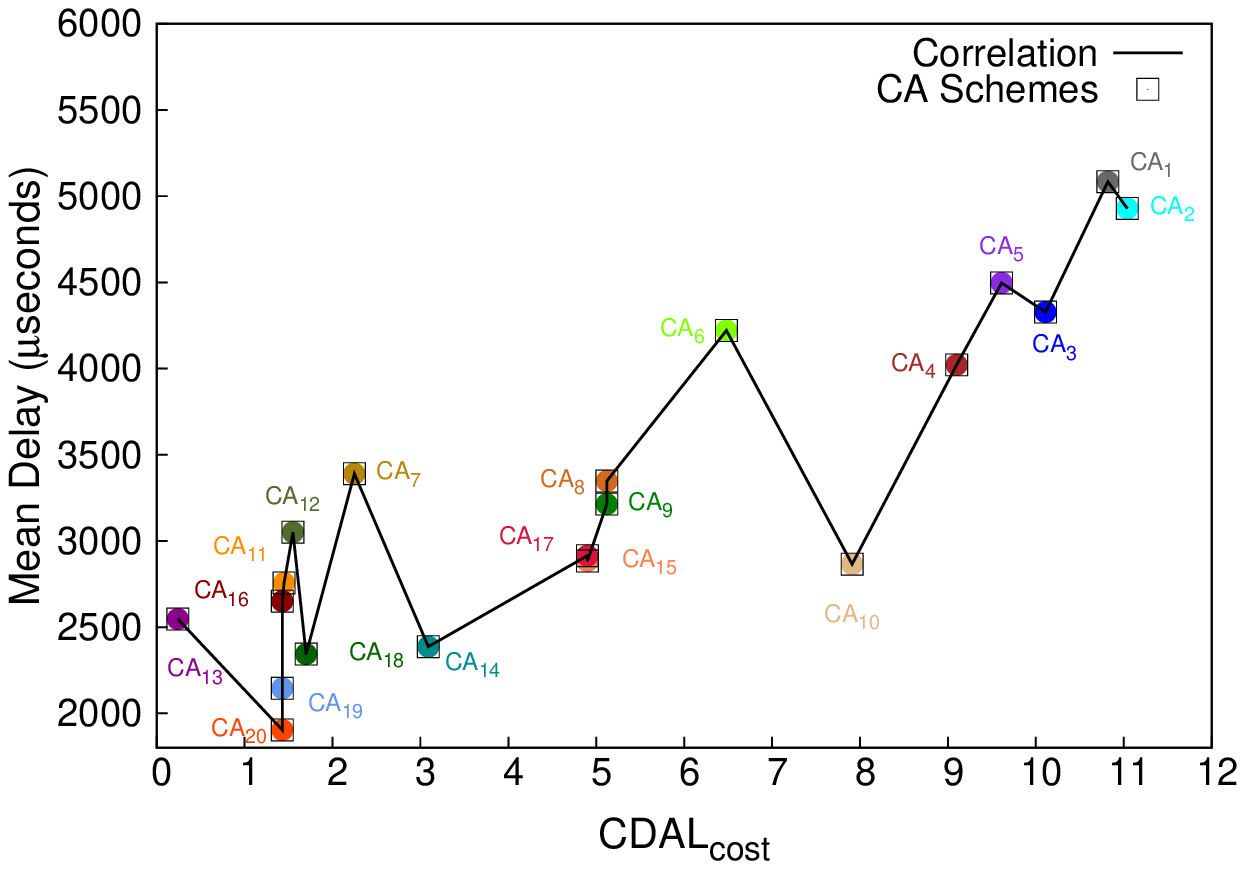}}\hfill%
   \subfloat[CXLS$_{wt}$ vs MD] {\includegraphics[width=.25\linewidth ]{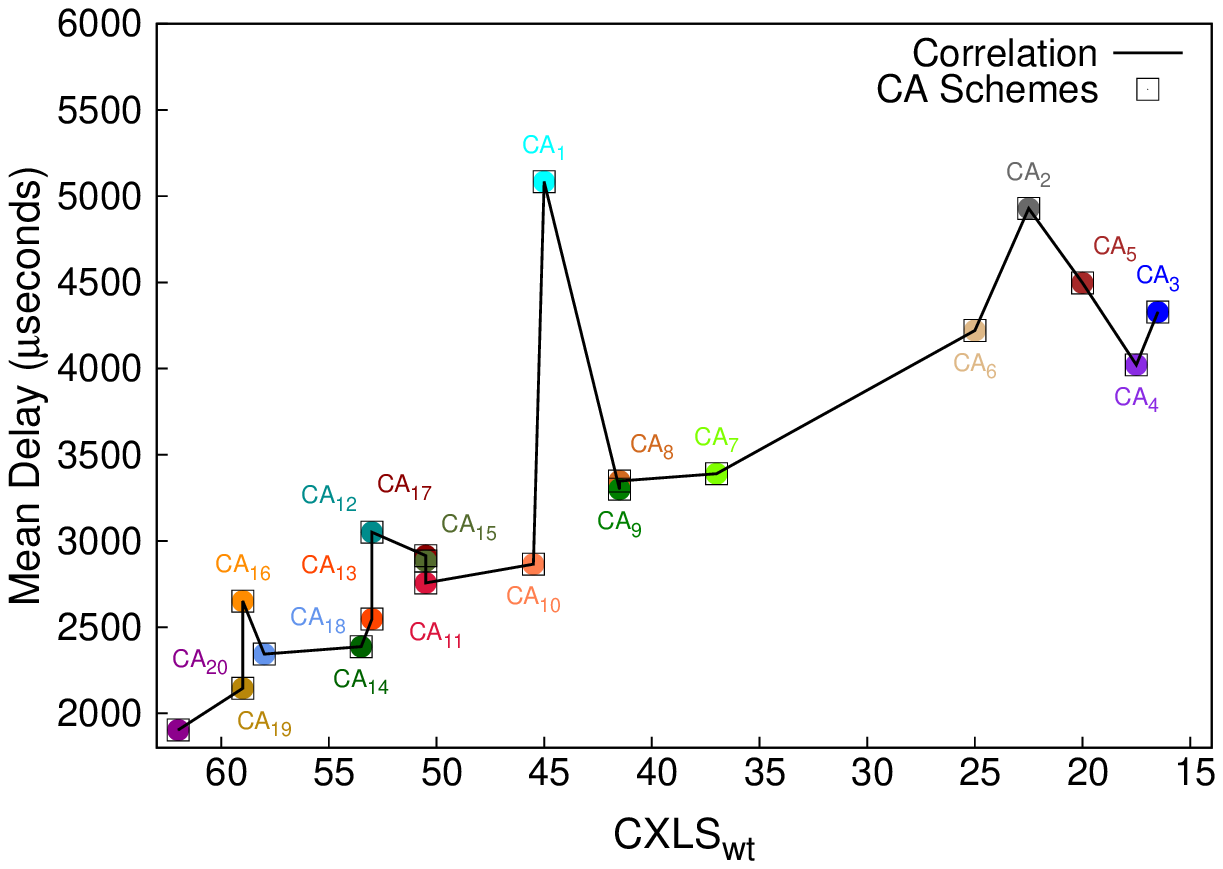}}\hfill%
   \subfloat[CALM vs MD] {\includegraphics[width=.25\linewidth ]{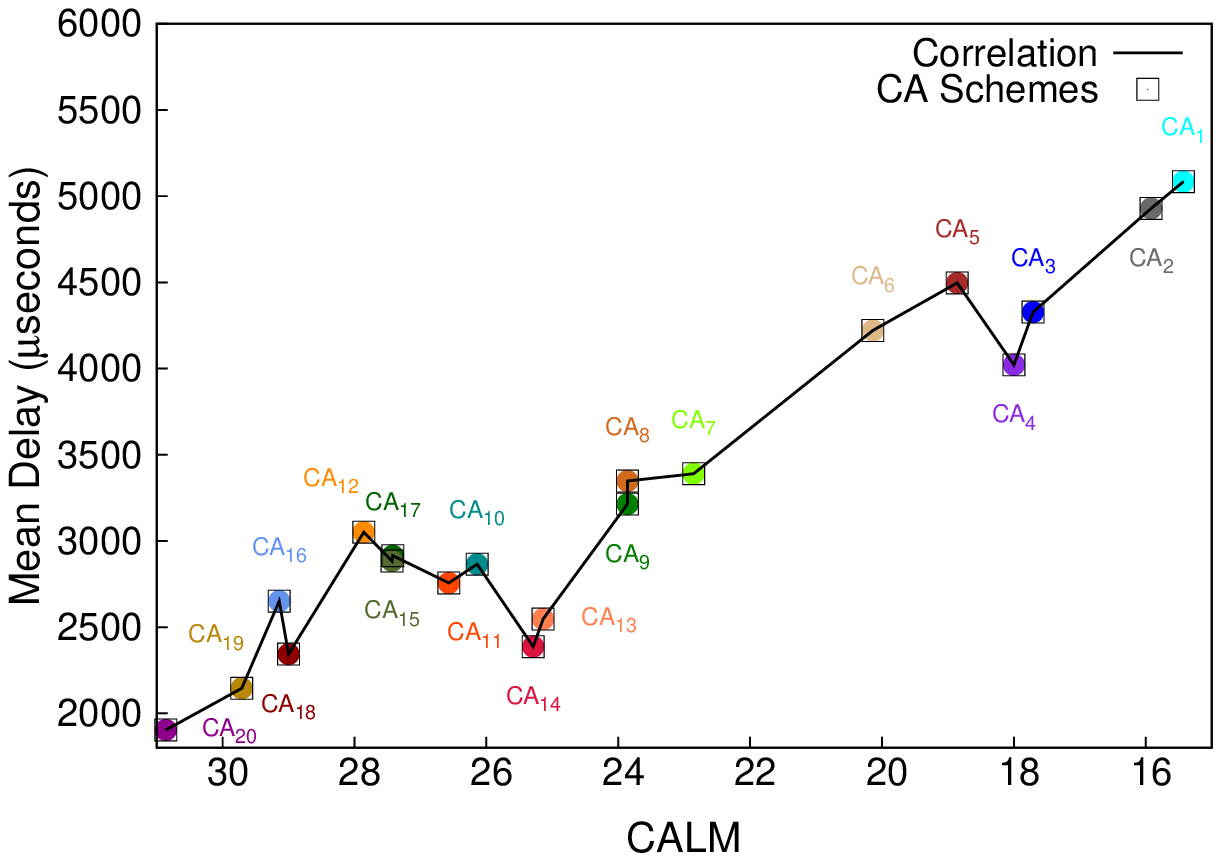}}%
    \end{tabular}
    \caption{WBT (54 Mbps) : Observed correlation of estimation metrics \& Mean Delay.} 
     \label{54M}
\end{figure*}
\subsubsection{WBT$_{9Mbps}$}
\begin{figure*}[ht!]
  \centering%
  \begin{tabular}{cc}
    \subfloat[WBT (9 Mbps)]{\includegraphics[width=.33\linewidth]{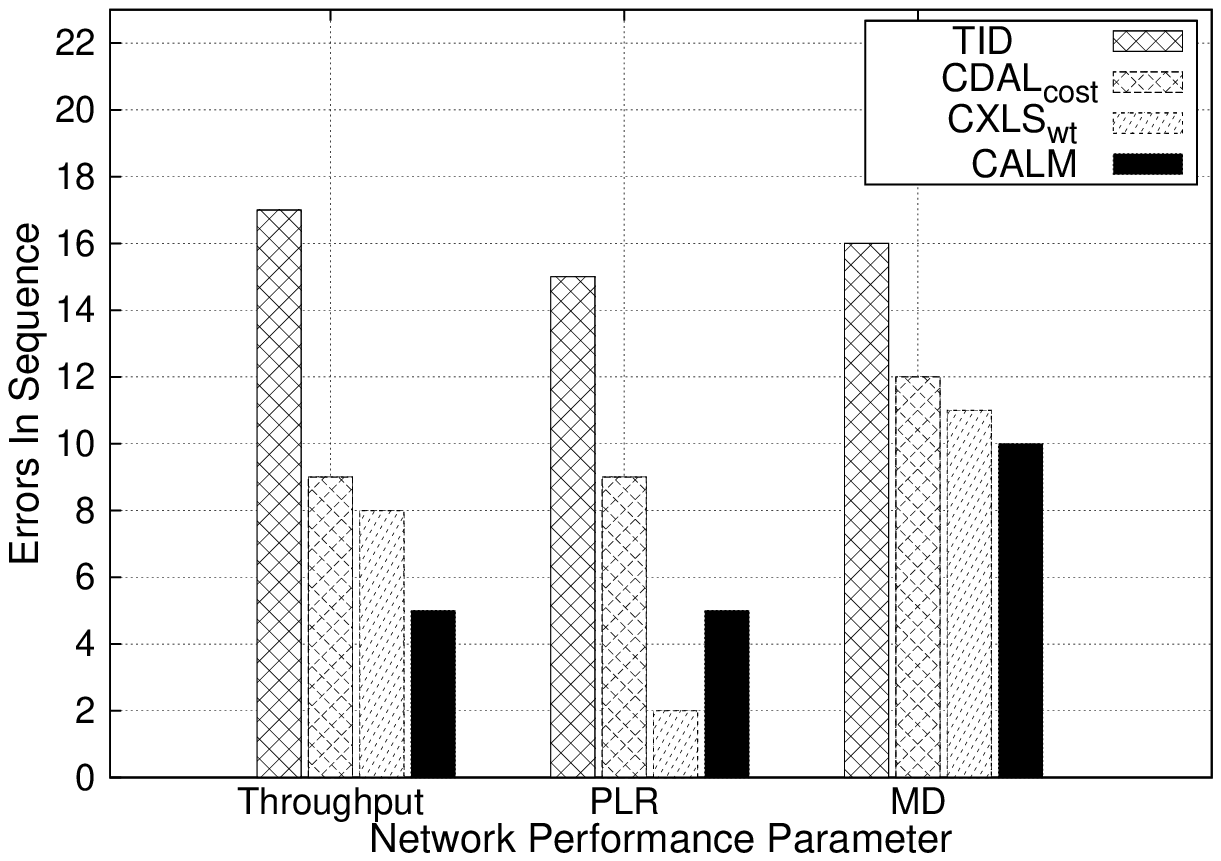}}\hfill%
    \subfloat[WBT (54 Mbps)]{\includegraphics[width=.33\linewidth]{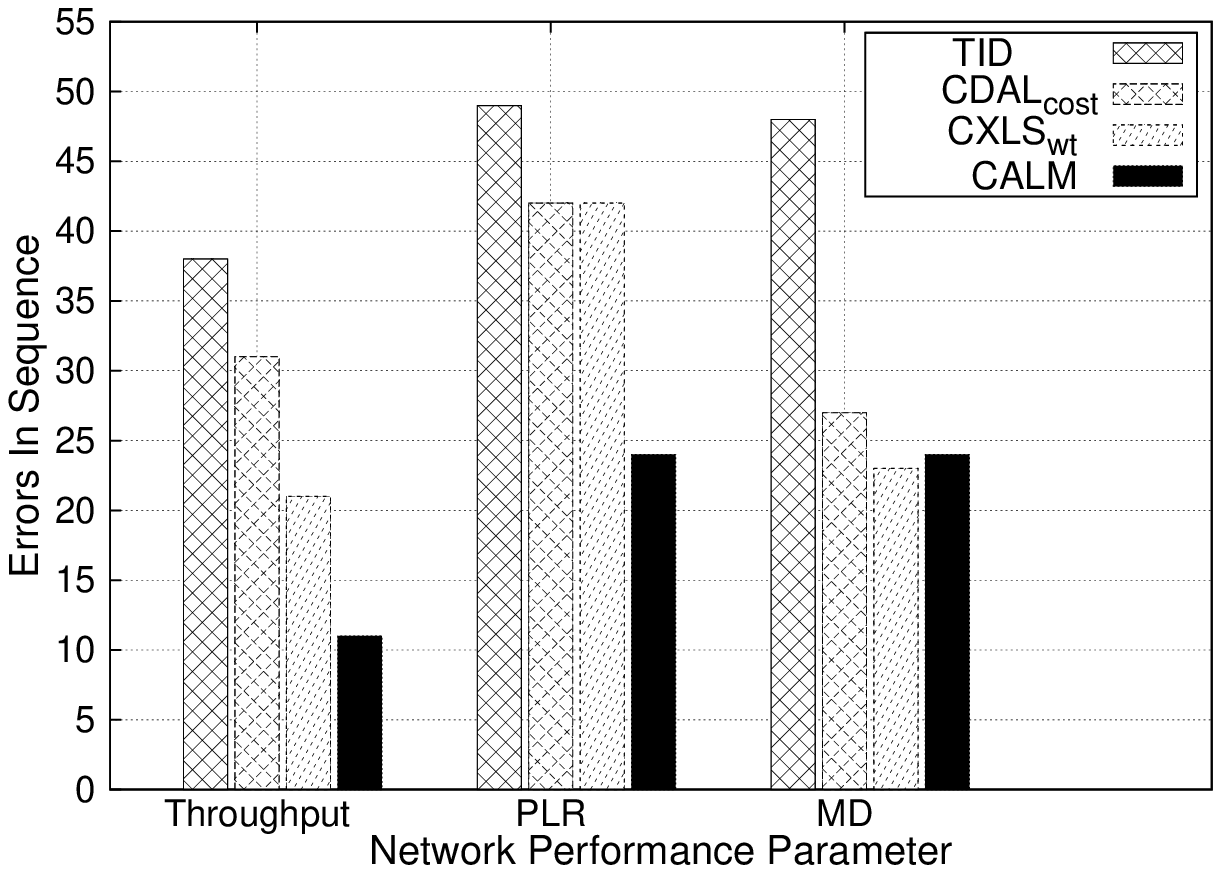}}\hfill%
   \subfloat[BBT (54 Mbps)] {\includegraphics[width=.33\linewidth]{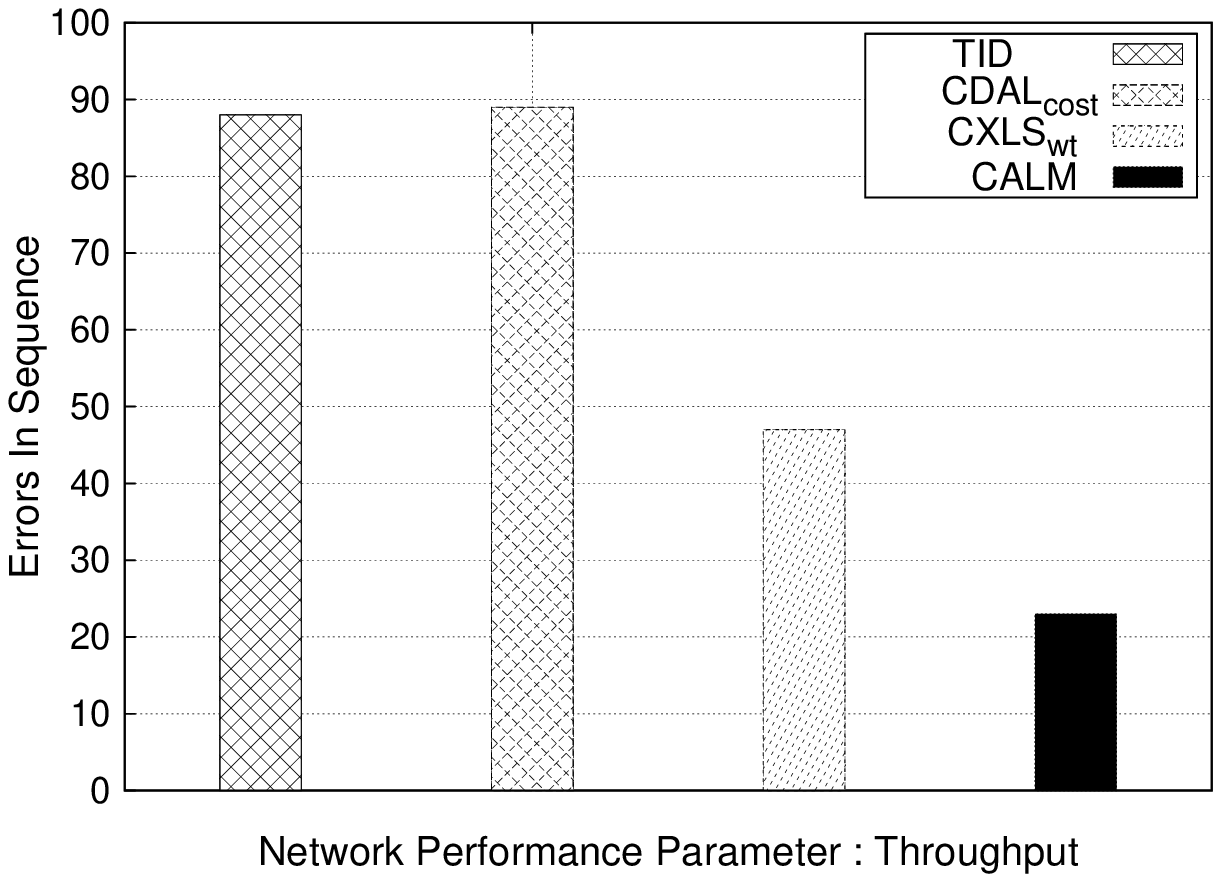}}\hfill%
       \end{tabular}
    \caption{EIS values of CA performance prediction metrics.} 
 \label{EIS}
\end{figure*}

Simulations are run for the 11 TPCA schemes denoted as $CA_n$, where $n \in \{{1\dots11}\}$. The relationship between the 4 theoretical estimates and the 3 network parameters \emph{viz.,} NAT, PLR, and MD are presented in Figure~\ref{9Th}, Figure~\ref{9P}, and Figure~\ref{9M}, respectively. Upon comparing the experimentally observed relationship with the expected correlation in Figure~\ref{cor}, we can discern that the level of conformance with the expected correlation is lowest for TID, and the highest for CALM, while CDAL$_{cost}$ plot gradients display higher conformance than TID, but lesser than CXLS$_{wt}$. Next, we compute EIS by first determining the \textit{reference CA sequences} based on experimental results, and then comparing the reference sequences with the \textit{expected CA sequences} generated by the four theoretical estimates. We illustrate this by considering NAT as the network parameter and CALM as the theoretical estimate. The reference CA sequence for NAT is created by ordering CAs 
in terms of increasing 
recorded NAT values :
 \textit{(CA$_1$ $<$ CA$_2$ $<$ CA$_3$ $<$ CA$_4$ $<$CA$_5$ $<$ CA$_6$ $<$ CA$_7$ $<$ CA$_8$ $<$ CA$_9$ $<$ CA$_{10}$ $<$ CA$_{11}$)}.
 The expected CA sequence generated by CALM estimates is : \textit{(CA$_1$ $<$ CA$_4$ $<$ CA$_2$ $<$ CA$_5$ $<$CA$_6$ $<$ CA$_3$ $<$ CA$_7$ $<$ CA$_8$ $<$ CA$_{10}$ $<$ CA$_{9}$ $<$ CA$_{11}$)}. Upon comparing the actual pairwise CA relationships with those predicted by CALM, we arrive at an EIS of $5$. Likewise, the EIS for TID, CDAL$_{cost}$ and CXLS$_{wt}$ with respect to NAT are, $17$, $9$ and $8$, respectively.  Similar reference CA sequences are determined for PLR and MD and compared to theoretical CA sequences to evaluate EIS values, which are depicted in Figure~\ref{EIS}~(a). Apart from PLR, where CXLS$_{wt}$ outperforms it, CALM proves to be the most reliable theoretical estimate.

\begin{figure*}[ht!]
  \centering%
  \begin{tabular}{cc}
    \subfloat[TID vs Throughput]{\includegraphics[width=.25\linewidth]{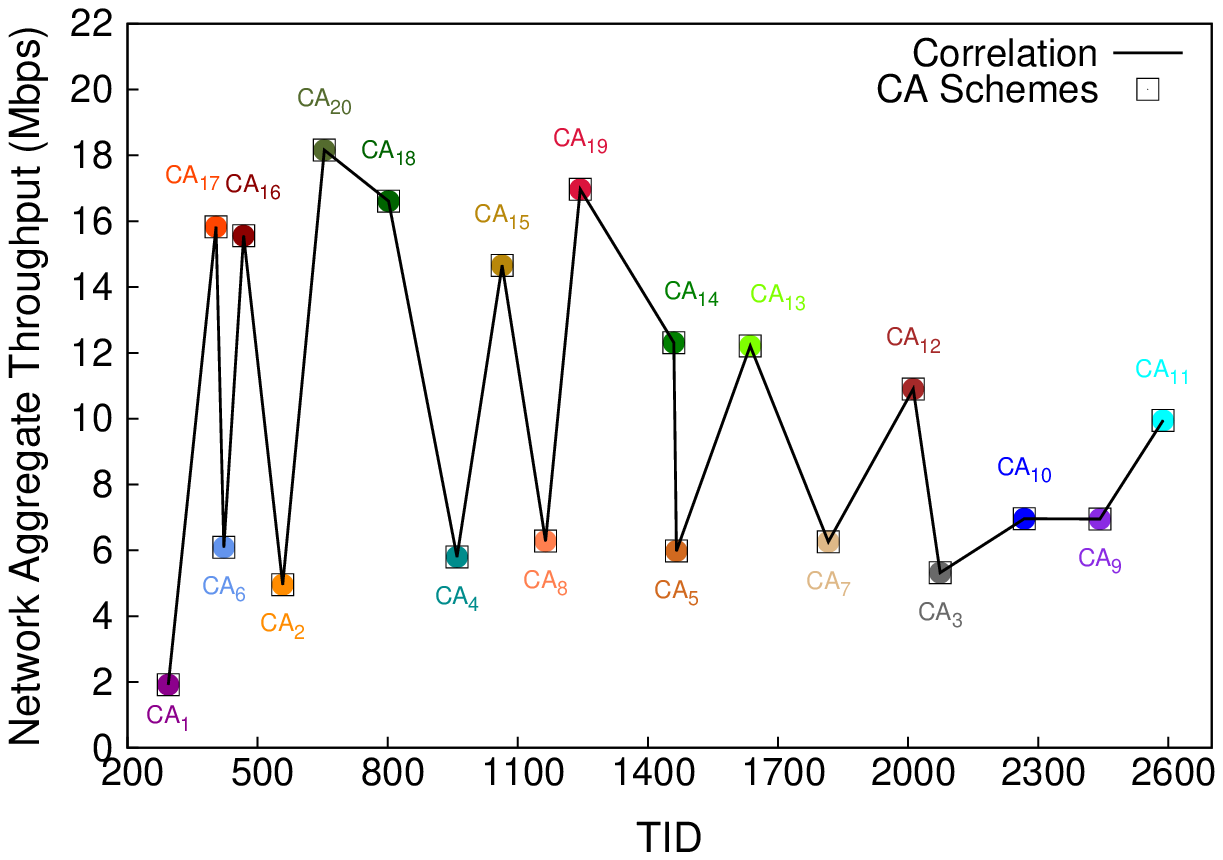}}\hfill%
    \subfloat[CDAL$_{cost}$ vs Throughput]{\includegraphics[width=.25\linewidth]{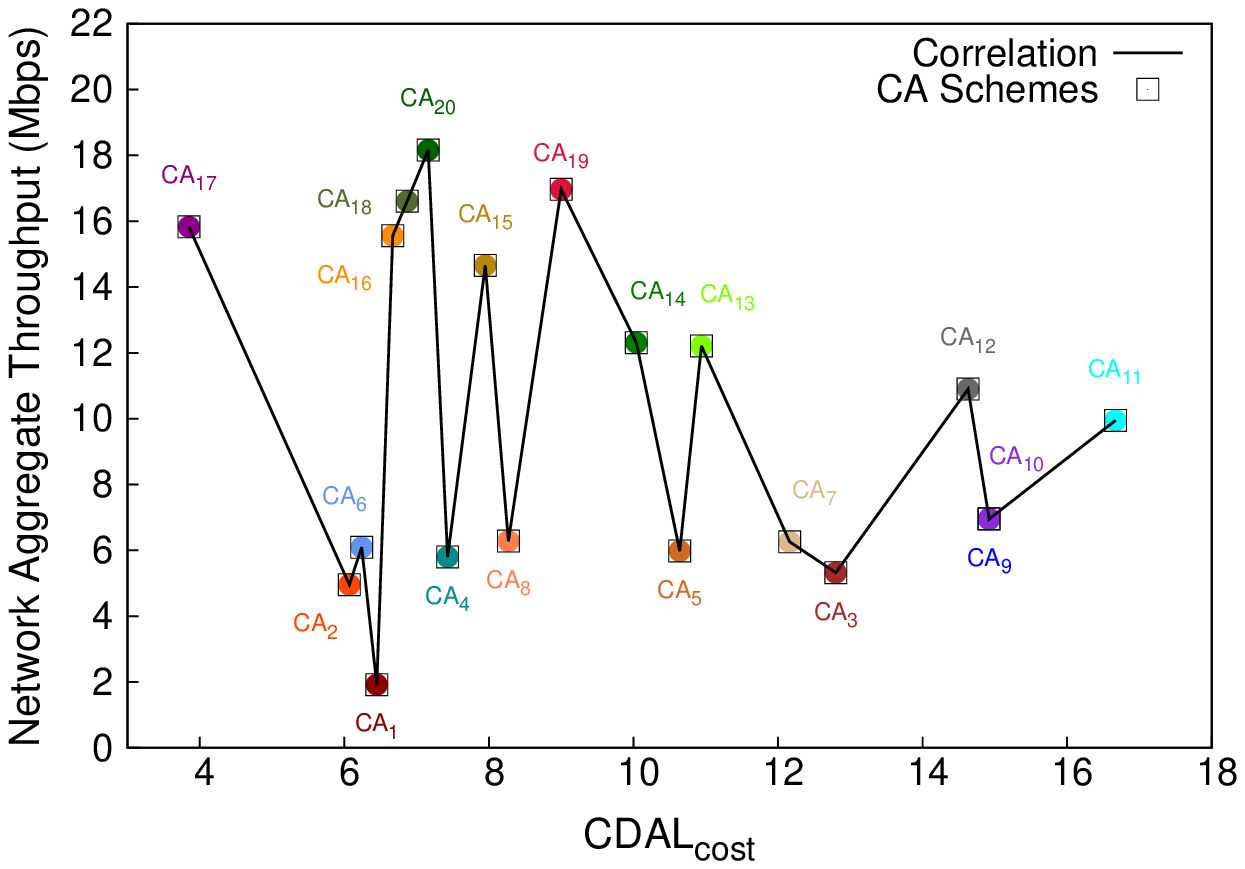}}\hfill%
    
   \subfloat[CXLS$_{wt}$ vs Throughput] {\includegraphics[width=.25\linewidth]{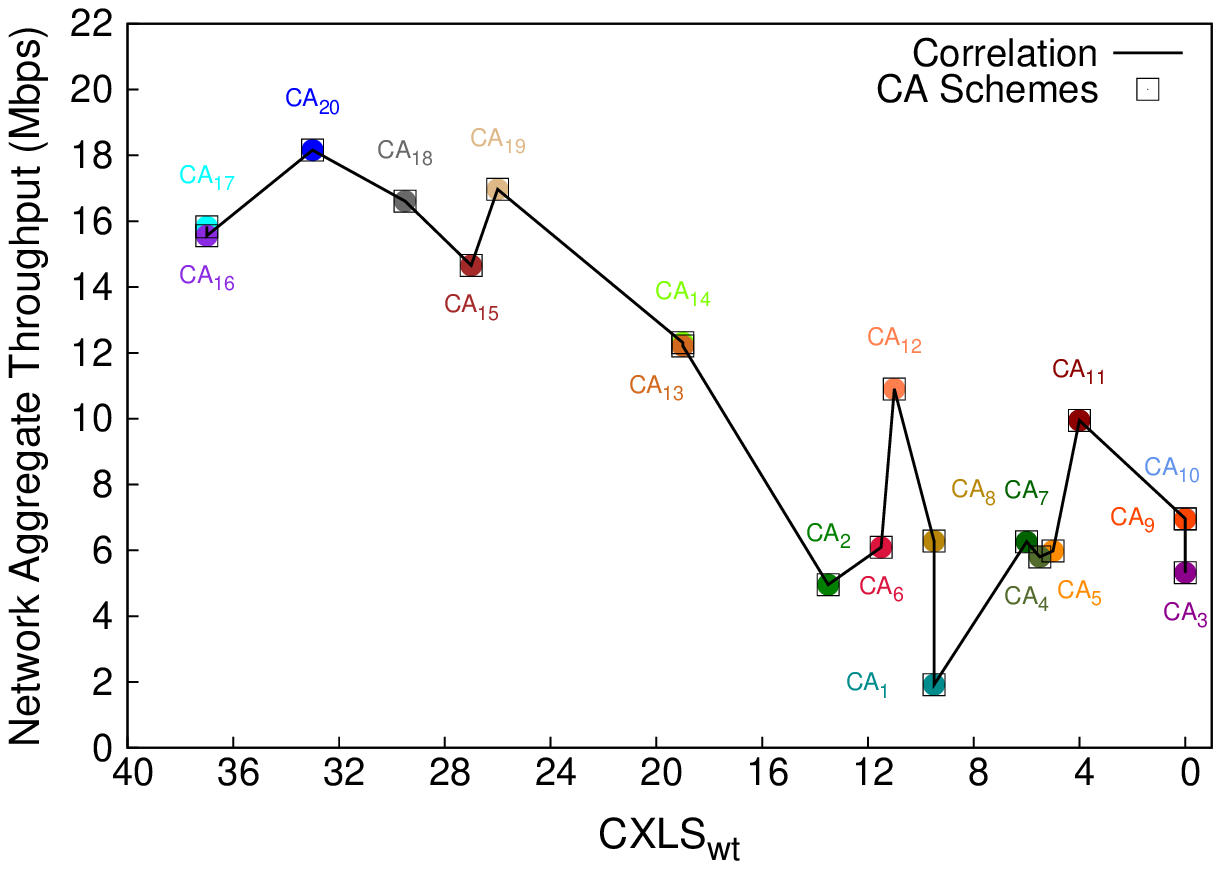}}\hfill%
   \subfloat[CALM vs Throughput] {\includegraphics[width=.25\linewidth]{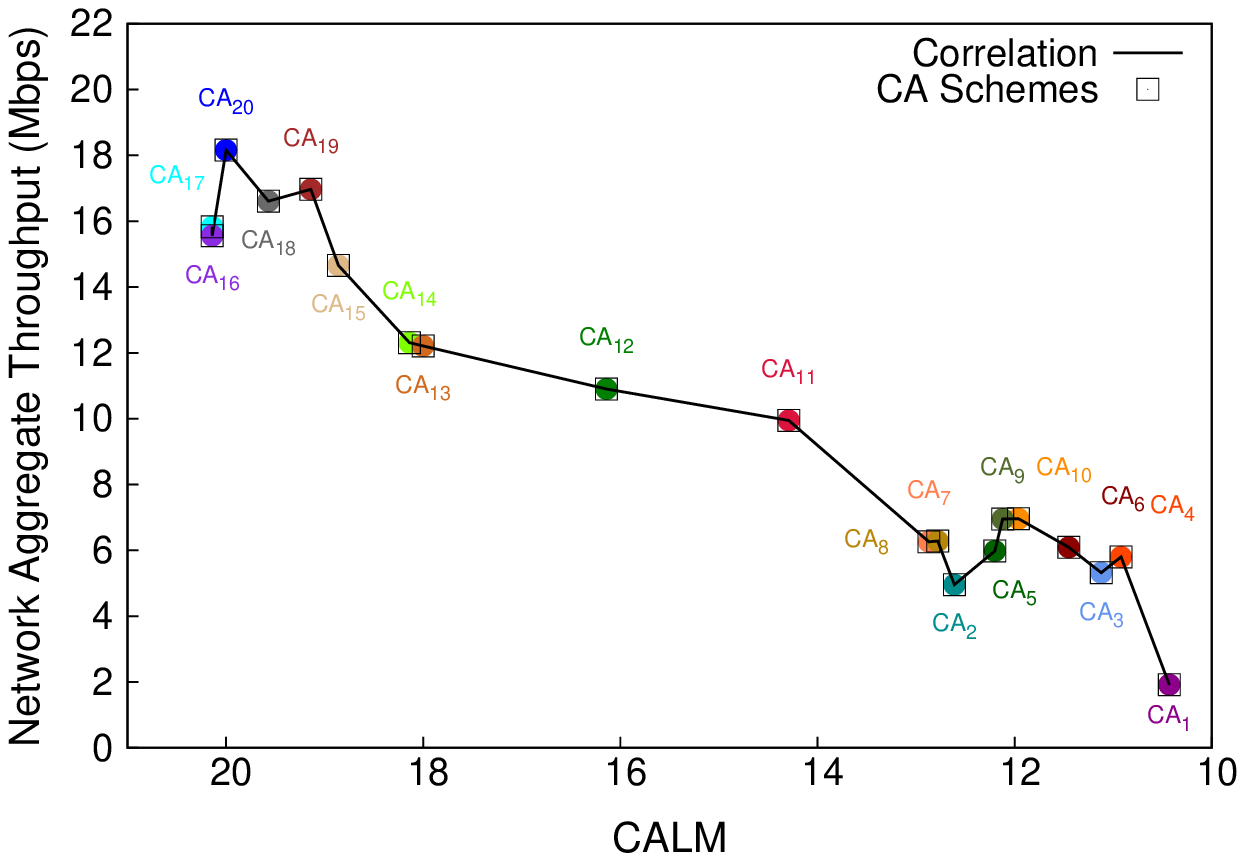}}%
    \end{tabular}
    \caption{BBT (54 Mbps) : Observed correlation of estimation metrics \& Throughput.} 
     \label{BBT}
\end{figure*}

\subsubsection{WBT$_{54Mbps}$}
With an enhanced CA test-set of 20 TPCAs, this test-case aptly highlights the performance of each of the 4 theoretical metrics. From the gradient of plots presented in Figure~\ref{54Th}, Figure~\ref{54P}, and Figure~\ref{54M}, it can be observed that the actual relationship of TID and CDAL$_{cost}$ with PLR and MD is erratic, in stark contrast to the expected correlation depicted in Figure~\ref{cor}, and it is only slightly better with NAT. CXLS$_{wt}$ plots depict a greater conformity but there exist steep abrupt deviations as well. CALM demonstrates a general overall conformance with the expected correlation despite a few distorting spikes. 
We apply the evaluation methodology to determine reference CA sequences for NAT, PLR, and MD, and compare them to theoretical CA sequences. The EIS values are presented in Figure~\ref{EIS}~(b). Again, except for MD, where CXLS$_{wt}$ performs marginally better, CALM continues to register very few CA performance prediction errors. In fact, EIS of CALM in terms of NAT is one-third that of TID, and half that of CXLS$_{wt}$.
 
\subsubsection{BBT$_{54Mbps}$}
The purpose of this test-case is to carry out a stress-test of CALM and other theoretical estimates, when the input is a set of GPCA and GDCA schemes. As discussed earlier, TID, CDAL$_{cost}$, and CXLS$_{wt}$ offer reliable interference estimates only when the CA schemes are topology preserving. Through BBT$_{54Mbps}$, we demonstrate that CALM's socio-inspired design is agnostic to the type of CA, and offers enhanced accuracy and reliability, even in case of graph disconnected CA schemes. We focus only on NAT in this test-case and the observed correlation of theoretical estimates with network capacity is depicted in Figure~\ref{BBT}. There is a greater degree of deviation from the expected correlation in plots of all four estimates, but this deviation varies. For TID and CDAL$_{cost}$, the plots are haphazard and there is no correspondence between theoretical estimates and observed NAT results. For example, CA$_{1}$ has the lowest throughput, but both TID and CDAL$_{cost}$ consider it to be a high 
performance CA, and TID expects it to have the highest NAT in the CA test-set. CXLS$_{wt}$ exhibits some adherence to the expected pattern, but for the three GDCAs \emph{viz.,} CA$_{3}$, CA$_{9}$, and CA$_{10}$, it generates an estimate of $0$. These results are anomalous, as the NATs of these three schemes are non-zero, and 3$^{rd}$, 9$^{th}$ and 10$^{th}$ highest, respectively. CALM demonstrates highest conformity to the expected correlation as it appropriately deals with link disruptions in the original WMN topology and accounts for them in its interference estimate. Thus, CA$_{1}$ is appropriately predicted to offer the lowest throughput. The EIS values presented in Figure~\ref{EIS}~(c) substantiate these arguments. Even though the EIS has doubled for CALM in the stress-test, this increase is marginal compared to the other three estimates.


%
\begin{table} [h!]
\caption{Reduction in EIS by CALM vis-a-vis other metrics.}
\centering
\tabcolsep=0.10cm
\begin{tabular}{|M{2cm}|M{1.1cm}|M{1.6cm}|M{1.5cm}|M{1.1cm}|M{1.6cm}|M{1.5cm}|M{1.1cm}|M{1.6cm}|M{1.5cm}|}
\hline 
&\multicolumn{9}{|c|}{\textbf{Reduction in EIS by CALM } ($\%$)}\\ \cline{2-10}
 \multicolumn{1}{|c|}{\textbf{Network}}&\multicolumn{3}{|c|}{\textbf{WBT$_{9Mbps}$}}&\multicolumn{3}{|c|}{\textbf{WBT$_{54Mbps}$}}&\multicolumn{3}{|c|}{\textbf{BBT$_{54Mbps}$}}\\ \cline{2-10}

\multicolumn{1}{|c|}{\textbf{Parameter}}&\textbf{TID}&\textbf{CDAL$_{cost}$}&\textbf{CXLS$_{wt}$}&\textbf{TID}&\textbf{CDAL$_{cost}$}&\textbf{CXLS$_{wt}$}&\textbf{TID}&\textbf{CDAL$_{cost}$}&\textbf{CXLS$_{wt}$}\\
\hline  
NAT	&	71	&	44 	&	38	&	71 	&	65 	&	48 &	74 	&	74 	&	51 	\\
\hline  
PLR	&	67	&	44 	&	-60	&	51 	&	43 	&	75 	&	NA	&	NA	&	NA\\
\hline  
MD	&	38	&	17	&	9	&	50 	&	11 	&	-4 	&	NA	&	NA	&	NA\\
\hline  
\end{tabular} 
\label{EIS1}

\end{table}

\begin{table} [ht!]
\caption{Accuracy of Estimation Metrics in Performance Testing.}
\centering
\tabcolsep=0.10cm
\begin{tabular}{|M{2.5cm}|M{1.3cm}|M{1.7cm}|M{1.5cm}|M{1.4cm}|M{1.3cm}|M{1.7cm}|M{1.5cm}|M{1.5cm}|M{1.5cm}|}
\hline 
&\multicolumn{8}{|c|}{\textbf{Measure of Accuracy} ($\%$)}\\ \cline{2-9}
 \multicolumn{1}{|c|}{\textbf{Network}}&\multicolumn{4}{|c|}{\textbf{WBT$_{9Mbps}$}}&\multicolumn{4}{|c|}{\textbf{WBT$_{54Mbps}$}}\\ \cline{2-9}

\multicolumn{1}{|c|}{\textbf{Parameter}}&\textbf{TID}&\textbf{CDAL$_{cost}$}&\textbf{CXLS$_{wt}$}&\textbf{CALM}&\textbf{TID}&\textbf{CDAL$_{cost}$}&\textbf{CXLS$_{wt}$}&\textbf{CALM}\\
\hline  
NAT	&	69	&	84	&	85	&	91	&	80	&	84	&	89	&	94	\\
\hline  
PLR	&	73	&	84	&	96	&	91	&	74	&	78	&	78	&	87	\\
\hline  
MD	&	71	&	78	&	80	&	82	&	75	&	86	&	88	&	87	\\

\hline  
\end{tabular} 
\label{DOC1}
\end{table}

\begin{table} [ht!]
\caption{Accuracy of Estimation Metrics in Stress Testing.}
\centering
\tabcolsep=0.10cm
\begin{tabular}{|M{2.5cm}|M{1.3cm}|M{1.7cm}|M{1.5cm}|M{1.5cm}|}
\hline 
\multicolumn{1}{|c|}{\textbf{Network}}&\multicolumn{4}{|c|}{\textbf{Measure of Accuracy} ($\%$)}\\ \cline{2-5}
\multicolumn{1}{|c|}{\textbf{Parameter}}&\textbf{TID}&\textbf{CDAL$_{cost}$}&\textbf{CXLS$_{wt}$}&\textbf{CALM}\\
\hline  
NAT&54&	53&	75&	88\\
\hline  
\end{tabular} 
\label{DOC2}
\end{table}

\subsubsection{Reliability Analysis}

If we consider the EIS tally for NAT in Figure~\ref{EIS}, CALM registers only about half the number of prediction errors than CXLS$_{wt}$, in all three test-cases. Further, we present the reduction in prediction errors demonstrated by CALM with respect to the other three metrics in Table~\ref{EIS1}. CALM offers significant reduction in EIS, registering one-fourth instances of EIS when compared to TID and CDAL$_{cost}$, and half that of CXLS$_{wt}$ in BBT$_{54Mbps}$. Finally, we compute the measure of accuracy for each estimate, which is the number of positive predictions as a percentage of total number of pairwise comparisons that are possible in the CA sequence ($\textsuperscript {n} C_2$). MoA for performance testing (WBT$_{9Mbps}$ and WBT$_{54Mbps}$) for the four estimation schemes is presented in Table~\ref{DOC1}, while for the stress-test MoA are listed in Table~\ref{DOC2}. It can be inferred that TID has the lowest accuracy in most cases while CDAL$_{cost}$ is only marginally better, and in case of 
stress testing as unreliable as TID. With respect to network capacity, CALM outperforms CXLS$_{wt}$ and the other two metrics, offering an accuracy of above 90\% in performance testing. Further, while the reliability of the three estimates drops drastically while dealing with CA schemes that don't preserve WMN topology, CALM still boasts of 88\% accuracy. In terms of PLR and MD as well, CALM offers high accuracy in predictions and is slightly behind CXLS$_{wt}$ only in two instances. Clearly, CALM is the most reliable and accurate CA performance prediction metric.

\subsection{NETCAP : Results and Analysis}
\begin{figure*}[ht!]
  \centering%
  \begin{tabular}{cc}
    \subfloat[WBT (9 Mbps)]{\includegraphics[width=.33\linewidth]{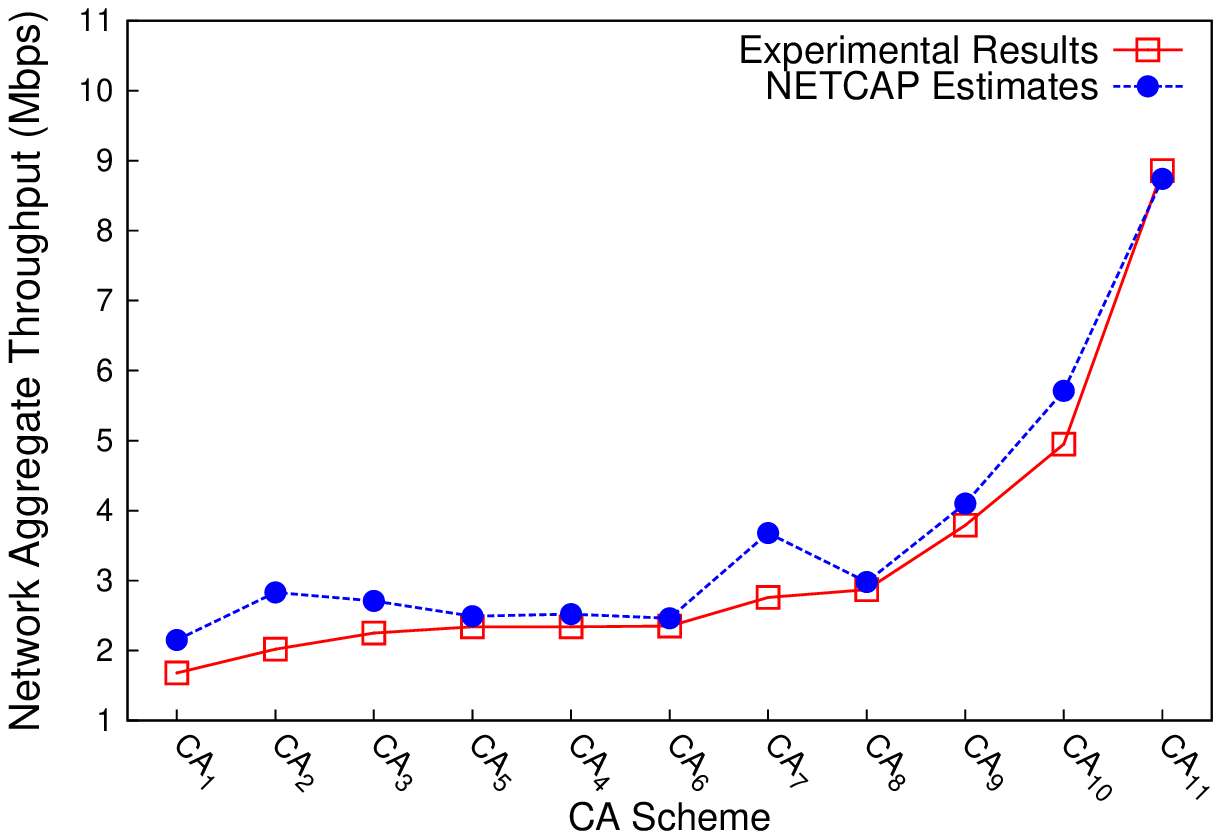}}\hfill%
    \subfloat[WBT (54 Mbps)]{\includegraphics[width=.33\linewidth]{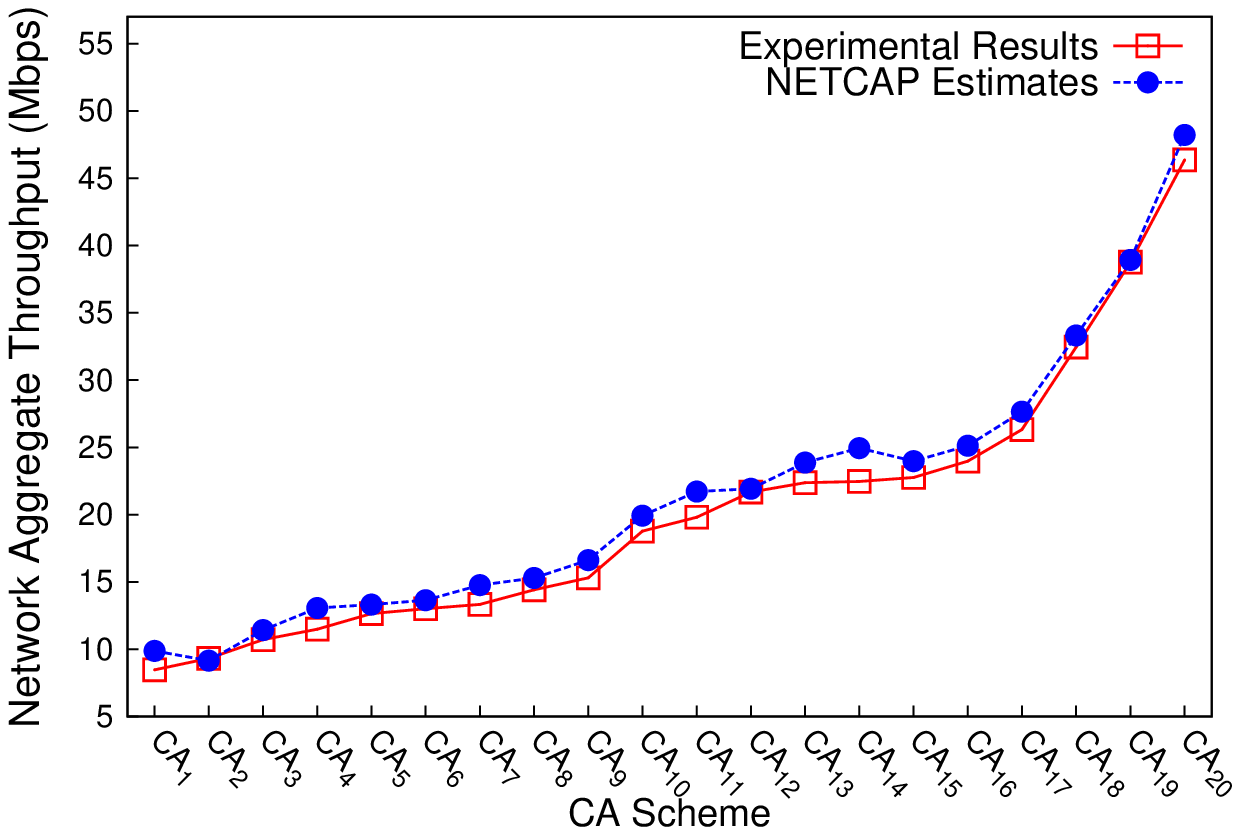}}\hfill%
   \subfloat[BBT (54 Mbps)] {\includegraphics[width=.33\linewidth]{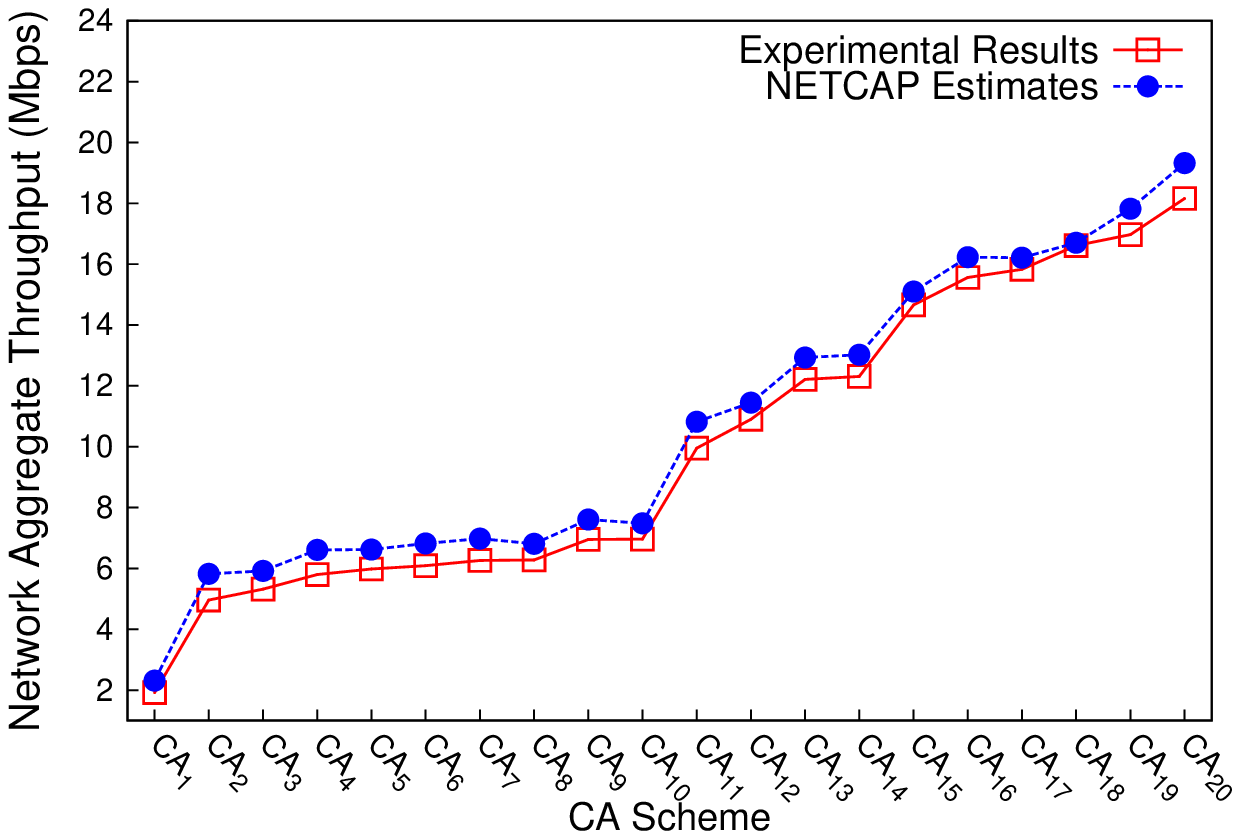}}\hfill%
      \end{tabular}
    \caption{Comparing NETCAP estimates with experimentally recorded NAT values} 
     \label{NETCAP}
\end{figure*}
\begin{table} [h!]
\caption{Statistical Analysis of NETCAP Spread$_{NC}$.}
\centering
\tabcolsep=0.10cm

\begin{tabular}{|M{2cm}|M{2.4cm}|M{2.4cm}|M{2.4cm}|}
\hline 

\multicolumn{1}{|c|}{\textbf{Test}}&\textbf{Mean}&\textbf{Standard}&\textbf{Absolute}\\
\multicolumn{1}{|c|}{\textbf{Case}}&\textbf{Spread$_{NC}$(\%)}&\textbf{Deviation(\%)}&\textbf{Range (\%)}\\
\hline  
WBT$_{9Mbps}$	&	15.1	&	12.9	&	[1.4 - 28.0]	\\
\hline  
WBT$_{54Mbps}$	&	6.4	&	4.3	&	[0.4 - 16.5]	\\
\hline  
BBT$_{54Mbps}$	&	8.5	&	4.9	&	[0.5 - 20.3]	\\

\hline  
\end{tabular} 
\label{NETCAP1}
\end{table}
The \textit{expected network capacity} estimates for each CA scheme generated by the NETCAP framework are compared to the experimentally observed NAT values for WBT$_{9Mbps}$, WBT$_{54Mbps}$, and BBT$_{54Mbps}$. The plots are presented in Figure~\ref{NETCAP}. It can be discerned that the NETCAP plot gradient is almost identical to that of recorded results, and the difference between the two, which we call the \textit{Spread$_{NC}$}, is marginal. For WBT$_{9Mbps}$, a few spikes in NETCAP estimates can be observed in Figure~\ref{NETCAP}~(a), however these anomalous surges are not seen in WBT$_{54Mbps}$ and BBT$_{54Mbps}$, which is evident from Figure~\ref{NETCAP}~(b) and Figure~\ref{NETCAP}~(c), respectively. Two aspects of these results demonstrate the ability of NETCAP to make highly accurate predictions about expected network capacity. First, the Spread$_{NC}$ is almost always positive and the two instances in which it is negative, the values are in the [0-2]\% range, which is insignificant. Second, from 
the 
analysis of Spread$_{NC}$ presented in Table~\ref{NETCAP1}, it is evident the average value of Spread$_{NC}$ is below 10\% for both, WBT$_{54Mbps}$ and BBT$_{54Mbps}$, and the dispersion in Spread$_{NC}$ is also under 5\%. The estimates of WBT$_{9Mbps}$ demonstrate greater deviations from observed results, but despite a higher average Spread$_{NC}$ the results are favorable. Further, for BBT$_{54Mbps}$ which implies an input of GPCA and GDCA schemes NETCAP exhibits high accuracy, which substantiates its estimation capability. The reason why theoretical NETCAP estimates are quite close to actual experimental results, are the Link-Weight values generated by CALM. The link quality estimates are precise in reflecting the adverse impact of interference, and assist NETCAP in approximately determining the \textit{effective link capacity}. This makes NETCAP a reliable framework for estimating maximal achievable network throughput offered by a CA scheme, and the link quality values generated by CALM play a crucial 
role in this process.

\section{Conclusions and Future Work} \label{9}
Interference estimation is an NP-hard problem which puts certain limitations on the role of a theoretical interference estimate. At best, a interference estimate may exhibit a high conformance to the expected correlation with experimentally observed performance of a CA deployed in a WMN. Further, due to the three dimensional nature of interference which includes the unpredictable temporal characteristics, it is not feasible to predict CA performance with absolute confidence. In this context, CALM proves to be a reliable CA prediction and interference estimation metric, by offering accuracy of over 90\% with respect to network capacity, over a large CA test-set of forty CA schemes. Its probabilistic link-conflict estimation mechanism, combined with an algorithm design inspired by sociological theory, enables it to produce fairly reliable estimates even for CA schemes that disrupt graph topology. Reduction in prediction errors and high accuracy of CALM substantiates the two corollaries proposed in Subsection~\
ref{D}. It also validates the proposed \textit{Sociological Ideas Borrowing Mechanism}, as the performance of CALM demonstrates a successful application of sociological concepts in wireless networks. Further, CALM enables NETCAP to generate an accurate estimate of network capacity as NETCAP offers above 90\% accuracy on an average, in two out of three test-cases. Thus, CALM and NETCAP create a robust and reliable framework for interference estimation and CA performance prediction, that works for all types of CA schemes. It simplifies and streamlines the task of appropriate CA selection for a WMN, and also offers an estimate of expected network capacity. Our work also paves the way for a greater and simplified borrowing of abstract ideas from Sociology, to wireless network paradigms in particular, and computer networks in general.

In this work, we have tested CALM and NETCAP on a planned GWMN environment, and we plan to extend this framework to a random or unplanned WMN. We also intend to introduce ideas of fairness through the use of Link-Weights generated by CALM. Although CDAL$_{cost}$ does offer a measure of absolute fairness in channel allocation and is used to test a fairness index \cite{cite4}, absolute fairness is not truly reflective of network performance \cite{tradeoff}. Thus, we plan to assess \textit{Qualified Fairness} of a CA scheme wherein fairness in channel allocation to links is dovetailed with the impact of interference. Further, In 5G network the cellular operators make use of the unlicensed spectrum \emph{i.e.,} LAA/LTE-U in the $5$ $Ghz$ band to serve the indoor users. Hence, we intend to use both CALM and NETCAP efficiently in 5G deployments as well.

\section*{References}

\bibliography{ref}

\end{document}